\pdfoutput=1
\documentclass{article}
\usepackage{iclr2027_conference,times}
\iclrfinalcopy
\usepackage{amsmath,amsfonts,bm}

\def\eqref#1{equation~\ref{#1}}
\def\1{\bm{1}}

\DeclareMathAlphabet{\mathsfit}{\encodingdefault}{\sfdefault}{m}{sl}
\SetMathAlphabet{\mathsfit}{bold}{\encodingdefault}{\sfdefault}{bx}{n}

\usepackage{hyperref}
\usepackage{url}
\usepackage{graphicx}
\usepackage{booktabs}
\usepackage{tabularx}
\usepackage{amsmath,amssymb}
\usepackage{xcolor}
\usepackage{multicol}
\usepackage{multirow}
\usepackage{longtable}
\usepackage{adjustbox}
\usepackage{newtxtext}
\usepackage{placeins}

\newcommand{\FLIP}{\textsc{flip}}
\newcommand{\STAY}{\textsc{stay}}

\newcommand{\accfs}{\mathrm{Acc}_{\mathrm{FS}}}
\newcommand{\fitwidth}[1]{\resizebox{\ifdim\width>\linewidth\linewidth\else\width\fi}{!}{#1}}
\newcommand{\rev}[1]{#1}

\title{MusicListenBench: Can Audio LLMs\\Hear the Difference Yet?}
\author{Abhinaba Roy$^{1}$, Ambuj Mehrish$^{2}$ \& Dorien Herremans$^{1}$ \\
$^{1}$Singapore University of Technology and Design \quad
$^{2}$University of Venice \\
\texttt{\{abhinaba\_roy,dorien\_herremans\}@sutd.edu.sg}, \texttt{ambuj.mehrish@unive.it}}

\begin{document}
\maketitle

\begin{abstract}
Audio large language models (LLMs) are typically evaluated on song genre, mood, and instrumentation. \rev{A more basic skill is telling whether two short clips play the same melody, chord or instrument.} We introduce \emph{MusicListenBench}, a benchmark of paired listening tests for melody, harmony, timbre, and tempo. Each test item comes with two variants. A \FLIP{} variant changes the music, so the answer must change. A \STAY{} variant changes only the sound, such as room echo or hiss, so the answer must stay the same. The two variants have opposite answers, so a model that ignores the audio scores 50\%, whatever letter it prefers. \rev{The benchmark has a training split and two test sets, one rendered from symbolic music with exact labels and one built from real recordings. On the rendered test,} seven of eight open audio LLMs score within 4 points of this floor. Of four commercial models, both GPT-Audio models are at chance, and the best, Gemini 2.5 Pro, reaches 74.2\%, 12.8 points below human listeners (87.0\%). Post-training with GRPO on 10k \rev{rendered} clean pairs raises Qwen2.5-Omni from 61.2\% to 99.0\% on held-out clean pairs, and the gain transfers to a task left out of training. \rev{On \FLIP/\STAY, however, it only reaches the level of our listeners (86.0\%).} It now misses almost no change (1.9\%, against 24.7\% for Gemini 2.5 Pro), \rev{yet when only the sound changes, it turns a correct answer into a wrong one 25.7\% of the time. More than half of its false flips come from a quiet background hiss.} Models learn to hear a difference much faster than they learn which differences matter. \rev{On real recordings, open models again stay near chance, and the trained model calls 27 of 42 pairs of two strums of the same chord different. Adding sound-only pairs to training lowers false flips on both test sets, but no training run teaches relative pitch.} We release the data\footnote{\url{https://huggingface.co/datasets/amaai-lab/MusicListenBench}}, code\footnote{\url{https://github.com/AMAAI-Lab/MusicListenBench}}, and an open leaderboard\footnote{\url{https://amaai-lab.github.io/MusicListenBench/}}.
\end{abstract}

\section{Introduction}
\label{sec:intro}

Audio large language models (audio LLMs) can describe aspects like the genre, mood and instruments of a recording \citep{yang2024airbench,sakshi2025mmau,weck2024muchomusic}. A more basic skill sits under these descriptions: telling whether aspects of two sounds are the same. Do two short clips play the same melody? The same chord? The same instrument? Which one is faster? Such same/different questions are a standard tool of hearing research \citep{hautus2021detection}, and tests of human musical ability are built from them \citep{law2012assessing,wallentin2010musical}. Our human listeners answer them correctly 87.5\% of the time. If a model cannot answer them, it is unclear what its descriptions of music are based on. \rev{MusicListenBench targets this level of listening, the level of the Musical Ear Test and PROMS, rather than the analysis of a whole song, such as its key, form or genre. This level comes first: a model that cannot tell whether two short motifs are the same cannot be trusted to describe the melody of a full song.}

Recent benchmarks find audio LLMs at or near chance on basic music and sound perception, such as telling whether a melody was shifted in pitch \citep{carone2026muse,carone2025evaluating,sun2026sonicbench}. Two problems remain. First, a same/different score mixes two skills: noticing a change that matters and ignoring a change that does not. Such scores are also inflated by text priors and by a preference for one answer letter \citep{weck2024muchomusic,zang2025you,lopez2025robustness,zheng2024large}. Second, these benchmarks only evaluate models as they are, so they cannot tell whether training fixes the problem.

\begin{figure}
    \centering
    \includegraphics[width=\textwidth]{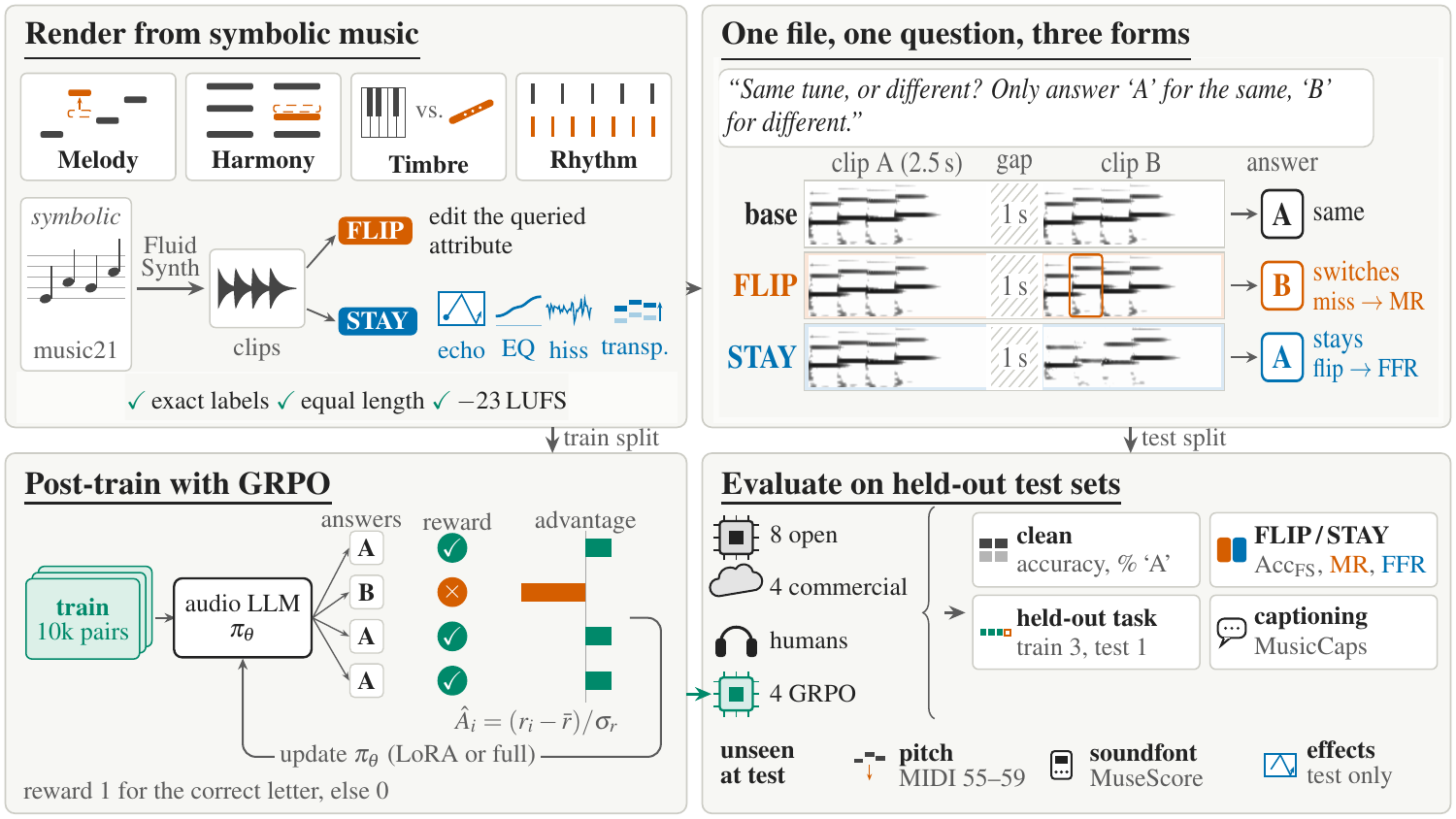}
    \caption{\textbf{MusicListenBench overview.} Items are rendered from symbolic music, so every answer is exact\rev{; a second test set builds the same items from real recordings (Section~\ref{sec:splits})}. A \FLIP{} version changes the queried attribute and the answer; a \STAY{} version changes only the sound and keeps the answer. We test open and commercial audio LLMs and human listeners under held-out conditions, and post-train open models with GRPO on clean pairs.}
    \label{fig:pipeline}
\end{figure}

We introduce MusicListenBench (Figure~\ref{fig:pipeline}). It is built on a simple idea. A model that hears the difference must do two things: change its answer when the music changes, and keep its answer when only the sound changes. So every test item comes with two variants. In the \FLIP{} variant we change the queried attribute, for example one note of the melody, and the correct answer switches. In the \STAY{} variant we change only how the music sounds, for example by adding room echo or noise, and the correct answer stays. The two variants ask the same question and have opposite answers. A model that ignores the audio therefore scores 50\%, whatever letter it prefers (exactly on the three same/different tasks and on average on rhythm; Section~\ref{sec:flipstay}). \STAY{} pairs also block a shortcut that clean pairs allow. On clean same/different pairs, a model can answer by checking whether the two clips sound alike, without knowing what a melody or a chord is. In a \STAY{} pair the clips sound different even when the answer is `same'. \rev{The main test set is rendered from symbolic music, so every answer is known exactly, and it uses pitch ranges, a soundfont and sound effects that never appear in training. A second test set builds the same kinds of item from real recordings.}

We use the benchmark to answer the title question in \rev{four} steps.
(1)~\emph{Not yet.} On \FLIP/\STAY, seven of eight open audio LLMs and both GPT-Audio models are within 4 points of the 50\% floor. The best commercial model, Gemini 2.5 Pro (74.2\%), is 12.8 points below our listeners (87.0\%; Section~\ref{sec:res-base}).
(2)~\emph{Training \rev{lifts} clean accuracy.} GRPO on our training split raises Qwen2.5-Omni from 61.2\% to 99.0\% on held-out clean pairs. It also improves a task left out of training (harmony: 59.2\% to 96.4\%) and \rev{keeps captioning close to the base model} (Sections~\ref{sec:res-grpo} and~\ref{sec:res-transfer}).
(3)~\emph{Models learn to notice change much more than to ignore it.} On \FLIP/\STAY{} the same model only reaches the level of our listeners (86.0\%). Its miss rate falls from 36.7\% to 1.9\%, but its false-flip rate only falls from 41.0\% to 26.1\%, no lower than that of Gemini 2.5 Pro without training (26.9\%)\rev{, and nearly all of them are on items it answers correctly when clean}. The model learns to ignore EQ changes it never saw in training (false flips fall from 33.7\% to 1.4\%), but not background hiss or transposition. On timbre and rhythm, where our listeners made at most one error each, hiss causes 85 of its 94 false flips. \rev{Supervised fine-tuning (SFT) gives the same pattern. The false flips follow the frozen audio encoder: on rendered clips, a quiet hiss moves its output further than a real change in the music. Adding the sound-only pairs of our training split lowers them, even for changes never seen in training (Section~\ref{sec:res-asym}).}
\rev{(4)~\emph{The same problems appear on real audio.} On the real test set, open models again stay near chance, and training on our rendered pairs still helps (52.1\% to 62.8\%). The trained model detects 38 of 41 moved thirds in real strummed chords, but also calls 27 of 42 pairs of two strums of the same chord different.}

Our contributions are (i)~MusicListenBench: 10{,}000 training items \rev{and two test sets, 3{,}000 rendered items} with exact labels, graded difficulty, checks against shortcuts and held-out test conditions\rev{, and 1{,}800 items from real recordings}, released with the generator, a scoring script and a public leaderboard (Section~\ref{sec:leaderboard}); (ii)~the \FLIP/\STAY{} protocol, which separates noticing a change that matters from ignoring one that does not; (iii)~an evaluation of eight open and four commercial audio LLMs against human listeners; and (iv)~a \rev{training} study on four models\rev{, with GRPO and SFT,} covering transfer to an unseen task and side effects on captioning\rev{, the frozen audio encoder, training on sound-only pairs, and a repeat on Slakh2100 (Appendix~\ref{app:slakh})}.

\section{Related work}
\label{sec:related}

\paragraph{Benchmarks for audio LLMs.} AIR-Bench, MMAU and MMAR test audio LLMs on thousands of questions about speech, sound and music \citep{yang2024airbench,sakshi2025mmau,ma2026mmar}, and MuChoMusic focuses on music \citep{weck2024muchomusic}. 
They use real audio and questions written by people or LLMs, which makes them broad but lets text priors and letter bias enter the scores \citep{weck2024muchomusic,rouditchenko2025omni,zang2025you,lopez2025robustness}.
\rev{PitchBench~\citep{pitchbench} tests pitch hearing in 28 controlled experiments that vary loudness, note length, sound source, time stretching and background noise. CMI-Bench~\citep{ma2025cmibench} turns annotations of music information retrieval (MIR) datasets into instruction tasks on real recordings, and scores them with standard MIR metrics.} Three benchmarks are closest to ours. MUSE tests ten music perception skills, including same/different judgments, on 200 recorded stimuli with 200 human listeners \citep{carone2026muse}. \citet{carone2025evaluating} compare MIDI and audio input on syncopation, transposition and chord quality. SonicBench renders controlled sounds for twelve physical attributes, including timbre and tempo, and puts both clips in one file as we do \citep{sun2026sonicbench}. All three evaluate models only as they are, without training. We differ in three ways. Every item has a \FLIP{} and a \STAY{} variant, so we can separate noticing a change from ignoring one, and a model that ignores the audio scores 50\%. All 3{,}000 \rev{rendered} test items have exact labels and graded difficulty\rev{, and a second test set from real recordings uses the same construction}. And we provide a training split with held-out test conditions, so the benchmark can measure what post-training changes. 
Live arenas such as Music Arena \citep{kim2026music} rank text-to-music systems by human
votes, which scores what a model produces rather than what it hears.

\paragraph{Paired tests.} Same/different and two-choice designs are standard in hearing research \citep{hautus2021detection}. The Musical Ear Test asks listeners whether two short melodies are the same when one note may have changed \citep{wallentin2010musical}, much like our melody task, and PROMS uses same/different trials for melody, tempo and timbre \citep{law2012assessing}. For models, MelodySim and MERIT learn whether two audio clips are similar in melody or in other musical aspects \citep{lu2025melodysim,roy2026merit}; we instead ask an audio LLM to make this judgement in words. In language and vision, minimal pairs \citep{warstadt2020blimp}, contrast sets \citep{gardner2020evaluating} and paired image-caption tests \citep{thrush2022winoground} check whether a model's answer follows the right change in the input. MMVP and NaturalBench pair images that need different answers and give credit only when all answers in a pair or group are right \citep{tong2024eyes,li2024naturalbench}. CheckList separates directional tests, where an edit should change the output, from invariance tests, where it should not \citep{ribeiro2020beyond}. \FLIP{} and \STAY{} are the listening versions of these two test types. Pairing them also cancels the option-letter bias that language models show on multiple-choice questions \citep{zheng2024large}.

\paragraph{Reinforcement learning for audio and music models.} \citet{herremans2026aligning} review preference alignment for music generation and its open problems, among them how reliable a preference model can be when musical quality is subjective, and inference-time alignment improves symbolic music generation \citep{roy2026text2midi}. Such work optimizes what a model writes; we ask whether a model can hear, the skill any such judge needs first. Group relative policy optimization (GRPO; \citealp{shao2024deepseekmath}) has improved audio question answering, for example for Qwen2-Audio on MMAU \citep{li2025reinforcement}.

\section{MusicListenBench}
\label{sec:bench}

\subsection{Items and tasks}
Each item is one audio file and one question. The file contains clip A, 1.0\,s of silence and clip B. Putting both clips in one file, as SonicBench does \citep{sun2026sonicbench}, lets us test models that accept only one audio input. After GRPO, one model answers 99.0\% of clean items correctly in this format (Section~\ref{sec:res-grpo}), so the format itself does not stop models from comparing the clips. The question is one of 8 to 10 hand-written wordings per task. It ends with a fixed instruction that says which letter means which answer, and the model answers with one letter. Table~\ref{tab:tasks} lists the four tasks. Melody, harmony and timbre ask whether the two clips are the same. Rhythm asks which clip is faster (some wordings ask which is slower). Melody, harmony and rhythm have several difficulty levels, and test items are spread evenly across them.
\begin{table}[t]
  \caption{The four evaluation tasks. \emph{Difference}: how the clips differ when they differ in the queried attribute (for rhythm, how the faster clip differs). \emph{Clip}: length of one clip. Each task has 2{,}500 training and 250 clean test items.}
  \label{tab:tasks}
  \centering
  \small
  \begin{tabularx}{\linewidth}{@{}l l >{\raggedright\arraybackslash}X >{\raggedright\arraybackslash}p{1.15in} c@{}}
    \toprule
    Task    & Question              & Difference                                       & Levels                            & Clip \\
    \midrule
    Melody  & Same melody?          & one note of a four-note motif moves              & 100, 200, 400, 700 cents          & 2.5\,s \\
    Harmony & Same chord quality?   & the third of the chord moves                     & 100, 200, 300 cents               & 2.0\,s \\
    Timbre  & Same instrument?      & instrument family (piano, guitar, violin, flute) & one (categorical)                 & 2.5\,s \\
    Rhythm  & Which clip is faster? & click rate of a hi-hat track                     & rate ratio 1.10, 1.20, 1.35, 1.50 & 4.0\,s \\
    \bottomrule
  \end{tabularx}
\end{table}

\subsection{Exact labels and no easy cues}
We write each item as symbolic music with music21 \citep{cuthbert2010music21} and render it with FluidSynth \citep{fluidsynth} and General MIDI soundfonts. From each generated item we render a small set of files: the base pair, one edited clip that changes the answer, and two or three clips that change only the sound. Because the audio is generated, every answer is known exactly, and no human labeling is needed. We also remove easy cues. All clips in a task have exactly the same length, so a faster rhythm clip has more clicks, not a shorter file. All clips have the same loudness ($-23$\,LUFS). \rev{As in the standard same/different design, the two clips of a clean `same' pair are identical, so clean pairs are the easiest version of the test.} Appendix~\ref{app:construction} gives all settings.
 For rhythm we leave out onset count, because counting clicks is the task itself. These checks cover cues within the file. A cue in the relation between the two clips, such as how alike they sound, is what the \STAY{} variants test (Section~\ref{sec:flipstay}).

\subsection{\texorpdfstring{\FLIP{} and \STAY: two ways to fail}{FLIP and STAY: two ways to fail}}
\label{sec:flipstay}
For each of the 1{,}000 clean test items we build two variants by replacing clip B.
\begin{itemize}
  \item \FLIP: clip B is replaced by an edited clip that changes the queried attribute, such as one moved note, a moved third or another instrument. For rhythm, the two clips swap places. The correct answer switches.
  \item \STAY: clip B is replaced by a clip that sounds different but keeps the queried attribute. We use four such changes: room echo (reverberation time 0.2, 0.5 or 1.0\,s, simulated with pyroomacoustics; \citealp{scheibler2018pyroomacoustics}), a mild EQ change ($\pm 3$\,dB shelves at 200\,Hz and 4\,kHz), background hiss at $-50$\,dBFS and, for melody and harmony only, a transposition of up to 200 cents. Room echo, EQ and noise are also among the degradations that music restoration models learn to remove \citep{melechovsky2025sonicmaster}. The correct answer stays.
\end{itemize}
Transposition tests relative pitch. A melody moved up as a whole is still the same melody \citep{dowling1978scale}, but a melody with one moved note is not. The four \STAY{} changes are spread evenly over the items.

We report two error rates. The miss rate (MR) is the share of \FLIP{} items answered wrongly: the model keeps the old answer although the music changed. The false-flip rate (FFR) is the share of \STAY{} items answered wrongly: the model changes its answer although only the sound changed. Their average gives the \FLIP/\STAY{} accuracy, $\accfs = 1 - (\mathrm{MR} + \mathrm{FFR})/2$. This score has a useful property. The two variants of an item have opposite answers. On melody, harmony and timbre, where `A' always means same, they also have opposite letters, so a model that ignores the audio gets exactly one of them right and scores exactly 50\%, whatever letter it prefers. Rhythm questions ask for the faster or the slower clip, and each variant draws its own wording, so in 131 of 250 rhythm items both variants share the correct letter. There, a model that ignores the audio scores 50\% only on average: always answering `A' gives 47.8\% on rhythm and 49.45\% overall, against 52.8\% on the clean test.

The two rates catch different failures. On the three same/different tasks, a model that answers `different' whenever the two clips sound different rarely misses a \FLIP{} change, but it false-flips on every \STAY{} pair whose answer is `same'. Each rate alone also moves with letter bias (always answering `A' gives an MR of 53.0\% and an FFR of 48.1\%), but their average stays near 50\%. \rev{FFR counts every wrong \STAY{} answer, including errors that a model already makes on the clean pair. For example, a model that answers `same' to a clean pair that differs will likely answer `same' again after hiss is added, although the hiss changed nothing. To count only new errors, we also report $\mathrm{FFR}_c = P(\text{\STAY{} wrong} \mid \text{clean right})$, the false-flip rate on items whose clean version is answered correctly. Appendix~\ref{app:paired} gives the same measure for misses, and a score that counts an item only when its clean, \FLIP{} and \STAY{} versions are all right.}

\subsection{Splits and held-out conditions}
\label{sec:splits}
The training split has 10{,}000 items (2{,}500 per task)\rev{, each with sound-only versions (Appendix~\ref{app:construction})}. We use it only for post-training. The \rev{rendered test set} has the 1{,}000 clean items and the 2{,}000 \FLIP/\STAY{} items. Files made from one generated item never appear in both splits. The test \rev{set} also differs from training in three ways.
\begin{itemize}
  \item \emph{Pitch range.} Melody and timbre items start from MIDI notes 60 to 72 in training and 55 to 59 at test. Harmony roots come from MIDI 48 to 60 in training and 62 to 67 at test.
  \item \emph{Soundfont.} Training audio is rendered with FluidR3\_GM \citep{fluidr3gm} or GeneralUser GS \citep{generalusergs}, and all test audio with MuseScore\_General \citep{musescoregeneral}.
  \item \emph{Sound effects.} Room echo, EQ and hiss appear only at test\rev{. Our main runs train on clean pairs only; one control run adds sound-only versions, but never room echo, EQ or hiss (Section~\ref{sec:res-asym}).}
\end{itemize}
In test items the letters are fixed (A means same), and 528 of the 1{,}000 answers are `A'. In training items we swap the meaning of the letters for about half of the items, so the letter never predicts the answer. The four tasks also allow a leave-one-task-out test: train on three tasks and test on the fourth.

\rev{\paragraph{Real test set.} A second test set is built the same way from real recordings and is never used for training (600 clean and 1{,}200 \FLIP/\STAY{} items). Melody and harmony use guitar recordings from GuitarSet \citep{xi2018guitarset}: we pitch-shift one note, or only the strings that play the third of a chord. Timbre uses solo recordings of seven instruments from Medley-solos-DB \citep{lostanlen2018medley}, and the two clips of every pair come from different recordings. Rhythm uses time-stretched drum stems from MUSDB18-HQ \citep{rafii2019musdb18hq}. In melody and harmony `same' pairs, the audio passes through the pitch shifter at zero cents, so both kinds of pair are processed alike. (Appendix~\ref{app:real})}

\subsection{Protocol and public leaderboard}
\label{sec:leaderboard}
MusicListenBench comes with a public leaderboard\footnote{\url{https://amaai-lab.github.io/MusicListenBench/}}, and anyone can submit a model. The site reports clean accuracy per task, \FLIP/\STAY{} accuracy, MR, FFR and the share of `A' answers \rev{for both test sets}, so letter bias is always visible. It lists zero-shot models and models that use our training split separately (Appendix~\ref{app:leaderboard}). Because the \rev{rendered} test items come from a generator, we can draw a fresh test set with new seeds if the public one becomes stale.

\section{Experimental setup}
\label{sec:setup}

\paragraph{Models.} We test eight open audio LLMs: Qwen2-Audio \citep{chu2024qwen2}, Qwen2.5-Omni \citep{xu2025qwen3}, Audio Flamingo 3 \citep{ghosh2026audio}, Phi-4-multimodal \citep{abouelenin2025phi}, Kimi-Audio \citep{ding2025kimi}, MiMo-Audio \citep{xiaomi2025mimo}, Fun-Audio-Chat \citep{team2025fun} and Step-Audio~2 \rev{mini} \citep{wu2025step}. We also test four commercial models through their APIs: GPT-Audio-1.5 \citep{openai2026gptaudio15}, GPT-Audio-mini \citep{openai2026gptaudio15}, Gemini 2.5 Pro \citep{comanici2025gemini} and Gemini 3.8 Flash \citep{gemini38flash}. For open models we read the answer by comparing the probabilities of the tokens `A' and `B'; reading the generated letter instead gives the same answer on every item. For the API models we use the generated letter. \rev{A reply without a letter counts as `A' in Table~\ref{tab:base} and as an error on the real test set. Appendix~\ref{app:paired} scores the rendered test both ways; no conclusion changes.} Appendix~\ref{app:models} lists all checkpoints and API settings.

\paragraph{Human reference.} \rev{The test was open to anyone, whatever their musical background.} Twelve listeners, from casual listeners to musicians with years of practice, answered 24 items each: 288 answers in total, 72 per task and 96 each from the clean, \FLIP{} and \STAY{} tests. They heard the same audio and questions as the models, over headphones (Appendix~\ref{app:human}).

\paragraph{GRPO.} We post-train four open models with GRPO: Qwen2-Audio, Qwen2.5-Omni, Audio Flamingo 3 and Phi-4-multimodal. We use LoRA \citep{hu2021lora} or full fine-tuning. The reward is $1$ for the correct letter and $0$ otherwise, plus a small bonus ($0.2$) for producing a parseable option \rev{and a smaller term (at most $0.1$) that favours the letter the model currently gives less often, against letter collapse} (group size 8; learning rate $10^{-5}$ for LoRA and $10^{-6}$ for full fine-tuning; 8 prompts per step, i.e., 64 sampled rollouts\rev{, and 24 prompts for Qwen2.5-Omni with LoRA}). We report the checkpoint at step 1{,}000, except for Audio Flamingo 3 with full fine-tuning, where validation accuracy peaks at step 500 before the model over-optimizes the binary reward (rising KL and entropy collapse), so we stop there. Appendix~\ref{app:grpo} gives all settings.

\paragraph{Leave-one-task-out.} For Qwen2-Audio and Qwen2.5-Omni we train LoRA GRPO for 500 steps on three tasks (7{,}500 items) and test on the 250 clean items of the fourth task. We use the final checkpoint and do not look at test accuracy to choose it.

\paragraph{Captioning.} To check for side effects, Qwen2-Audio and Qwen2.5-Omni caption the 2{,}735 available clips of the MusicCaps evaluation split \citep{agostinelli2023musiclm} before and after GRPO; dedicated music captioning models are trained for this task \citep{chopra2025sonicverse}. We compare each caption with the reference using BLEU-4, METEOR, ROUGE-L and BERTScore-F1 \citep{papineni2002bleu,banerjee2005meteor,lin2004rouge,zhang2019bertscore}, on the same clips for every model (Appendix~\ref{app:caption}).

\paragraph{Chance level.} A score is not significantly different from 50\% (two-sided, $p > 0.05$, normal approximation) between 43.8\% and 56.2\% for 250 items, and between 45.6\% and 54.4\%, 46.9\% and 53.1\%, and 47.8\% and 52.2\% for 500, 1{,}000 and 2{,}000 items.

\section{Results}
\label{sec:results}

\subsection{Can audio LLMs hear the difference?}
\label{sec:res-base}

\begin{table}[t]
  \caption{Results without training (\%). \emph{Clean}: accuracy on 250 items per task. \emph{\FLIP/\STAY}: accuracy ($\accfs$), miss rate (MR) and false-flip rate (FFR) on 2{,}000 items; lower is better for MR and FFR. Bold: best open model per column; underlined: best commercial model. The last two rows are references; the human row has 24 clean answers per task and 96 \FLIP{} and 96 \STAY{} answers in total.}
  \label{tab:base}
  \centering
  \small
  \setlength{\tabcolsep}{4.5pt}
  \begin{tabular}{@{}l rrrrr rrr@{}}
    \toprule
                       & \multicolumn{5}{c}{Clean accuracy} & \multicolumn{3}{c}{\FLIP/\STAY} \\
    \cmidrule(lr){2-6}\cmidrule(l){7-9}
    Model              & Mel. & Har. & Tim. & Rhy. & All & $\accfs$ & MR & FFR \\
    \midrule
    \multicolumn{9}{@{}l}{\emph{Open models}} \\
    Qwen2-Audio        & 46.0 & 50.4 & 43.2 & 49.2 & 47.2 & 50.7 & 47.7 & 50.9 \\
    Qwen2.5-Omni       & \textbf{68.8} & \textbf{59.2} & \textbf{66.8} & 50.0 & \textbf{61.2} & \textbf{61.2} & \textbf{36.7} & \textbf{41.0} \\
    Audio Flamingo 3   & 51.2 & 48.8 & 58.0 & 52.8 & 52.7 & 51.4 & 49.6 & 47.6 \\
    Phi-4-multimodal   & 54.4 & 47.2 & 56.8 & 50.8 & 52.3 & 49.1 & 53.5 & 48.3 \\
    Kimi-Audio         & 57.6 & 47.2 & 53.2 & 52.8 & 52.7 & 51.0 & 52.3 & 45.7 \\
    MiMo-Audio         & 54.0 & 54.4 & 58.0 & \textbf{59.6} & 56.5 & 52.2 & 49.4 & 46.1 \\
    Fun-Audio-Chat     & 50.0 & 51.6 & 60.0 & 49.6 & 52.8 & 53.6 & 46.6 & 46.1 \\
    Step-Audio 2       & 52.4 & 50.0 & 62.8 & 52.4 & 54.4 & 52.7 & 49.1 & 45.5 \\
    \midrule
    \multicolumn{9}{@{}l}{\emph{Commercial models (API)}} \\
    GPT-Audio-mini     & 47.6 & 51.6 & 54.4 & 50.8 & 51.1 & 49.6 & 50.6 & 50.3 \\
    GPT-Audio-1.5      & 45.6 & 47.6 & 52.0 & 44.0 & 47.3 & 52.1 & 45.8 & 50.1 \\
    Gemini 2.5 Pro     & \underline{80.0} & \underline{68.8} & \underline{88.4} & \underline{79.6} & \underline{79.2} & \underline{74.2} & \underline{24.7} & \underline{26.9} \\
    Gemini 3.8 Flash   & 53.6 & 54.4 & 85.6 & 72.4 & 66.5 & 64.0 & 33.0 & 39.1 \\
    \midrule
    Human (12)         & 79.2 & 87.5 & 95.8 & 87.5 & 87.5 & 87.0 & 5.2  & 20.8 \\
    Always `A'         &      &      &      &      & 52.8 & 49.5 & 53.0 & 48.1 \\
    \bottomrule
  \end{tabular}
\end{table}

Table~\ref{tab:base} answers the title question: not yet. On clean pairs, \rev{the easiest version of the test,} seven of the eight open models score between 47.2\% and 56.5\%, close to the 52.8\% of always answering `A'. \FLIP/\STAY{} removes the benefit of preferring a letter. Five open models land inside the chance range (49.1\% to 52.2\%), Fun-Audio-Chat and Step-Audio~2 are just above it (53.6\% and 52.7\%), and only Qwen2.5-Omni is far above it (61.2\%). Both GPT-Audio models are inside the chance range (49.6\% and 52.1\%). Gemini 3.8 Flash reaches 64.0\% and Gemini 2.5 Pro 74.2\%, the best score of any untrained model, yet still 12.8 points below human listeners (87.0\%; $p < 0.001$, Fisher exact test).

\rev{Several open models mostly give one letter. Qwen2-Audio answers `B' for 98.6\% of clean items, and Phi-4-multimodal and Audio Flamingo 3 answer `A' for 95.9\% and 87.7\% (Figure~\ref{fig:bias} and Table~\ref{tab:app-bias} in Appendix~\ref{app:full}). If we average their accuracy over the two answer classes, all three are at chance (49.9\%, 49.7\% and 50.6\%).}

Per-task clean scores also carry this bias, so we compare tasks on \FLIP/\STAY{} (Table~\ref{tab:app-task-fs} in Appendix~\ref{app:full}). Timbre is where most models rise above chance: three open and three commercial models are clearly above 50\% on timbre, but only Qwen2.5-Omni and the two Gemini models are on melody. Gemini 3.8 Flash reaches 84.2\% on timbre but only 57.2\% on melody and 51.2\% on harmony, where our listeners score 85.4\% and 66.7\%. Gemini 2.5 Pro is more even (73.2\% on melody, 63.8\% on harmony), but it trails our listeners most on the two tasks they find easiest, timbre (82.4\%) and rhythm (77.4\%, against 97.9\% for both). Most of its errors are `same' answers to clips that differ: it detects about half of the moved notes, whether a note moves by 100 or by 700 cents, and hears piano and guitar as the same instrument in 34 of 36 pairs (Appendix~\ref{app:api-items}).

\subsection{Can post-training close the gap?}
\label{sec:res-grpo}

\begin{table}[t]
  \caption{GRPO results (\%) on the held-out test sets. \emph{none}: no training. $^\dagger$Step 500; all other runs step 1{,}000. Letter bias for the same runs is in Figure~\ref{fig:bias} and Appendix~\ref{app:full}.}
  \label{tab:grpo}
  \centering
  \small
  \setlength{\tabcolsep}{4pt}
  \begin{tabular}{@{}ll rrrrr rrr@{}}
    \toprule
                     &                & \multicolumn{5}{c}{Clean accuracy} & \multicolumn{3}{c}{\FLIP/\STAY} \\
    \cmidrule(lr){3-7}\cmidrule(l){8-10}
    Model            & GRPO           & Mel. & Har. & Tim. & Rhy. & All & $\accfs$ & MR & FFR \\
    \midrule
    Qwen2-Audio        & none           & 46.0 & 50.4 & 43.2 & 49.2 & 47.2 & 50.7 & 47.7 & 50.9 \\
                       & LoRA           & 70.8 & 60.8 & 92.4 & 51.6 & 68.9 & 63.8 & 29.7 & 42.7 \\
                       & full           & 70.4 & 61.6 & 80.8 & 52.0 & 66.2 & 61.5 & 31.7 & 45.4 \\
    \midrule
    Qwen2.5-Omni       & none           & 68.8 & 59.2 & 66.8 & 50.0 & 61.2 & 61.2 & 36.7 & 41.0 \\
                       & LoRA           & 99.6 & 96.8 & 99.6 & 100.0 & 99.0 & 86.0 & 1.9 & 26.1 \\
                       & full           & 100.0 & 85.2 & 100.0 & 99.6 & 96.2 & 83.5 & 4.2 & 28.9 \\
    \midrule
    Audio Flamingo 3   & none           & 51.2 & 48.8 & 58.0 & 52.8 & 52.7 & 51.4 & 49.6 & 47.6 \\
                       & LoRA           & 57.2 & 74.8 & 97.6 & 50.8 & 70.1 & 64.4 & 29.7 & 41.4 \\
                       & full$^\dagger$ & 68.0 & 80.8 & 99.6 & 69.2 & 79.4 & 70.4 & 22.5 & 36.7 \\
    \midrule
    Phi-4-multimodal   & none           & 54.4 & 47.2 & 56.8 & 50.8 & 52.3 & 49.1 & 53.5 & 48.3 \\
                       & LoRA & 52.0 & 49.6 & 56.8 & 50.8 & 52.3 & 49.1 & 53.4 & 48.4 \\
                       & full           & 55.6 & 54.0 & 88.8 & 50.8 & 62.3 & 58.9 & 42.9 & 39.3 \\
    \bottomrule
  \end{tabular}
\end{table}

GRPO raises clean accuracy in seven of the eight runs, by 10.0 to 37.8 points (Table~\ref{tab:grpo}). Qwen2.5-Omni reaches 99.0\% with LoRA and 96.2\% with full fine-tuning. Audio Flamingo 3 reaches 79.4\% and Qwen2-Audio 68.9\%. The test audio uses other pitch ranges and another soundfont than the training audio, so these gains do not come from memorizing training clips\rev{; for Qwen2.5-Omni they also carry over to real recordings (Section~\ref{sec:real})}. Training also reduces letter bias: the gap between accuracy on items whose answer is `A' and on items whose answer is `B' falls from 75.3 to 17.0 points for Audio Flamingo 3 (full) and from 97.2 to 39.6 points for Qwen2-Audio (LoRA; Figure~\ref{fig:bias} in Appendix~\ref{app:full}). \rev{Only} Qwen2.5-Omni (100.0\% and 99.6\%) and Audio Flamingo 3 with full fine-tuning (69.2\%) learn the rhythm task; all other runs stay between 50.8\% and 52.0\%.

\subsection{Models learn to notice change more than to ignore it}
\label{sec:res-asym}

\begin{figure}[t]
  \centering
  \begin{minipage}[c]{0.56\linewidth}
    \centering
    \includegraphics[width=\linewidth]{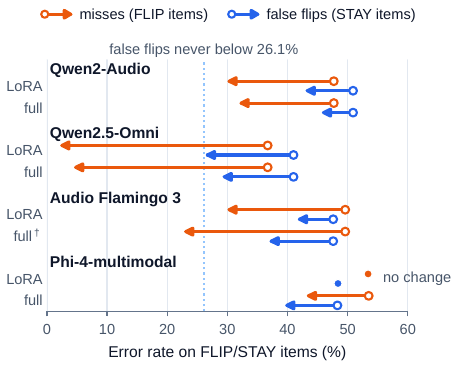}
  \end{minipage}\hfill
  \begin{minipage}[c]{0.41\linewidth}
    \caption{Misses and false flips on \FLIP/\STAY{} before and after GRPO, one row per run. Each arrow starts at the error rate without training (open circle) and ends at the rate after GRPO. Orange: miss rate (\FLIP{} items); blue: false-flip rate (\STAY{} items). Orange arrows are 2.2 to 3.2 times longer than blue ones in six of the seven runs that learned, and \rev{none of these runs} gets below 26.1\% false flips (dotted line). $^\dagger$Step 500. Values are in Table~\ref{tab:grpo}.}
    \label{fig:mrffr}
  \end{minipage}
\end{figure}

Clean accuracy hides what GRPO teaches. Figure~\ref{fig:mrffr} shows how the miss rate and the false-flip rate of each model change with training. Training lowers the miss rate much more than the false-flip rate. In six of the seven runs that learned, the miss rate fell 2.2 to 3.2 times more than the false-flip rate ($p < 0.01$ for each run; Table~\ref{tab:app-drop}). The exception is Phi-4-multimodal with full fine-tuning, where both rates fell by similar amounts (10.6 and 9.0 points). For Qwen2.5-Omni with LoRA, the miss rate falls from 36.7\% to 1.9\%, but the false-flip rate only falls from 41.0\% to 26.1\%. The trained model now misses 13 times fewer changes than Gemini 2.5 Pro (1.9\% against 24.7\%), but its false-flip rate is no lower (26.1\% against 26.9\%). \rev{But most of Gemini 2.5 Pro's false flips (155 of 209 on melody, harmony and timbre) miss a musical change rather than react to the sound. Counted only on items whose clean version is right ($\mathrm{FFR}_c$), its rate falls to 20.8\%, while that of the trained model stays at 25.7\% (Appendix~\ref{app:paired}).} Figure~\ref{fig:plane} in Appendix~\ref{app:full} places every model and our listeners on these two axes.

Human listeners also make more false flips than misses (20.8\% against 5.2\%), but at least 11 of their 20 false flips are on harmony (Table~\ref{tab:app-human}), and on timbre and rhythm they made at most one error each in 48 \FLIP/\STAY{} answers. Trained Qwen2.5-Omni misses fewer changes than our listeners (1.9\% against 5.2\%), yet false-flips on 94 of the 500 timbre and rhythm \STAY{} pairs.

\paragraph{False flips by sound change.} Each \STAY{} pair applies one of four sound changes, and \rev{our main} GRPO \rev{runs train} on none of them (Table~\ref{tab:app-ffr-change} in Appendix~\ref{app:stay-change}). Qwen2.5-Omni with LoRA learns to ignore EQ, where its false-flip rate falls from 33.7\% to 1.4\%, and partly room echo (37.2\% to 16.6\%). It does not learn to ignore hiss (49.0\% before training, 52.4\% after) or transposition (48.4\% and 45.2\%). On timbre and rhythm, the gap to our listeners comes from hiss alone: hiss causes 85 of the 94 false flips, and on the other 334 \STAY{} pairs of these tasks (EQ or room echo) the model errs on 9 (2.7\%). The model also detects a single moved note in 99.6\% of melody \FLIP{} pairs, yet errs on 40.3\% of melody pairs that are only transposed (46.8\% before training). It hears a change of pitch but has not learned relative pitch. Full fine-tuning, Qwen2-Audio and Audio Flamingo~3 show the same pattern (Figure~\ref{fig:ffr-change} in Appendix~\ref{app:stay-change}).

This ranking holds before training too. Without GRPO, Qwen2.5-Omni and Gemini 3.8 Flash err least on EQ and most on hiss and transposition, and Gemini 2.5 Pro, which we did not train, errs most on transposition and hiss (46.8\% and 32.5\%, against 22.4\% for EQ and 17.2\% for room echo). So the weakness to hiss is not caused by our training: GRPO sharpens the ranking but does not change it. The reward only checks the letter on clean pairs, so any invariance the model gains is a side effect.

\rev{\paragraph{Where the false flips come from.} The audio encoder is frozen in every run, so training only changes how the language model reads its output. On our rendered clips, hiss moves this output 2.75 times as far as a real change in the music, and EQ only 0.04 times as far (Appendix~\ref{app:encoder}). This distance predicts which `same' \STAY{} pairs the trained model calls different (AUROC 0.92, against 0.74 before training). Training still teaches more than a threshold on this distance, which makes far more false flips under EQ and room echo. The encoder output also holds much of what is needed to ignore hiss: a linear probe on it, trained with room-echo, EQ and hiss examples, cuts false flips under hiss from 55.0\% to 32.5\%.}

\rev{\paragraph{Training algorithm and data.} SFT on the same clean pairs gives the same pattern: misses fall by 33.7 points and false flips by 13.2 (Appendix~\ref{app:trainbase}). Adding the sound-only versions of the training split (codec, soundfont swap and transposition) to GRPO lowers false flips on rendered clips under room echo from 16.6\% to 4.5\% and under hiss from 52.4\% to 37.7\%, although neither change is in training. \FLIP/\STAY{} accuracy rises from 86.0\% to 88.2\%. Under transposition, false flips stay at 45.2\%, although transposed pairs are now in training.}

\subsection{Transfer to an unseen task, and side effects}
\label{sec:res-transfer}

\begin{figure}[t]
  \centering
  \includegraphics[width=\linewidth]{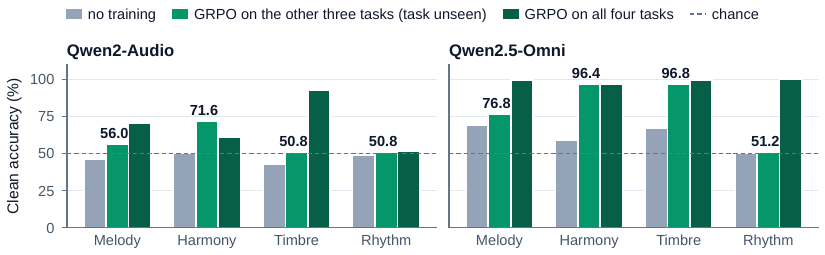}
  \caption{Transfer to a task left out of training (clean accuracy, 250 items per bar). Grey: no training. Green, with values: LoRA GRPO for 500 steps on the other three tasks, so the task itself is never seen. Dark green: LoRA GRPO for 1{,}000 steps on all four tasks.}
  \label{fig:transfer}
\end{figure}

\paragraph{Transfer to an unseen task.} Training on three tasks improves the fourth, with a clear pattern (Figure~\ref{fig:transfer}). Harmony improves in both models. Qwen2-Audio goes from 50.4\% to 71.6\%, higher than the 60.8\% it reaches when harmony is in training. Qwen2.5-Omni goes from 59.2\% to 96.4\%, close to its 96.8\% with harmony in training. Timbre improves for Qwen2.5-Omni (66.8\% to 96.8\%), and melody gains 8 to 10 points. For Qwen2-Audio, timbre only moves from 43.2\% to 50.8\%, which is still chance. Rhythm does not improve (1.2 and 1.6 points), although Qwen2.5-Omni reaches 100.0\% on rhythm when rhythm is in training. Rhythm is the only task about timing and the only one that is not a same/different question. 
\rev{So the models seem to learn a general same/different skill that helps any such task but does not say which clip is faster.}

\paragraph{Side effects on captioning.} GRPO on \rev{rendered} pairs keeps captioning close to the base model (Table~\ref{tab:caption} in Appendix~\ref{app:caption}). On 2{,}735 MusicCaps clips, BERTScore-F1 changes by $-1.28$ to $+1.95$ points and BLEU-4 by $-0.22$ to $+1.73$. Mean caption length changes by less than one word, and no run produces degenerate captions (paired tests in Appendix~\ref{app:caption}).

\begingroup
\subsection{Real recordings}
\label{sec:real}

\paragraph{Open models stay near chance.} On the real test set, the eight open models score 48.3\% to 55.5\% on \FLIP/\STAY, and only Audio Flamingo 3 and Fun-Audio-Chat are significantly above chance (Table~\ref{tab:real_untrained}). Gemini 2.5 Pro is again the best untrained model (73.8\%, against 73.7\% on the rendered test under the same scoring). Over all 12 untrained models, the ranking agrees with the rendered test (Spearman $\rho = 0.80$, $p = 0.002$).

\paragraph{Training carries over.} Trained on our rendered pairs, Qwen2.5-Omni rises from 52.1\% to 62.8\% (exact McNemar test, $p < 0.001$). SFT gives 62.9\%, and adding sound-only pairs gives 70.1\%. The other trained models stay within 6 points of chance. Trained on Slakh2100 pairs (Appendix~\ref{app:slakh}), Qwen2.5-Omni reaches 75.0\%, level with Gemini 2.5 Pro ($p = 0.50$).

\paragraph{Two strums of the same chord.} Two strums of the same chord always differ a little, but a listener should call them the same (Table~\ref{tab:real_take}). The trained Qwen2.5-Omni detects 38 of 41 moved thirds, but also calls 27 of 42 such same-chord pairs different. As in Section~\ref{sec:res-asym}, it tends to treat any difference between two recordings as a change. With sound-only pairs in training, it calls only 7 of 42 different ($p < 0.001$) and detects 26 of 41 moved thirds. Gemini 2.5 Pro errs the other way: it calls 37 of 42 same-chord pairs the same, but detects only 12 of 41 moved thirds.

\paragraph{False flips on real audio.} After training, false flips stay at or above misses in every Qwen2.5-Omni run, on all three test sets. On real recordings the model also misses more changes, so the two rates end close together (37.0\% and 37.3\%); real melodies are hard for every open model (at most 55.3\% clean accuracy, against 72.7\% for Gemini 2.5 Pro). Of the three sound effects, hiss causes the most false flips on rendered clips (52.4\%) but the fewest on real recordings (20.9\%, against 46.6\% for EQ), perhaps because real recordings already contain some noise. Every trained Qwen2.5-Omni run, on all three test sets, still gets 38\% to 49\% of transposed pairs wrong.

\paragraph{Further checks.} Appendices~\ref{app:paired} to~\ref{app:timbre} add paired metrics and per-task tests, a control with each edited clip played twice, 12- and 24-note melodies, and finer timbre. With longer melodies, untrained open models other than Qwen2.5-Omni stay near chance (48.5\% to 56.8\%) and the trained model's false flips stay high. On close instrument pairs, the trained model detects only 10\% of violin/viola changes.
\endgroup

\section{Discussion}
\label{sec:discussion}

\paragraph{Clean accuracy is not enough.} On our clean test, always answering `A' scores 52.8\%\rev{, and on} \FLIP/\STAY{} 49.45\%. \rev{GRPO lifts Qwen2.5-Omni to 99.0\% on clean pairs but to 86.0\% on \FLIP/\STAY, and SFT to 97.4\% and 84.6\%. Clean accuracy cannot tell a model that compares the music from one that reacts to any difference in sound, but the paired variants can, so benchmarks should report both.} This matters beyond benchmarking: audio LLMs are already used to label, filter and rank music, and a judge that changes its verdict when a quiet hiss is added scores the recording, not the music. Every clip in our tests holds one instrument or one stem, so full mixes remain untested.

\section{Conclusion}
\label{sec:conclusion}
With exact labels and paired \FLIP/\STAY{} variants, MusicListenBench shows that most open audio LLMs cannot hear the difference between two musical clips\rev{, on rendered and real audio,} and that the best commercial model, Gemini 2.5 Pro, stays 12.8 points below human listeners. After a little GRPO, the best model at most matches \rev{our listeners} on \FLIP/\STAY: it learns to notice change and\rev{, on rendered clips,} to ignore a mild EQ change, but not a quiet hiss or a transposition. \rev{Training on clean pairs never asks the model to ignore a change in sound, so invariance has to be asked for, and asking for it is cheap: the reward only has to require the same answer for an item’s clean and \STAY{} versions, which come from the same item. Adding sound-only pairs to training already helps, but does not teach relative pitch. We consider the title question answered when a model, never trained on a \STAY{}-type change, reaches our listeners' interval on \FLIP/\STAY{} (82.3\% to 91.1\%; Appendix~\ref{app:human}) and, like them, almost never false-flips on timbre and rhythm. Our trained Qwen2.5-Omni meets the first condition (86.0\%) but not the second (20.8\% and 16.8\% false flips).} Teaching models which differences to ignore is the open problem.
\section*{LLM usage statement}
We used an LLM assistant (Claude Sonnet) to paraphrase sentences and carry out grammatical checks. All research ideas, experiments and conclusions are the authors', who take full responsibility for the content.

\section*{Ethics statement}
\paragraph{Human listeners.} Twelve adults took part in the listening test. They were told what the task was, that participation was voluntary and that they
could stop at any time, and they gave consent before starting. We recorded only the answers and a coarse self-report of musical experience. No names, contact
details, audio recordings or other identifying information were collected, so the released answers cannot be traced to a participant. Listeners set their own
comfortable playback level before the test.

\paragraph{Audio and third-party material.} All audio \rev{of the rendered set} is synthesized from generated symbolic music with freely available soundfonts, so it contains no recordings of people and no copyrighted musical works. We follow the licence terms of each soundfont \citep{fluidr3gm,generalusergs,musescoregeneral} and state them with the release. MusicCaps clips are used only to measure captioning side effects. We do not redistribute them; we release captions and scores, and the clips stay with their original source. \rev{The real test set (Section~\ref{sec:splits}) and the Slakh2100 study (Appendix~\ref{app:slakh}) use public research datasets under their licences: GuitarSet (CC BY-NC 4.0), Medley-solos-DB (CC BY 4.0), MUSDB18-HQ (research use, access on request) and Slakh2100 (CC BY 4.0). Unlike the rendered set, they contain recordings of performers and existing songs. For the real test set we release the item lists and the build script, and no MUSDB18-HQ audio.}

\section*{Reproducibility statement}
The leaderboard is available at \url{https://amaai-lab.github.io/MusicListenBench/}. We release the generator, all audio and item files \rev{of the rendered set, the item lists and build script of the real test set}, the scoring script, and the evaluation code at \url{https://github.com/AMAAI-Lab/MusicListenBench}; the dataset is hosted at \url{https://huggingface.co/datasets/amaai-lab/MusicListenBench}. 

Appendix~\ref{app:construction} gives all generation parameters, Appendix~\ref{app:models} all model versions, and Appendix~\ref{app:grpo} all training settings.

\bibliography{iclr2027_conference}

\begin{thebibliography}{63}
\providecommand{\natexlab}[1]{#1}
\providecommand{\url}[1]{\texttt{#1}}
\expandafter\ifx\csname urlstyle\endcsname\relax
  \providecommand{\doi}[1]{doi: #1}\else
  \providecommand{\doi}{doi: \begingroup \urlstyle{rm}\Url}\fi

\bibitem[Abouelenin et~al.(2025)Abouelenin, Ashfaq, Atkinson, Awadalla, Bach,
  Bao, Benhaim, Cai, Chaudhary, Chen, et~al.]{abouelenin2025phi}
Abdelrahman Abouelenin, Atabak Ashfaq, Adam Atkinson, Hany Awadalla, Nguyen
  Bach, Jianmin Bao, Alon Benhaim, Martin Cai, Vishrav Chaudhary, Congcong
  Chen, et~al.
\newblock Phi-4-mini technical report: Compact yet powerful multimodal language
  models via mixture-of-loras.
\newblock \emph{arXiv preprint arXiv:2503.01743}, 2025.

\bibitem[Agostinelli et~al.(2023)Agostinelli, Denk, Borsos, Engel, Verzetti,
  Caillon, Huang, Jansen, Roberts, Tagliasacchi, Sharifi, Zeghidour, and
  Frank]{agostinelli2023musiclm}
Andrea Agostinelli, Timo~I. Denk, Zal{\'a}n Borsos, Jesse Engel, Mauro
  Verzetti, Antoine Caillon, Qingqing Huang, Aren Jansen, Adam Roberts, Marco
  Tagliasacchi, Matt Sharifi, Neil Zeghidour, and Christian Frank.
\newblock Musiclm: Generating music from text.
\newblock \emph{arXiv preprint arXiv:2301.11325}, 2023.

\bibitem[Banerjee \& Lavie(2005)Banerjee and Lavie]{banerjee2005meteor}
Satanjeev Banerjee and Alon Lavie.
\newblock Meteor: An automatic metric for mt evaluation with improved
  correlation with human judgments.
\newblock In \emph{Proceedings of the acl workshop on intrinsic and extrinsic
  evaluation measures for machine translation and/or summarization}, pp.\
  65--72, 2005.

\bibitem[Carone et~al.(2025)Carone, Roman, and
  Ripoll{\'e}s]{carone2025evaluating}
Brandon~James Carone, Iran~R Roman, and Pablo Ripoll{\'e}s.
\newblock Evaluating multimodal large language models on core music perception
  tasks.
\newblock \emph{arXiv preprint arXiv:2510.22455}, 2025.

\bibitem[Carone et~al.(2026)Carone, Roman, and Ripoll{\'e}s]{carone2026muse}
Brandon~James Carone, Iran~R Roman, and Pablo Ripoll{\'e}s.
\newblock The muse benchmark: Probing music perception and auditory relational
  reasoning in audio llms.
\newblock In \emph{ICASSP 2026-2026 IEEE International Conference on Acoustics,
  Speech and Signal Processing (ICASSP)}, pp.\  16097--16101. IEEE, 2026.

\bibitem[Chopra et~al.(2025)Chopra, Roy, and Herremans]{chopra2025sonicverse}
Anuradha Chopra, Abhinaba Roy, and Dorien Herremans.
\newblock Sonicverse: Multi-task learning for music feature-informed
  captioning.
\newblock \emph{arXiv preprint arXiv:2506.15154}, 2025.

\bibitem[Chu et~al.(2024)Chu, Xu, Yang, Wei, Wei, Guo, Leng, Lv, He, Lin,
  et~al.]{chu2024qwen2}
Yunfei Chu, Jin Xu, Qian Yang, Haojie Wei, Xipin Wei, Zhifang Guo, Yichong
  Leng, Yuanjun Lv, Jinzheng He, Junyang Lin, et~al.
\newblock Qwen2-audio technical report.
\newblock \emph{arXiv preprint arXiv:2407.10759}, 2024.

\bibitem[Collins(2020)]{musescoregeneral}
S.~Christian Collins.
\newblock {MuseScore\_General}: the default {MuseScore} {SoundFont}.
\newblock \url{https://musescore.org/en/node/317991}, 2020.
\newblock Derived from {FluidR3Mono}, MIT license; Accessed: September 20,
  2026.

\bibitem[Collins(2024)]{generalusergs}
S.~Christian Collins.
\newblock {GeneralUser GS}: a {General MIDI} {SoundFont} bank.
\newblock \url{https://github.com/mrbumpy409/GeneralUser-GS}, 2024.
\newblock Accessed: September 20, 2026.

\bibitem[Comanici et~al.(2025)Comanici, Bieber, Schaekermann, Pasupat,
  Sachdeva, Dhillon, Blistein, Ram, Zhang, Rosen, et~al.]{comanici2025gemini}
Gheorghe Comanici, Eric Bieber, Mike Schaekermann, Ice Pasupat, Noveen
  Sachdeva, Inderjit Dhillon, Marcel Blistein, Ori Ram, Dan Zhang, Evan Rosen,
  et~al.
\newblock Gemini 2.5: Pushing the frontier with advanced reasoning,
  multimodality, long context, and next generation agentic capabilities.
\newblock \emph{arXiv preprint arXiv:2507.06261}, 2025.

\bibitem[Cuthbert \& Ariza(2010)Cuthbert and Ariza]{cuthbert2010music21}
Michael~Scott Cuthbert and Christopher Ariza.
\newblock music21: A toolkit for computer-aided musicology and symbolic music
  data.
\newblock 2010.

\bibitem[Ding et~al.(2025)Ding, Ju, Leng, Liu, Liu, Shang, Shen, Song, Tan,
  Tang, et~al.]{ding2025kimi}
Ding Ding, Zeqian Ju, Yichong Leng, Songxiang Liu, Tong Liu, Zeyu Shang, Kai
  Shen, Wei Song, Xu~Tan, Heyi Tang, et~al.
\newblock Kimi-audio technical report.
\newblock \emph{arXiv preprint arXiv:2504.18425}, 2025.

\bibitem[Dowling(1978)]{dowling1978scale}
W~Jay Dowling.
\newblock Scale and contour: Two components of a theory of memory for melodies.
\newblock \emph{Psychological review}, 85\penalty0 (4):\penalty0 341, 1978.

\bibitem[Dujardin et~al.(2026)Dujardin, Yu, Thomas-Smith, Chan, and
  Nguyen]{pitchbench}
Milan~Liessens Dujardin, Song-Ze Yu, Craver~Corbyn Thomas-Smith, David~M Chan,
  and Karina Nguyen.
\newblock Pitchbench: Measuring pitch hearing in audio-language models.
\newblock \emph{arXiv preprint arXiv:2605.26176}, 2026.

\bibitem[EBU-Recommendation(2011)]{ebu2011loudness}
R~EBU-Recommendation.
\newblock Loudness normalisation and permitted maximum level of audio signals.
\newblock \emph{Eur. Broadcast. Union}, 2011.

\bibitem[{FluidSynth Developers}(2000)]{fluidsynth}
{FluidSynth Developers}.
\newblock {FluidSynth}: a real-time {SoundFont} synthesizer.
\newblock \url{https://www.fluidsynth.org}, 2000.
\newblock Accessed: September 20, 2026.

\bibitem[Gardner et~al.(2020)Gardner, Artzi, Basmov, Berant, Bogin, Chen,
  Dasigi, Dua, Elazar, Gottumukkala, et~al.]{gardner2020evaluating}
Matt Gardner, Yoav Artzi, Victoria Basmov, Jonathan Berant, Ben Bogin, Sihao
  Chen, Pradeep Dasigi, Dheeru Dua, Yanai Elazar, Ananth Gottumukkala, et~al.
\newblock Evaluating models’ local decision boundaries via contrast sets.
\newblock In \emph{Findings of the Association for Computational Linguistics:
  EMNLP 2020}, pp.\  1307--1323, 2020.

\bibitem[Ghosh et~al.(2026)Ghosh, Goel, Kim, Kumar, Kong, Lee, Yang,
  Duraiswami, Manocha, Valle, et~al.]{ghosh2026audio}
Sreyan Ghosh, Arushi Goel, Jaehyeon Kim, Sonal Kumar, Zhifeng Kong, Sang-gil
  Lee, Chao-Han Yang, Ramani Duraiswami, Dinesh Manocha, Rafael Valle, et~al.
\newblock Audio flamingo 3: Advancing audio intelligence with fully open large
  audio language models.
\newblock \emph{Advances in Neural Information Processing Systems},
  38:\penalty0 41819--41886, 2026.

\bibitem[{Google DeepMind}(2026)]{gemini38flash}
{Google DeepMind}.
\newblock {Gemini 3.8 Flash} model card.
\newblock
  \url{https://storage.googleapis.com/deepmind-media/Model-Cards/Gemini-3-8-Flash-Model-Card.pdf},
  2026.
\newblock Accessed 26 September 2026.

\bibitem[Hautus et~al.(2021)Hautus, Macmillan, and
  Creelman]{hautus2021detection}
Michael~J Hautus, Neil~A Macmillan, and C~Douglas Creelman.
\newblock \emph{Detection theory: A user's guide}.
\newblock Routledge, 2021.

\bibitem[Herremans \& Roy(2026)Herremans and Roy]{herremans2026aligning}
Dorien Herremans and Abhinaba Roy.
\newblock Aligning generative music ai with human preferences: methods and
  challenges.
\newblock In \emph{Proceedings of the AAAI Conference on Artificial
  Intelligence}, volume~40, pp.\  39699--39706, 2026.

\bibitem[Hu et~al.(2021)Hu, Shen, Wallis, Allen-Zhu, Li, Wang, Wang, and
  Chen]{hu2021lora}
Edward~J Hu, Yelong Shen, Phillip Wallis, Zeyuan Allen-Zhu, Yuanzhi Li, Shean
  Wang, Lu~Wang, and Weizhu Chen.
\newblock Lora: Low-rank adaptation of large language models.
\newblock \emph{arXiv preprint arXiv:2106.09685}, 2021.

\bibitem[Kim et~al.(2026)Kim, Chi, Angelopoulos, Chiang, Saito, Watanabe,
  Mitsufuji, and Donahue]{kim2026music}
Yonghyun Kim, Wayne Chi, Anastasios Angelopoulos, Wei-Lin Chiang, Koichi Saito,
  Shinji Watanabe, Yuki Mitsufuji, and Chris Donahue.
\newblock Music arena: Live evaluation for text-to-music.
\newblock \emph{Advances in Neural Information Processing Systems}, 38, 2026.

\bibitem[Law \& Zentner(2012)Law and Zentner]{law2012assessing}
Lily~NC Law and Marcel Zentner.
\newblock Assessing musical abilities objectively: Construction and validation
  of the profile of music perception skills.
\newblock \emph{PloS one}, 7\penalty0 (12):\penalty0 e52508, 2012.

\bibitem[Li et~al.(2024)Li, Lin, Peng, Nyandwi, Jiang, Ma, Khanuja, Krishna,
  Neubig, and Ramanan]{li2024naturalbench}
Baiqi Li, Zhiqiu Lin, Wenxuan Peng, Jean De~Dieu Nyandwi, Daniel Jiang, Zixian
  Ma, Simran Khanuja, Ranjay Krishna, Graham Neubig, and Deva Ramanan.
\newblock Naturalbench: Evaluating vision-language models on natural
  adversarial samples.
\newblock \emph{Advances in Neural Information Processing Systems},
  37:\penalty0 17044--17068, 2024.

\bibitem[Li et~al.(2025)Li, Liu, Dinkel, Niu, Zhang, and
  Luan]{li2025reinforcement}
Gang Li, Jizhong Liu, Heinrich Dinkel, Yadong Niu, Junbo Zhang, and Jian Luan.
\newblock Reinforcement learning outperforms supervised fine-tuning: A case
  study on audio question answering.
\newblock \emph{arXiv preprint arXiv:2503.11197}, 2025.

\bibitem[Lin(2004)]{lin2004rouge}
Chin-Yew Lin.
\newblock Rouge: A package for automatic evaluation of summaries.
\newblock In \emph{Text summarization branches out}, pp.\  74--81, 2004.

\bibitem[L{\'o}pez et~al.(2025)L{\'o}pez, Kesiraju, and
  Luque]{lopez2025robustness}
Fernando L{\'o}pez, Santosh Kesiraju, and Jordi Luque.
\newblock Robustness assessment of large audio language models in
  multiple-choice evaluation.
\newblock \emph{arXiv preprint arXiv:2510.04584}, 2025.

\bibitem[Lostanlen et~al.(2018)Lostanlen, Cella, Bittner, and
  Essid]{lostanlen2018medley}
Vincent Lostanlen, Carmine-Emanuele Cella, Rachel Bittner, and Slim Essid.
\newblock Medley-solos-db: a cross-collection dataset for musical instrument
  recognition.
\newblock \emph{Zenodo: Meyrin, Switzerland}, 2018.

\bibitem[Lu et~al.(2025)Lu, Geist, Melechovsky, Roy, and
  Herremans]{lu2025melodysim}
Tongyu Lu, Charlotta-Marlena Geist, Jan Melechovsky, Abhinaba Roy, and Dorien
  Herremans.
\newblock Melodysim: measuring melody-aware music similarity for plagiarism
  detection.
\newblock \emph{arXiv preprint arXiv:2505.20979}, 2025.

\bibitem[Ma et~al.(2025)Ma, Li, Yu, Benetos, and Maezawa]{ma2025cmibench}
Yinghao Ma, Siyou Li, Juntao Yu, Emmanouil Benetos, and Akira Maezawa.
\newblock Cmi-bench: A comprehensive benchmark for evaluating music instruction
  following.
\newblock \emph{arXiv preprint arXiv:2506.12285}, 2025.

\bibitem[Ma et~al.(2026)Ma, Ma, Zhu, Yang, Chao, Xu, Chen, Chen, Chen, Cong,
  et~al.]{ma2026mmar}
Ziyang Ma, Yinghao Ma, Yanqiao Zhu, Chen Yang, Yi-Wen Chao, Ruiyang Xu, Wenxi
  Chen, Yuanzhe Chen, Zhuo Chen, Jian Cong, et~al.
\newblock Mmar: A challenging benchmark for deep reasoning in speech, audio,
  music, and their mix.
\newblock \emph{Advances in Neural Information Processing Systems}, 38, 2026.

\bibitem[Manilow et~al.(2019)Manilow, Wichern, Seetharaman, and
  Le~Roux]{manilow2019slakh}
Ethan Manilow, Gordon Wichern, Prem Seetharaman, and Jonathan Le~Roux.
\newblock Cutting music source separation some slakh: A dataset to study the
  impact of training data quality and quantity.
\newblock In \emph{2019 IEEE Workshop on Applications of Signal Processing to
  Audio and Acoustics (WASPAA)}, pp.\  45--49. IEEE, 2019.

\bibitem[McAdams et~al.(1995)McAdams, Winsberg, Donnadieu, De~Soete, and
  Krimphoff]{mcadams1995perceptual}
Stephen McAdams, Suzanne Winsberg, Sophie Donnadieu, Geert De~Soete, and Jochen
  Krimphoff.
\newblock Perceptual scaling of synthesized musical timbres: Common dimensions,
  specificities, and latent subject classes.
\newblock \emph{Psychological research}, 58\penalty0 (3):\penalty0 177--192,
  1995.

\bibitem[Melechovsky et~al.(2025)Melechovsky, Mehrish, Roy, and
  Herremans]{melechovsky2025sonicmaster}
Jan Melechovsky, Ambuj Mehrish, Abhinaba Roy, and Dorien Herremans.
\newblock Sonicmaster: Towards controllable all-in-one music restoration and
  mastering.
\newblock \emph{arXiv preprint arXiv:2508.03448}, 2025.

\bibitem[{OpenAI}(2026)]{openai2026gptaudio15}
{OpenAI}.
\newblock {GPT-Audio-1.5 Model Documentation}.
\newblock \url{https://developers.openai.com/api/docs/models/gpt-audio-1.5},
  2026.
\newblock Accessed: September 20, 2026.

\bibitem[{OpenRouter Team}(2026)]{openrouter}
{OpenRouter Team}.
\newblock Openrouter: A unified api for ai models.
\newblock \url{https://openrouter.ai}, 2026.
\newblock Accessed: 2026-09-26.

\bibitem[Papineni et~al.(2002)Papineni, Roukos, Ward, and
  Zhu]{papineni2002bleu}
Kishore Papineni, Salim Roukos, Todd Ward, and Wei-Jing Zhu.
\newblock Bleu: a method for automatic evaluation of machine translation.
\newblock In \emph{Proceedings of the 40th annual meeting of the Association
  for Computational Linguistics}, pp.\  311--318, 2002.

\bibitem[Rafii et~al.(2019)Rafii, Liutkus, Stöter, Mimilakis, and
  Bittner]{rafii2019musdb18hq}
Zafar Rafii, Antoine Liutkus, Fabian-Robert Stöter, Stylianos~Ioannis
  Mimilakis, and Rachel Bittner.
\newblock Musdb18-hq - an uncompressed version of musdb18, August 2019.
\newblock URL \url{https://doi.org/10.5281/zenodo.3338373}.

\bibitem[Ribeiro et~al.(2020)Ribeiro, Wu, Guestrin, and
  Singh]{ribeiro2020beyond}
Marco~Tulio Ribeiro, Tongshuang Wu, Carlos Guestrin, and Sameer Singh.
\newblock Beyond accuracy: Behavioral testing of nlp models with checklist.
\newblock In \emph{Proceedings of the 58th annual meeting of the association
  for computational linguistics}, pp.\  4902--4912, 2020.

\bibitem[Rouditchenko et~al.(2025)Rouditchenko, Bhati, Araujo, Thomas, Kuehne,
  Feris, and Glass]{rouditchenko2025omni}
Andrew Rouditchenko, Saurabhchand Bhati, Edson Araujo, Samuel Thomas, Hilde
  Kuehne, Rogerio Feris, and James Glass.
\newblock Omni-r1: Do you really need audio to fine-tune your audio llm?
\newblock In \emph{2025 IEEE Automatic Speech Recognition and Understanding
  Workshop (ASRU)}, pp.\  1--7. IEEE, 2025.

\bibitem[Roy et~al.(2026{\natexlab{a}})Roy, Liang, and Herremans]{roy2026merit}
Abhinaba Roy, Junyi Liang, and Dorien Herremans.
\newblock Merit: Learning disentangled music representations for audio
  similarity.
\newblock \emph{arXiv preprint arXiv:2605.27346}, 2026{\natexlab{a}}.

\bibitem[Roy et~al.(2026{\natexlab{b}})Roy, Puri, and
  Herremans]{roy2026text2midi}
Abhinaba Roy, Geeta Puri, and Dorien Herremans.
\newblock Text2midi-inferalign: Improving symbolic music generation with
  inference-time alignment.
\newblock In \emph{ICASSP 2026-2026 IEEE International Conference on Acoustics,
  Speech and Signal Processing (ICASSP)}, pp.\  14817--14821. IEEE,
  2026{\natexlab{b}}.

\bibitem[Sakshi et~al.(2025)Sakshi, Tyagi, Kumar, Seth, Selvakumar, Nieto,
  Duraiswami, Ghosh, and Manocha]{sakshi2025mmau}
Sakshi Sakshi, Utkarsh Tyagi, Sonal Kumar, Ashish Seth, Ramaneswaran
  Selvakumar, Oriol Nieto, Ramani Duraiswami, Sreyan Ghosh, and Dinesh Manocha.
\newblock Mmau: A massive multi-task audio understanding and reasoning
  benchmark.
\newblock In \emph{International Conference on Learning Representations},
  volume 2025, pp.\  84929--84964, 2025.

\bibitem[Scheibler et~al.(2018)Scheibler, Bezzam, and
  Dokmani{\'c}]{scheibler2018pyroomacoustics}
Robin Scheibler, Eric Bezzam, and Ivan Dokmani{\'c}.
\newblock Pyroomacoustics: A python package for audio room simulation and array
  processing algorithms.
\newblock In \emph{2018 IEEE international conference on acoustics, speech and
  signal processing (ICASSP)}, pp.\  351--355. IEEE, 2018.

\bibitem[Shao et~al.(2024)Shao, Wang, Zhu, Xu, Song, Bi, Zhang, Zhang, Li, Wu,
  et~al.]{shao2024deepseekmath}
Zhihong Shao, Peiyi Wang, Qihao Zhu, Runxin Xu, Junxiao Song, Xiao Bi, Haowei
  Zhang, Mingchuan Zhang, YK~Li, Yang Wu, et~al.
\newblock {Deepseekmath}: Pushing the limits of mathematical reasoning in open
  language models.
\newblock \emph{arXiv preprint arXiv:2402.03300}, 2024.

\bibitem[Sun et~al.(2026)Sun, Chen, Qiu, Zhang, Chen, Wu, Li, Yang, Zhu, Zhang,
  et~al.]{sun2026sonicbench}
Yirong Sun, Yanjun Chen, Xin Qiu, Gang Zhang, Hongyu Chen, Daokuan Wu,
  Chengming Li, Min Yang, Dawei Zhu, Wei Zhang, et~al.
\newblock Sonicbench: Dissecting the physical perception bottleneck in large
  audio language models.
\newblock \emph{arXiv preprint arXiv:2601.11039}, 2026.

\bibitem[Team et~al.(2025)Team, Chen, Cheng, Deng, Li, Liu, Tan, Wang, Xu, Ye,
  et~al.]{team2025fun}
Tongyi~Fun Team, Qian Chen, Luyao Cheng, Chong Deng, Xiangang Li, Jiaqing Liu,
  Chao-Hong Tan, Wen Wang, Junhao Xu, Jieping Ye, et~al.
\newblock Fun-audio-chat technical report.
\newblock \emph{arXiv preprint arXiv:2512.20156}, 2025.

\bibitem[Thrush et~al.(2022)Thrush, Jiang, Bartolo, Singh, Williams, Kiela, and
  Ross]{thrush2022winoground}
Tristan Thrush, Ryan Jiang, Max Bartolo, Amanpreet Singh, Adina Williams, Douwe
  Kiela, and Candace Ross.
\newblock Winoground: Probing vision and language models for visio-linguistic
  compositionality.
\newblock In \emph{2022 IEEE/CVF Conference on Computer Vision and Pattern
  Recognition (CVPR)}, pp.\  5228--5238. IEEE, 2022.

\bibitem[Tong et~al.(2024)Tong, Liu, Zhai, Ma, LeCun, and Xie]{tong2024eyes}
Shengbang Tong, Zhuang Liu, Yuexiang Zhai, Yi~Ma, Yann LeCun, and Saining Xie.
\newblock Eyes wide shut? exploring the visual shortcomings of multimodal llms.
\newblock In \emph{2024 IEEE/CVF Conference on Computer Vision and Pattern
  Recognition (CVPR)}, pp.\  9568--9578. IEEE, 2024.

\bibitem[Valin et~al.(2012)Valin, Vos, and Terriberry]{valin2012definition}
Jean-Marc Valin, Koen Vos, and Timothy Terriberry.
\newblock Definition of the opus audio codec.
\newblock Technical report, 2012.

\bibitem[Wallentin et~al.(2010)Wallentin, Nielsen, Friis-Olivarius, Vuust, and
  Vuust]{wallentin2010musical}
Mikkel Wallentin, Andreas~H{\o}jlund Nielsen, Morten Friis-Olivarius, Christian
  Vuust, and Peter Vuust.
\newblock The musical ear test, a new reliable test for measuring musical
  competence.
\newblock \emph{Learning and Individual Differences}, 20\penalty0 (3):\penalty0
  188--196, 2010.

\bibitem[Warstadt et~al.(2020)Warstadt, Parrish, Liu, Mohananey, Peng, Wang,
  and Bowman]{warstadt2020blimp}
Alex Warstadt, Alicia Parrish, Haokun Liu, Anhad Mohananey, Wei Peng, Sheng-Fu
  Wang, and Samuel~R Bowman.
\newblock Blimp: The benchmark of linguistic minimal pairs for english.
\newblock \emph{Transactions of the Association for Computational Linguistics},
  8:\penalty0 377--392, 2020.

\bibitem[Weck et~al.(2024)Weck, Manco, Benetos, Quinton, Fazekas, and
  Bogdanov]{weck2024muchomusic}
Benno Weck, Ilaria Manco, Emmanouil Benetos, Elio Quinton, George Fazekas, and
  Dmitry Bogdanov.
\newblock Muchomusic: Evaluating music understanding in multimodal
  audio-language models.
\newblock \emph{arXiv preprint arXiv:2408.01337}, 2024.

\bibitem[Wen(2000)]{fluidr3gm}
Frank Wen.
\newblock {FluidR3\_GM}: a {General MIDI} {SoundFont}.
\newblock \url{https://musical-artifacts.com/artifacts/738}, 2000.
\newblock MIT license; Accessed: September 20, 2026.

\bibitem[Wu et~al.(2025)Wu, Yan, Hu, Yi, Feng, Tian, Shen, Yu, Zhang, Li,
  et~al.]{wu2025step}
Boyong Wu, Chao Yan, Chen Hu, Cheng Yi, Chengli Feng, Fei Tian, Feiyu Shen,
  Gang Yu, Haoyang Zhang, Jingbei Li, et~al.
\newblock Step-audio 2 technical report.
\newblock \emph{arXiv preprint arXiv:2507.16632}, 2025.

\bibitem[Xi et~al.(2018)Xi, Bittner, Pauwels, Ye, and Bello]{xi2018guitarset}
Qingyang Xi, Rachel~M Bittner, Johan Pauwels, Xuzhou Ye, and Juan~Pablo Bello.
\newblock Guitarset: A dataset for guitar transcription.
\newblock Proceedings of the 19th International Society for Music Information
  Retrieval Conference (ISMIR), 2018.

\bibitem[Xiaomi(2025)]{xiaomi2025mimo}
LCT Xiaomi.
\newblock Mimo-audio: Audio language models are few-shot learners.
\newblock \emph{CoRR, abs/2512.23808}, 2025.

\bibitem[Xu et~al.(2025)Xu, Guo, He, Hu, He, Bai, Chen, Wang, Fan, Dang, Zhang,
  Wang, Chu, and Lin]{xu2025qwen3}
Jin Xu, Zhifang Guo, Jinzheng He, Hangrui Hu, Ting He, Shuai Bai, Keqin Chen,
  Jialin Wang, Yang Fan, Kai Dang, Bin Zhang, Xiong Wang, Yunfei Chu, and
  Junyang Lin.
\newblock {Qwen2.5-Omni} technical report.
\newblock \emph{arXiv preprint arXiv:2503.20215}, 2025.

\bibitem[Yang et~al.(2024)Yang, Xu, Liu, Chu, Jiang, Zhou, Leng, Lv, Zhao,
  Zhou, et~al.]{yang2024airbench}
Qian Yang, Jin Xu, Wenrui Liu, Yunfei Chu, Ziyue Jiang, Xiaohuan Zhou, Yichong
  Leng, Yuanjun Lv, Zhou Zhao, Chang Zhou, et~al.
\newblock Air-bench: Benchmarking large audio-language models via generative
  comprehension.
\newblock In \emph{Proceedings of the 62nd Annual Meeting of the Association
  for Computational Linguistics (Volume 1: Long Papers)}, pp.\  1979--1998,
  2024.

\bibitem[Zang et~al.(2025)Zang, O'Brien, Berg-Kirkpatrick, McAuley, and
  Novack]{zang2025you}
Yongyi Zang, Sean O'Brien, Taylor Berg-Kirkpatrick, Julian McAuley, and Zachary
  Novack.
\newblock Are you really listening? boosting perceptual awareness in music-qa
  benchmarks.
\newblock \emph{arXiv preprint arXiv:2504.00369}, 2025.

\bibitem[Zhang et~al.(2019)Zhang, Kishore, Wu, Weinberger, and
  Artzi]{zhang2019bertscore}
Tianyi Zhang, Varsha Kishore, Felix Wu, Kilian~Q Weinberger, and Yoav Artzi.
\newblock Bertscore: Evaluating text generation with bert.
\newblock \emph{arXiv preprint arXiv:1904.09675}, 2019.

\bibitem[Zheng et~al.(2024)Zheng, Zhou, Meng, Zhou, and Huang]{zheng2024large}
Chujie Zheng, Hao Zhou, Fandong Meng, Jie Zhou, and Minlie Huang.
\newblock Large language models are not robust multiple choice selectors.
\newblock In \emph{International Conference on Learning Representations},
  volume 2024, pp.\  19426--19454, 2024.

\end{thebibliography}
\bibliographystyle{iclr2027_conference} 

\clearpage
\appendix

\FloatBarrier
\section{Benchmark construction}
\label{app:construction}

\paragraph{Generation.} Table~\ref{tab:app-gen} gives the generator settings. All settings, including the split rules, were fixed in configuration files before any model was trained.

\begin{table}[!htbp]
  \caption{Generator settings. MIDI note 60 is middle C. Training and test pitch ranges do not overlap.}
  \label{tab:app-gen}
  \centering
  \small
  \begin{tabularx}{\linewidth}{@{}l X@{}}
    \toprule
    Task    & Setting \\
    \midrule
    Melody  & Four-note motif, 0.4\,s per note, in a 2.5\,s clip. When the melodies differ, one note moves by 100, 200, 400 or 700 cents. Starting pitch from MIDI 60 to 72 (train) or 55 to 59 (test). \\
    Harmony & Triad in a 2.0\,s clip. When the chord qualities differ, the third moves by 100, 200 or 300 cents. Major thirds move down and minor thirds move up, so a 100-cent move turns major into minor and minor into major. The moved third stays at least 50 cents away from the root and the fifth. Root from MIDI 48 to 60 (train) or 62 to 67 (test). \\
    Timbre  & Short melody in a 2.5\,s clip on piano, guitar, violin or flute. Two different instruments always come from different families; close pairs such as acoustic and electric piano are excluded \rev{here and tested in Appendix~\ref{app:timbre}}. Starting pitch as for melody. \\
    Rhythm  & Hi-hat click track in a 4.0\,s clip, 50\,ms clicks. The click rates of the two clips differ by a ratio of 1.10, 1.20, 1.35 or 1.50. The time of the first click is drawn independently of the answer. \\
    \bottomrule
  \end{tabularx}
\end{table}

\paragraph{Audio versions of each item.} From each generated item we render a base pair (clips A and B), one edited clip B that changes the answer, and two or three clips B that only change the sound. Table~\ref{tab:app-nuis} lists the sound-only changes. Each keeps the queried attribute fixed, so the correct answer stays the same. \rev{As in the standard same/different design, clips A and B of a clean `same' pair of melody, harmony and timbre are identical; the \STAY{} versions give the same pairs a difference in sound only (Section~\ref{sec:flipstay}).}

\begin{table}[!htbp]
  \caption{Sound-only changes by split and task.}
  \label{tab:app-nuis}
  \centering
  \small
  \begin{tabularx}{\linewidth}{@{}l >{\raggedright\arraybackslash}X l l@{}}
    \toprule
    Change          & Setting                                                                 & Split       & Tasks \\
    \midrule
    Room echo       & pyroomacoustics simulation, reverberation time 0.2, 0.5 or 1.0\,s        & test        & all \\
    EQ              & shelving filters, $\pm 3$\,dB at 200\,Hz and 4\,kHz                      & test        & all \\
    Hiss            & added noise at $-50$\,dBFS                                               & test        & all \\
    Codec           & Opus at 48\,kbps \citep{valin2012definition} or MP3 at 128\,kbps               & train       & all \\
    Soundfont swap  & same instrument, other soundfont                                         & train       & melody, harmony, rhythm \\
    New melody, key & melody and key drawn again, same instrument                              & train       & timbre \\
    Transposition   & 0 to 200 cents                                                           & train, test & melody, harmony \\
    \bottomrule
  \end{tabularx}
\end{table}

\paragraph{Length, loudness and format.} After every change we decode (for the codecs), trim and check that the number of samples equals the clip length of the task. A test file (clip A, 1.0\,s of silence, clip B) is therefore 6.0\,s long for melody and timbre, 5.0\,s for harmony and 9.0\,s for rhythm, with the same number of samples for every file in a task. Every clip is normalized to $-23$\,LUFS \citep{ebu2011loudness} and stored as 16-bit PCM WAV at 44.1\,kHz, mono.

\paragraph{\texorpdfstring{\FLIP{} and \STAY{} items}{FLIP and STAY items}.} These items use no new audio; they recombine test renders. Each clean test item gives one \FLIP{} and one \STAY{} item. Every \STAY{} item keeps the original clip A and takes clip B from a sound-only change. For melody, harmony and timbre, the edit changes clip B only, and clip A of the edited render is byte-identical to the original clip A. The rhythm edit swaps the two clips, so rhythm \FLIP{} items use the swapped pair. The correct answer of every \FLIP{} item differs from that of its base item (1{,}000 of 1{,}000). For melody, harmony and timbre the correct letter changes too. For rhythm the faster clip changes, but each version draws its own wording (faster or slower), so the correct letter changes in 130 of the 250 rhythm \FLIP{} items (Section~\ref{sec:flipstay}). The sound-only changes rotate over items (Table~\ref{tab:app-flipstay}).

\begin{table}[!htbp]
  \caption{Composition of the \FLIP/\STAY{} test (2{,}000 items).}
  \label{tab:app-flipstay}
  \centering
  \small
  \begin{tabular}{@{}lrrrrrr@{}}
    \toprule
            &          & \multicolumn{4}{c}{\STAY{} change} & \\
    \cmidrule(lr){3-6}
    Task    & \FLIP{}  & EQ  & Hiss & Room & Transposition & Total \\
    \midrule
    Melody  & 250      & 63  & 63   & 62   & 62            & 500 \\
    Harmony & 250      & 63  & 63   & 62   & 62            & 500 \\
    Timbre  & 250      & 84  & 83   & 83   & n/a           & 500 \\
    Rhythm  & 250      & 84  & 83   & 83   & n/a           & 500 \\
    \midrule
    Total   & 1{,}000  & 294 & 292  & 290  & 124           & 2{,}000 \\
    \bottomrule
  \end{tabular}
\end{table}

\paragraph{Questions.} Each task has 8 to 10 hand-written wordings. Rhythm has 10 wordings, some asking for the faster clip and some for the slower one, and the correct letter follows the wording. Every question ends with a fixed instruction that says what each letter means, for example ``Only answer `A' for the same, `B' for different \ldots''. One wording is drawn per test file with a fixed seed, so the base, \FLIP{} and \STAY{} versions of an item usually have different wordings. Test items keep a fixed letter meaning. Training items swap the meaning for about half of the items, balanced so that the letter carries no information about the answer. Tables~\ref{tab:wordings} and~\ref{tab:instructions} list every wording and
the exact instruction under both letter meanings.

\begingroup
\renewcommand{\arraystretch}{1.2}

\begin{longtable}{@{}
    >{\raggedright\arraybackslash}p{1.6cm}
    >{\raggedright\arraybackslash}p{8.9cm}
    >{\raggedright\arraybackslash}p{2.6cm}@{}}

\caption{All 34 question wordings, verbatim. Each prompt is
\emph{wording}~+~\emph{fixed instruction} (Table~\ref{tab:instructions})
followed by the audio placeholder \texttt{<audio>}, to which the item's
concatenated clip is attached. Gold letters are balanced (train: 49.6\%~A;
test: 52.8\%~A).}\label{tab:wordings}\\

\toprule
Task & Question wording & Correct letter is the\\
\midrule
\endfirsthead

\multicolumn{3}{@{}l}{\itshape Table~\ref{tab:wordings} (continued)}\\
\toprule
Task & Question wording & Correct letter is the\\
\midrule
\endhead

\midrule
\multicolumn{3}{r@{}}{\footnotesize\itshape continued on next page}\\
\endfoot

\bottomrule
\endlastfoot

Melody
  & Two melodic clips are joined by a brief silence. Do they play the same melody, or a different one?
  & -- \\
  & You will hear two short melodies separated by a short pause. Is the melody the same in both, or different?
  & -- \\
  & Listen to two melodic phrases with a brief gap between them. Same tune, or different?
  & -- \\
  & A pair of melodies is presented back to back with a short silence in between. Are they the same melody?
  & -- \\
  & Two musical phrases follow one another after a brief pause. Is the second phrase the same melody as the first, or different?
  & -- \\
  & You are given two clips of melodic material with a short silent gap. Do they share the same melody?
  & -- \\
  & Listen carefully to the two melodic clips, divided by a brief silence. Same melody, or not?
  & -- \\
  & Two short tunes are played in sequence with a pause between them. Is it the same melody twice, or two different melodies?
  & -- \\
\midrule

Harmony
  & Two chords are played in sequence with a brief silence between them. Is the chord the same in both, or different?
  & -- \\
  & You will hear two chords separated by a short pause. Same chord, or different?
  & -- \\
  & Listen to two chords joined by a brief gap of silence. Do they sound like the same kind of chord, or different?
  & -- \\
  & A pair of chords is presented back to back with a short silence in between. Is the harmony the same, or different?
  & -- \\
  & Two short chords follow each other after a brief pause. Same chord type, or different?
  & -- \\
  & You are given two chord clips with a short silent gap. Do they share the same chord type?
  & -- \\
  & Listen to the two chords, divided by a brief silence. Same harmonic quality, or not?
  & -- \\
  & Two chords are played one after the other with a pause between them. Is it the same chord twice, or two different qualities?
  & -- \\
\midrule

Timbre
  & Two clips are joined by a brief silence. Are they played by the same instrument, or different instruments?
  & -- \\
  & You will hear two short clips separated by a short pause. Same instrument in both, or different?
  & -- \\
  & Listen to two clips with a brief gap between them. Is it the same instrument, or a different one?
  & -- \\
  & A pair of clips is presented back to back with a short silence in between. Are they performed on the same instrument?
  & -- \\
  & Two clips follow one another after a brief pause. Is the second clip the same instrument as the first, or different?
  & -- \\
  & You are given two clips with a short silent gap. Do they share the same instrument?
  & -- \\
  & Listen carefully to the two clips, divided by a brief silence. Same instrument, or not?
  & -- \\
  & Two clips are played in sequence with a pause between them. Is it the same instrument twice, or two different instruments?
  & -- \\
\midrule

Rhythm
  & Two audio segments are given, with a brief silence in between. Which segment moves faster?
  & faster clip \\
  & Two audio segments are given, with a brief silence in between. Which segment moves slower?
  & slower clip \\
  & You are presented with two sound clips, separated by a short silent interval. Which clip carries more speed?
  & faster clip \\
  & You are presented with two sound clips, separated by a short silent interval. Which part proceeds at a slower pace?
  & slower clip \\
  & Listen to a pair of audio segments with a brief silent space separating them. Which audio feels slower?
  & slower clip \\
  & You are given two clips with a small silent gap in between. Which segment moves faster?
  & faster clip \\
  & Two auditory recordings are given, separated by a small pause. Which part has a faster rhythm?
  & faster clip \\
  & You are provided with two audio clips separated by a short silent gap. Which part proceeds at a faster pace?
  & faster clip \\
  & Two sound recordings are presented, separated by a brief silent segment. Which clip runs more leisurely?
  & slower clip \\
  & You have two auditory segments, separated by a short pause. Which segment plays more rapidly?
  & faster clip \\

\end{longtable}
\endgroup

\begin{table}[!htbp]
\centering
\footnotesize
\caption{Fixed instruction appended verbatim to every wording (before the
audio placeholder), under both letter meanings. The swapped order applies
only to same/different tasks, to 51.4\% of their training items; the test
split always uses the original order.}\label{tab:instructions}
\begin{tabular}{@{}p{4.6cm}p{8.9cm}@{}}
\toprule
Applies to & Instruction (verbatim) \\
\midrule
Melody/Harmony/Timbre, original order & ``Only answer `A' for the same, `B' for different. Do not add any explanation, punctuation, or extra text.'' \\
Melody/Harmony/Timbre, swapped order & ``Only answer `B' for the same, `A' for different. Do not add any explanation, punctuation, or extra text.'' \\
Rhythm, both splits & ``Only answer `A' for the first, `B' for the second. Do not add any explanation, punctuation, or extra text.'' \\
\bottomrule
\end{tabular}
\end{table}

\paragraph{Statistics.} The benchmark contains 11{,}825 generated items: 3{,}025 each for melody, harmony and timbre and 2{,}750 for rhythm. Melody, harmony and timbre include extra catch trials with two identical clips, which we do not use in this paper. The items used here are 10{,}000 training items, 1{,}000 clean test items and 2{,}000 \FLIP/\STAY{} items (250 of each kind per task in the test split). Of the 1{,}000 clean test items, 528 have the answer `A' and 472 the answer `B'. \rev{The real test set adds 600 clean and 1{,}200 \FLIP/\STAY{} items (Appendix~\ref{app:real}).}

\paragraph{Soundfont overlap.} MuseScore\_General is derived from FluidR3Mono, a mono version of FluidR3. Its documentation states that instruments without separate credit still use FluidR3Mono samples, and its changelog lists new piano samples \footnote{\url{https://github.com/CarlGao4/Muse-Sounds}}. For instruments that keep FluidR3Mono samples, the test soundfont changes the file and its instrument settings, but the source recordings overlap with FluidR3\_GM, one of the two training soundfonts. The soundfont change is therefore weaker for those instruments than for piano. \rev{The real test set (Appendix~\ref{app:real}) shares no audio with any soundfont.}

\FloatBarrier
\section{Human reference study}
\label{app:human}
12 people took part with various levels of musical expertise. Each answered 24 test items, so we have 288 answers. They heard the same audio files and questions as the models. We removed three test runs and one empty submission. Accuracy was 87.5\% on clean items (84 of 96; 95\% interval 79.4 to 92.7), 94.8\% on \FLIP{} items (91 of 96; 88.4 to 97.8) and 79.2\% on \STAY{} items (76 of 96; 70.0 to 86.1). By task, over all items, it was 97.2\% for timbre (70 of 72), 94.4\% for rhythm (68 of 72), 83.3\% for melody (60 of 72) and 73.6\% for harmony (53 of 72). Per person, accuracy ranged from 66.7\% to 95.8\%. \rev{The intervals are Wilson intervals over answers. Because each person gave 24 answers, we also resample listeners: drawing the 12 listeners with replacement gives a 95\% interval of 82.3\% to 91.1\% for \FLIP/\STAY accuracy.} \rev{The test was open to anyone. We did not select listeners by musical background or control who took part, and each listener chose one of four levels of musical experience: casual listener; amateur musician (plays or sings, little formal training); trained musician (5 or more years of lessons or a music degree); music or audio researcher. Four listeners are casual listeners, two amateur musicians, four trained musicians and two music researchers. On \FLIP/\STAY, the casual listeners score 82.8\% (53/64) and the eight with musical experience 89.1\% (114/128); on clean items, both groups score 87.5\% (28/32 and 56/64). All 12 listeners heard the same 24 items, and agreement between them is high (Fleiss' $\kappa$ 0.71 over all items, 0.76 over the \FLIP/\STAY{} items).}

\begin{table}[!htbp]
  \caption{Human answers by task. Clean: 24 answers per task. \FLIP/\STAY: 48 answers per task, both variants together. Over all tasks there are 5 misses and 20 false flips, so at least 11 of the false flips are on harmony and at most one each on timbre and rhythm.}
  \label{tab:app-human}
  \centering
  \small
  \begin{tabular}{@{}l rr rr@{}}
    \toprule
            & \multicolumn{2}{c}{Clean} & \multicolumn{2}{c}{\FLIP/\STAY} \\
    \cmidrule(lr){2-3}\cmidrule(l){4-5}
    Task    & Correct    & \%   & Correct     & \%   \\
    \midrule
    Melody  & 19 of 24   & 79.2 & 41 of 48    & 85.4 \\
    Harmony & 21 of 24   & 87.5 & 32 of 48    & 66.7 \\
    Timbre  & 23 of 24   & 95.8 & 47 of 48    & 97.9 \\
    Rhythm  & 21 of 24   & 87.5 & 47 of 48    & 97.9 \\
    \midrule
    All     & 84 of 96   & 87.5 & 167 of 192  & 87.0 \\
    \bottomrule
  \end{tabular}
\end{table}

\FloatBarrier
\section{Models and inference}
\label{app:models}
\paragraph{Open models.} All open models are tested without training and with their default chat templates. Each item is one audio file and one text question. We compare the log-probabilities of the tokens `A' and `B' at the first answer position and take the larger. Greedy decoding gives the same letter on every item. Table~\ref{tab:open_models} lists the exact checkpoint (Hugging Face name and revision), parameter count, and audio sampling rate for each of the eight models.

\begin{table}[!htbp]
\centering
\footnotesize
\caption{Open-model checkpoints. Revisions are the local Hugging Face cache
commits actually used for evaluation; parameter counts are computed from the
checkpoint safetensors headers.}
\label{tab:open_models}
\begin{adjustbox}{max width=\columnwidth}
\begin{tabular}{@{}llrr@{}}
\toprule
Model & Hugging Face checkpoint (revision) & Params & Audio sr \\
\midrule
Qwen2-Audio      & \texttt{Qwen/Qwen2-Audio-7B-Instruct} (\texttt{0a09522}) & 8.4B & 16\,kHz \\
Qwen2.5-Omni     & \texttt{Qwen/Qwen2.5-Omni-7B} (\texttt{ae9e169}) & 10.7B\textsuperscript{$\dagger$} & 16\,kHz \\
Audio Flamingo 3 & \texttt{nvidia/audio-flamingo-3-hf} (\texttt{7d4bae6}) & 16.5B & 16\,kHz\textsuperscript{$\ddagger$} \\
Phi-4-MM         & \texttt{microsoft/Phi-4-multimodal-instruct} (\texttt{93f923e}) & 5.6B & 16\,kHz \\
Kimi-Audio       & \texttt{moonshotai/Kimi-Audio-7B-Instruct} (\texttt{9a82a84}) & $\sim$7.9B\textsuperscript{$\S$} & 16\,kHz \\
MiMo-Audio       & \texttt{XiaomiMiMo/MiMo-Audio-7B-Instruct} (\texttt{c359441}) & 8.0B\textsuperscript{$\P$} & 24\,kHz \\
Fun-Audio-Chat   & \texttt{FunAudioLLM/Fun-Audio-Chat-8B} (\texttt{7bf72dc}) & 9.5B & 16\,kHz \\
Step-Audio 2 mini & \texttt{stepfun-ai/Step-Audio-2-mini} (\texttt{e36fdd5}) & 8.3B\textsuperscript{$\|$} & 16\,kHz \\
\bottomrule
\end{tabular}
\end{adjustbox}

\raggedright\scriptsize
\textsuperscript{$\dagger$}Full checkpoint (thinker 8.9B + talker 1.4B + token2wav 0.5B); evaluation instantiates the thinker only.
\textsuperscript{$\ddagger$}From the model card; the cached snapshot contains no processor config to confirm.
\textsuperscript{$\S$}Model-card figure; local count not completed.
\textsuperscript{$\P$}Excludes the separate MiMo-Audio-Tokenizer (1.3B, rev \texttt{5df9914}).
\textsuperscript{$\|$$|$}Excludes the token2wav vocoder weights.
\end{table}

\paragraph{Commercial models.} GPT-Audio-1.5 is called through the OpenAI API with the model \texttt{gpt-audio-1.5}. Gemini 3.8 Flash is called through the Gemini API with the model \texttt{gemini-3.8-flash} (high thinking). Both are called at temperature 0. Each trial's two clips are concatenated into a single mono 16-bit, 44.1\,kHz WAV and sent inline (base64) with the question. Neither API exposes option log-probabilities for these models, so both are scored by their generated letter. For Gemini we constrain decoding to the two option letters, so every response is a valid letter; no invalid answers occurred (0 in 3{,}000 calls). Chat Completions offers no such constraint for audio models, so GPT-Audio-1.5 returns a completion of at most 16 tokens, from which we read the first standalone option letter (word-bounded, case-insensitive). Responses containing no standalone letter (10.3\% clean, 6.9\% \rev{\FLIP/\STAY}) are recorded as a prediction of the first option; under the balanced answer key this fallback scores at chance, and all such items are flagged in the released logs. GPT-Audio-mini and Gemini 2.5 Pro are called through the OpenRouter API \citep{openrouter} with the model IDs \texttt{openai/gpt-audio-mini} and \texttt{google/gemini-2.5-pro}, on the same WAV files and prompts. We read the letter from the generated text. \rev{A reply without an answer letter is recorded as the first option (`A'), as for GPT-Audio-1.5. Such replies make up 1{,}385 of the 3{,}000 replies of GPT-Audio-mini and 21 of those of Gemini 2.5 Pro. Appendix~\ref{app:paired} also scores them as errors.} All API runs were conducted in September 2026.
\FloatBarrier
\section{GRPO details}
\label{app:grpo}
GRPO uses only the clean base pairs of the training split; none of the sound-only versions in Table~\ref{tab:app-nuis} is used for training. \rev{The one exception is the control run R1 of Appendix~\ref{app:trainbase}, which adds them.} We report the checkpoint at step 1{,}000 for all runs except Audio Flamingo 3 with full fine-tuning, which uses step 500. Leave-one-task-out runs use LoRA for 500 steps on 7{,}500 items and the final checkpoint. 

\begin{table}[!htbp]
\centering
\caption{GRPO training details. Batch size is given as prompts per GPU per step
($B$) and rollouts per prompt (group size $G$); all runs use one gradient update
per step. Wall-clock times are elapsed times measured on a shared node
(\emph{cf.}~Footnote~\ref{fn:wallclock}). Q2A~= Qwen2-Audio-7B-Instruct,
Omni~= Qwen2.5-Omni-7B, AF3~= Audio-Flamingo-3, Phi-4~= Phi-4-Multimodal.}
\label{tab:grpo_details}
\small
\begin{tabular}{@{}p{3.5cm}p{4.7cm}p{4.7cm}@{}}
\toprule
 & LoRA & Full fine-tuning \\
\midrule
Group size $G$ (rollouts/prompt) & 8 & 8 \\
Sampling temperature / top-$p$ & $1.0$ / $1.0$ & $1.0$ / $1.0$ \\
Batch size $B$ (prompts/GPU/step) & 4 & 4 \\
Effective batch (rollouts/step) & $64$ ($8\times 8$)\textsuperscript{a} & $64$ ($8\times 8$) \\
Learning rate (schedule) & $1{\times}10^{-5}$, constant & $1{\times}10^{-6}$, constant \\
Optimizer & AdamW & Adafactor \\
Gradient clipping & $1.0$ & $1.0$ \\
KL coefficient (vs.\ frozen reference) & $0.10$ & $0.10$ \\
LoRA rank / $\alpha$ / dropout & 16 / 32 / 0 & ---  \\
Trainable parameters & LM-decoder linear projections only\textsuperscript{b} & LM decoder (+ lm\_head) \\
Max.\ answer length & 8 new tokens & 8 new tokens \\
Training steps & 1000 (all models) & 1000 (Q2A, Omni, Phi-4); 500 (AF3) \\
Hardware & 2 GPUs per run\textsuperscript{c} & $2\times$ NVIDIA A100 (80\,GB) \\
Wall-clock time (h) & Q2A $\approx$2.3, Omni $\approx$5.9, AF3 $\approx$3.9, Phi-4 $\approx$0.8 &
Q2A $\approx$2.7, Omni $\approx$2.8, AF3 $\approx$2.7, Phi-4 $\approx$25.5\textsuperscript{d} \\
\midrule
\multicolumn{3}{@{}p{13.7cm}@{}}{Reward (per rollout):
$r = 1.0\cdot\mathbb{1}[\text{correct}] + 0.2\cdot\mathbb{1}[\text{parses to A/B}]
+ 0.1\cdot s_{\text{task}}\cdot(\pm1)$, where the last term is an anti-collapse
nudge ($+$ if the rollout predicts the currently under-represented letter for
that task, $-$ otherwise; $s_{\text{task}}$ is an EMA with decay $0.95$ of the
per-task predicted-letter skew). Zero-advantage groups are resampled up to 4
extra times before being dropped.} \\
\bottomrule
\end{tabular}

\begin{flushleft}\scriptsize
\textsuperscript{a}Qwen2.5-Omni LoRA ran on 6 GPUs: 24 prompts $\times$ 8 = 192 rollouts/step.
\textsuperscript{b}Attention and MLP projections of the language-model decoder
($q/k/v/o/\text{gate}/\text{up}/\text{down}$; fused $qkv/o/\text{gate\_up}/\text{down}$ for Phi-4); audio encoders and projectors frozen.
\textsuperscript{c}Single 8-GPU node: $2\times$ A100 (80\,GB) $+$ $6\times$ A40 (46\,GB); the exact A100/A40 assignment per LoRA run was not recorded.
\textsuperscript{d}\label{fn:wallclock}Elapsed time between first and last checkpoint writes on a node shared with other jobs; the Phi-4 full-FT run overlapped several concurrent jobs and its elapsed time is substantially inflated relative to exclusive-GPU compute (cf.\ $\approx$2.7--2.8\,h for the other full-FT runs).
\end{flushleft}
\end{table}

\FloatBarrier
\section{Leaderboard submission format}
\label{app:leaderboard}
A submission is one JSON Lines file with one record per test item, holding the item ID and the answer letter (`A' or `B'). It also carries the model name and version, how the answer was read (letter probabilities or generated text), whether the model used our training split in any form, the date of the run and an optional link to code. Missing or invalid answers count as wrong. \rev{The same format applies to both test sets.} The scoring script is the one used for Tables~\ref{tab:base} and~\ref{tab:grpo}. Training on test items is not allowed.

\FloatBarrier
\section{Full results}
\label{app:full}
\begin{figure}[!htbp]
  \centering
  \includegraphics[width=4.0in]{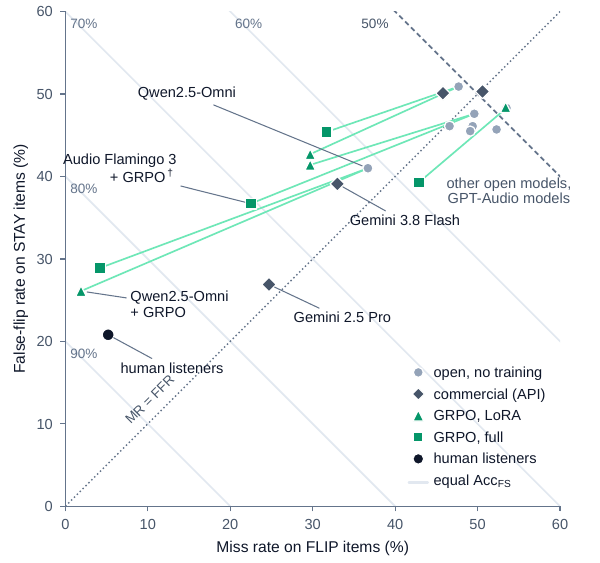}
  \caption{Every model on the two \FLIP/\STAY{} error rates, before and after GRPO. The accuracy $\accfs$ is constant along each grey diagonal; a model that ignores the audio lies on the dashed diagonal (50\%). Grey circles: open models without training. Dark diamonds: commercial models. Green: GRPO runs (triangles LoRA, squares full fine-tuning), joined to their untrained model by thin lines. Black: our listeners. Without training, the models that listen (Gemini 2.5 Pro, Gemini 3.8 Flash, Qwen2.5-Omni) lie near the dotted line of equal rates. GRPO moves models to the left, toward fewer misses, far more than down: the best run reaches 1.9\% misses but keeps 26.1\% false flips, the level of Gemini 2.5 Pro (26.9\%). $^\dagger$Step 500.}
  \label{fig:plane}
\end{figure}

\begin{figure}[!htbp]
  \centering
  \begin{minipage}[c]{0.56\linewidth}
    \centering
    \includegraphics[width=\linewidth]{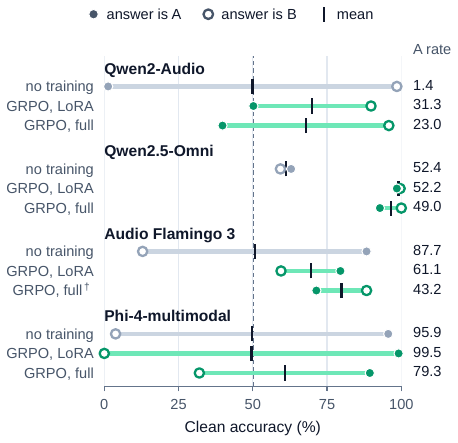}
  \end{minipage}\hfill
  \begin{minipage}[c]{0.41\linewidth}
    \caption{Letter bias on the clean test. Filled dot: accuracy on items whose correct answer is `A'. Open dot: accuracy on items whose answer is `B'. Tick: their mean, the class-balanced accuracy. Grey rows: no training; green rows: after GRPO. Right column: share of `A' answers (52.8\% of correct answers are `A'). $^\dagger$Step 500. Without training, three of the four models sit near the two ends of the scale, with a mean near 50\%. GRPO moves the dots closer together for Qwen2-Audio, Audio Flamingo 3 and Phi-4-multimodal with full fine-tuning, and lifts both dots of Qwen2.5-Omni above 92\%. }
    \label{fig:bias}
  \end{minipage}
\end{figure}

Figure~\ref{fig:plane} places every model on the two error rates of \FLIP/\STAY, and Figure~\ref{fig:bias} shows the letter bias of the four trained models. The tables below give the per-task numbers behind Tables~\ref{tab:base} and~\ref{tab:grpo}.

\begin{table}[!htbp]
  \caption{Per-task miss rate and false-flip rate (\%) without training. Commercial rows: the two models analyzed item by item in Appendix~\ref{app:api-items}.}
  \label{tab:app-base-task}
  \centering
  \small
  \begin{tabular}{@{}l rrrr rrrr@{}}
    \toprule
                     & \multicolumn{4}{c}{MR} & \multicolumn{4}{c}{FFR} \\
    \cmidrule(lr){2-5}\cmidrule(l){6-9}
    Model            & Mel. & Har. & Tim. & Rhy. & Mel. & Har. & Tim. & Rhy. \\
    \midrule
    Qwen2-Audio      & 46.0 & 50.4 & 46.0 & 48.4 & 54.0 & 49.6 & 52.8 & 47.2 \\
    Qwen2.5-Omni     & 38.0 & 36.0 & 24.8 & 48.0 & 42.0 & 45.6 & 33.6 & 42.8 \\
    Audio Flamingo 3 & 50.4 & 49.6 & 54.4 & 44.0 & 46.4 & 52.4 & 41.2 & 50.4 \\
    Phi-4-multimodal & 54.0 & 51.6 & 56.8 & 51.6 & 46.8 & 50.4 & 43.2 & 52.8 \\
    Kimi-Audio       & 53.2 & 49.2 & 56.0 & 50.8 & 41.2 & 52.4 & 42.8 & 46.4 \\
    MiMo-Audio       & 54.0 & 47.2 & 54.0 & 42.4 & 46.0 & 50.0 & 41.6 & 46.8 \\
    Fun-Audio-Chat   & 50.0 & 41.2 & 43.2 & 52.0 & 50.4 & 48.4 & 36.4 & 49.2 \\
    Step-Audio 2     & 55.2 & 49.2 & 42.0 & 50.0 & 49.2 & 46.4 & 38.0 & 48.4 \\
    \midrule
    GPT-Audio-mini   & 50.4 & 47.6 & 52.8 & 51.6 & 53.6 & 51.2 & 43.6 & 52.8 \\
    Gemini 2.5 Pro   & 23.2 & 35.2 & 19.2 & 21.2 & 30.4 & 37.2 & 16.0 & 24.0 \\
    \bottomrule
  \end{tabular}
\end{table}

\begin{table}[!htbp]
  \caption{Per-task miss rate and false-flip rate (\%) after GRPO. $^\dagger$Step 500.}
  \label{tab:app-grpo-task}
  \centering
  \small
  \begin{tabular}{@{}ll rrrr rrrr@{}}
    \toprule
                     &                & \multicolumn{4}{c}{MR} & \multicolumn{4}{c}{FFR} \\
    \cmidrule(lr){3-6}\cmidrule(l){7-10}
    Model            & GRPO           & Mel. & Har. & Tim. & Rhy. & Mel. & Har. & Tim. & Rhy. \\
    \midrule
    Qwen2-Audio      & LoRA           & 26.0 & 40.4 & 4.4  & 48.0 & 45.2 & 46.8 & 34.0 & 44.8 \\
                     & full           & 29.6 & 41.6 & 6.8  & 48.8 & 46.8 & 45.2 & 40.8 & 48.8 \\
    Qwen2.5-Omni     & LoRA           & 0.4  & 4.0  & 3.2  & 0.0  & 32.4 & 34.4 & 20.8 & 16.8 \\
                     & full           & 0.0  & 15.2 & 1.2  & 0.4  & 34.0 & 41.2 & 22.8 & 17.6 \\
    Audio Flamingo 3 & LoRA           & 40.4 & 24.4 & 2.4  & 51.6 & 50.8 & 41.2 & 20.4 & 53.2 \\
                     & full$^\dagger$ & 34.4 & 20.0 & 2.4  & 33.2 & 48.4 & 38.8 & 20.4 & 39.2 \\
    Phi-4-multimodal & LoRA  & 55.6 & 49.6 & 56.8 & 51.6 & 47.2 & 50.4 & 43.2 & 52.8 \\
                     & full           & 48.8 & 45.6 & 25.6 & 51.6 & 48.4 & 46.0 & 10.0 & 52.8 \\
    \bottomrule
  \end{tabular}
\end{table}

\begin{table}[!htbp]
  \caption{Letter bias on the clean test (\%). A rate: share of items answered `A' (52.8\% of correct answers are `A'). Acc$|$A and Acc$|$B: accuracy on items whose correct answer is `A' or `B'. Balanced: their mean. Gap: their difference. The last two rows are commercial models (Appendix~\ref{app:api-items}). \rev{The A rate of GPT-Audio-mini includes replies without an answer letter, which are recorded as `A' (Appendix~\ref{app:models}).}}
  \label{tab:app-bias}
  \centering
  \small
  \begin{tabular}{@{}ll rrrrr@{}}
    \toprule
    Model            & GRPO & A rate & Acc$|$A & Acc$|$B & Balanced & Gap \\
    \midrule
    Qwen2-Audio      & none & 1.4  & 1.3  & 98.5  & 49.9 & 97.2 \\
                     & LoRA & 31.3 & 50.2 & 89.8  & 70.0 & 39.6 \\
                     & full & 23.0 & 39.8 & 95.8  & 67.8 & 56.0 \\
    Qwen2.5-Omni     & none & 52.4 & 62.9 & 59.3  & 61.1 & 3.6 \\
                     & LoRA & 52.2 & 98.5 & 99.6  & 99.0 & 1.1 \\
                     & full & 49.0 & 92.8 & 100.0 & 96.4 & 7.2 \\
    Audio Flamingo 3 & none & 87.7 & 88.3 & 12.9  & 50.6 & 75.3 \\
                     & LoRA & 61.1 & 79.5 & 59.5  & 69.5 & 20.0 \\
                     & full & 43.2 & 71.4 & 88.3  & 79.8 & 17.0 \\
    Phi-4-multimodal & none & 95.9 & 95.6 & 3.8   & 49.7 & 91.8 \\
                     & LoRA & 99.5 & 99.1 & 0.0   & 49.5 & 99.1 \\
                     & full & 79.3 & 89.4 & 32.0  & 60.7 & 57.4 \\
    \midrule
    GPT-Audio-mini   & none & 81.1 & 80.5 & 18.2  & 49.4 & 62.3 \\
    Gemini 2.5 Pro   & none & 60.8 & 87.9 & 69.5  & 78.7 & 18.4 \\
    \bottomrule
  \end{tabular}
\end{table}

\begin{table}[!htbp]
  \caption{Per-task \FLIP/\STAY{} accuracy (\%) without training. Models: 500 items per task and 2{,}000 in All; on one task, a score between 45.6\% and 54.4\% is not significantly different from 50\%. Human listeners: 48 answers per task and 192 in All.}
  \label{tab:app-task-fs}
  \centering
  \small
  \begin{tabular}{@{}l rrrrr@{}}
    \toprule
    Model            & Mel. & Har. & Tim. & Rhy. & All \\
    \midrule
    Qwen2-Audio      & 50.0 & 50.0 & 50.6 & 52.2 & 50.7 \\
    Qwen2.5-Omni     & 60.0 & 59.2 & 70.8 & 54.6 & 61.2 \\
    Audio Flamingo 3 & 51.6 & 49.0 & 52.2 & 52.8 & 51.4 \\
    Phi-4-multimodal & 49.6 & 49.0 & 50.0 & 47.8 & 49.1 \\
    Kimi-Audio       & 52.8 & 49.2 & 50.6 & 51.4 & 51.0 \\
    MiMo-Audio       & 50.0 & 51.4 & 52.2 & 55.4 & 52.2 \\
    Fun-Audio-Chat   & 49.8 & 55.2 & 60.2 & 49.4 & 53.6 \\
    Step-Audio 2     & 47.8 & 52.2 & 60.0 & 50.8 & 52.7 \\
    \midrule
    GPT-Audio-mini   & 48.0 & 50.6 & 51.8 & 47.8 & 49.6 \\
    GPT-Audio-1.5    & 50.4 & 51.4 & 56.0 & 50.4 & 52.1 \\
    Gemini 2.5 Pro   & 73.2 & 63.8 & 82.4 & 77.4 & 74.2 \\
    Gemini 3.8 Flash & 57.2 & 51.2 & 84.2 & 63.2 & 64.0 \\
    \midrule
    Human (12)       & 85.4 & 66.7 & 97.9 & 97.9 & 87.0 \\
    \bottomrule
  \end{tabular}
\end{table}

\begin{table}[!htbp]
  \caption{Drop in miss rate and false-flip rate from no training to GRPO, in points. $z$ and $p$ (two-sided) test whether the miss rate fell more than the false-flip rate. The test treats the rates before and after training as independent samples of 1{,}000 items each, which overestimates the variance when they are positively correlated. $^\dagger$Step 500.}
  \label{tab:app-drop}
  \centering
  \small
  \begin{tabular}{@{}ll rrrrr@{}}
    \toprule
    Model            & GRPO           & $\Delta$MR & $\Delta$FFR & Ratio & $z$ & $p$ \\
    \midrule
    Qwen2-Audio      & LoRA           & 18.0 & 8.2 & 2.2 & 3.2 & 0.0015 \\
                     & full           & 16.0 & 5.5 & 2.9 & 3.4 & $<$0.001 \\
    Qwen2.5-Omni     & LoRA           & 34.8 & 14.9 & 2.3 & 7.6 & $<$0.001 \\
                     & full           & 32.5 & 12.1 & 2.7 & 7.6 & $<$0.001 \\
    Audio Flamingo 3 & LoRA           & 19.9 & 6.2 & 3.2 & 4.4 & $<$0.001 \\
                     & full$^\dagger$ & 27.1 & 10.9 & 2.5 & 5.4 & $<$0.001 \\
    Phi-4-multimodal & LoRA & 0.1 & $-0.1$ & n/a & 0.1 & 0.95 \\
                     & full           & 10.6 & 9.0 & 1.2 & 0.5 & 0.61 \\
    \bottomrule
  \end{tabular}
\end{table}

\FloatBarrier
\section{Captioning details}
\label{app:caption}
\begin{table}[!htbp]
  \caption{MusicCaps captioning before and after GRPO on 2{,}735 clips (scores $\times 100$). MET.: METEOR; R-L: ROUGE-L; BERTS.: BERTScore-F1; Words: mean caption length.}
  \label{tab:caption}
  \centering
  \small
  \begin{tabular}{@{}ll rrrrr@{}}
    \toprule
    Model        & GRPO & BLEU-4 & MET. & R-L & BERTS. & Words \\
    \midrule
      Qwen2-Audio  & none & 13.06 & 32.09 & 33.43 & 39.27 & 40.7 \\
                   & LoRA & 13.38 & 32.42 & 33.74 & 39.73 & 40.8 \\
                   & full & 14.79 & 34.22 & 35.28 & 41.22 & 41.3 \\
      \midrule
      Qwen2.5-Omni & none & 1.31  & 16.33 & 20.96 & 23.35 & 29.2 \\
                   & LoRA & 1.09  & 15.65 & 20.25 & 22.07 & 28.4 \\
                   & full & 1.60  & 16.69 & 21.23 & 23.50 & 30.0 \\
      \bottomrule
  \end{tabular}
\end{table}

We use the MusicCaps rows with \texttt{is\_audioset\_eval} set (2{,}858 clips). Of these, 2{,}735 could be downloaded and were captioned by every model version, so all comparisons use the same clips. Clips are 10\,s, 16\,kHz mono. Decoding is greedy, with at most 256 new tokens and repetition penalty 1.0. The prompt is ``Describe this music in a few sentences. Mention the genre, the instruments, the mood, and the tempo.'' Each clip has one reference caption. No model version produced degenerate captions.
\begin{table}[!htbp]
\centering
\caption{Paired difference from the base model on the 2{,}735 MusicCaps clips
($\Delta=\text{trained}-\text{base}$, in points $\times 100$); brackets give 95\%
bootstrap confidence intervals over clips (10{,}000 resamples). Stars denote
Wilcoxon signed-rank tests, Holm-corrected across metrics within each model
($^{*}p<.05$, $^{**}p<.01$, $^{***}p<.001$). BLEU-4$^{\dagger}$ is corpus-level,
so no per-clip Wilcoxon test applies. BERTScore-F1 uses \texttt{roberta-large}
(\texttt{lang=en}, default layer 17, \texttt{rescale\_with\_baseline=True}).
Checkpoints: GRPO step 1{,}000.}
\label{tab:musiccaps_paired}
\adjustbox{max width=\columnwidth}{%
\begin{tabular}{llccccc}
\toprule
Model & Variant & BLEU-4$^{\dagger}$ & sBLEU & METEOR & ROUGE-L & BERTScore-F1 \\
\midrule
\multirow{2}{*}{Qwen2-Audio} & LoRA
  & $+0.3\,[+0.2,+0.5]$
  & $+0.3\,[+0.2,+0.5]^{**}$
  & $+0.3\,[+0.1,+0.5]^{*}$
  & $+0.3\,[+0.1,+0.5]^{***}$
  & $+0.5\,[+0.3,+0.6]^{***}$ \\
& full FT
  & $+1.7\,[+1.5,+2.0]$
  & $+1.8\,[+1.5,+2.1]^{***}$
  & $+2.1\,[+1.8,+2.5]^{***}$
  & $+1.8\,[+1.6,+2.1]^{***}$
  & $+2.0\,[+1.7,+2.3]^{***}$ \\
\midrule
\multirow{2}{*}{Qwen2.5-Omni} & LoRA
  & $-0.2\,[-0.3,-0.1]$
  & $-0.1\,[-0.2,-0.1]^{***}$
  & $-0.7\,[-0.8,-0.5]^{***}$
  & $-0.7\,[-0.8,-0.6]^{***}$
  & $-1.3\,[-1.5,-1.0]^{***}$ \\
& full FT
  & $+0.3\,[+0.2,+0.4]$
  & $+0.1\,[+0.1,+0.2]^{***}$
  & $+0.4\,[+0.2,+0.5]^{***}$
  & $+0.3\,[+0.1,+0.4]^{***}$
  & $+0.2\,[-0.1,+0.4]$ \\
\bottomrule
\end{tabular}}
\end{table}

\FloatBarrier
\section{False flips by sound change}
\label{app:stay-change}

\begin{figure}[!htbp]
  \centering
  \includegraphics[width=\linewidth]{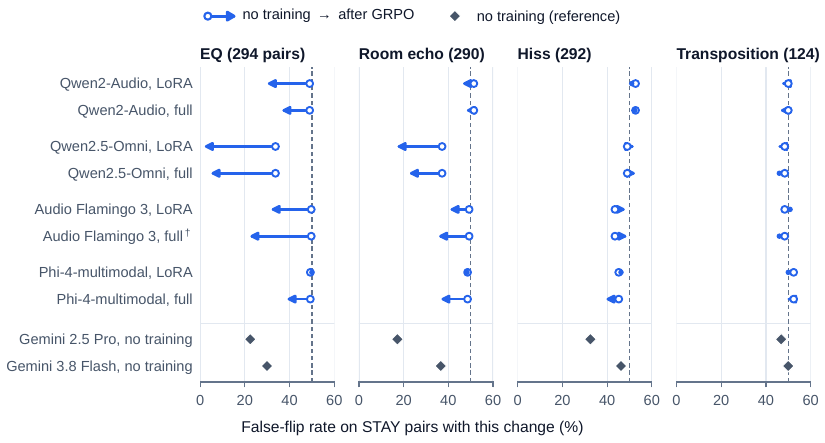}
  \caption{False-flip rate by kind of \STAY{} change, before and after GRPO, one row per run. Each blue arrow starts at the rate without training (open circle) and ends at the rate after GRPO; short changes are drawn as a line ending in a dot. Diamonds: the two Gemini models, without training. Dashed line: 50\%. GRPO lowers false flips under EQ in every run that learned (by 10.9 to 32.3 points), partly under room echo, and not under hiss or transposition. Gemini 2.5 Pro shows the same order without any training: room echo and EQ lowest, hiss higher, transposition near 50\%. $^\dagger$Step 500.}
  \label{fig:ffr-change}
\end{figure}

Each \STAY{} pair replaces clip B with a version that differs in one way (Section~\ref{sec:flipstay}): an EQ change ($\pm 3$\,dB shelves), room echo (reverberation time 0.2, 0.5 or 1.0\,s), background hiss ($-50$\,dBFS) or, for melody and harmony only, a transposition of up to 200 cents. The 1{,}000 \STAY{} pairs contain 294 EQ, 290 room-echo, 292 hiss and 124 transposition pairs. Table~\ref{tab:app-ffr-change} gives the false-flip rate for each change, pooled over the four tasks, and Table~\ref{tab:app-ffr-task} gives it per task. Both tables use the same answers as Tables~\ref{tab:base} and~\ref{tab:grpo}. Figure~\ref{fig:ffr-change} shows how GRPO changes the pooled rates.

\begin{table}[!htbp]
  \caption{False-flip rate (\%) for each kind of \STAY{} change, pooled over tasks; lower is better. Room: room echo. The number of pairs is given in brackets. \emph{All}: false-flip rate over all 1{,}000 \STAY{} pairs, as in Tables~\ref{tab:base} and~\ref{tab:grpo}. Transposition is used for melody and harmony only. $^\dagger$Step 500.}
  \label{tab:app-ffr-change}
  \centering
  \small
  \setlength{\tabcolsep}{5pt}
  \begin{tabular}{@{}ll rrrr r@{}}
    \toprule
    Model & GRPO & EQ & Room & Hiss & Transp. & All \\
          &      & (294) & (290) & (292) & (124) & (1{,}000) \\
    \midrule
    Qwen2-Audio & none & 49.0 & 51.4 & 52.7 & 50.0 & 50.9 \\
     & LoRA & 29.6 & 45.9 & 51.0 & 46.8 & 42.7 \\
     & full & 36.1 & 47.6 & 52.4 & 46.0 & 45.4 \\
    \midrule
    Qwen2.5-Omni & none & 33.7 & 37.2 & 49.0 & 48.4 & 41.0 \\
     & LoRA & 1.4 & 16.6 & 52.4 & 45.2 & 26.1 \\
     & full & 4.4 & 22.1 & 53.1 & 46.0 & 28.9 \\
    \midrule
    Audio Flamingo 3 & none & 49.7 & 49.3 & 43.5 & 48.4 & 47.6 \\
     & LoRA & 31.3 & 40.3 & 48.6 & 50.8 & 41.4 \\
     & full$^\dagger$ & 21.8 & 35.2 & 49.3 & 46.0 & 36.7 \\
    \midrule
    Phi-4-multimodal & none & 49.3 & 48.6 & 45.2 & 52.4 & 48.3 \\
     & LoRA  & 50.0 & 48.3 & 46.2 & 50.0 & 48.4 \\
     & full & 38.4 & 36.2 & 39.0 & 49.2 & 39.3 \\
    \midrule
    \multicolumn{7}{@{}l}{\emph{Other open models, no training}} \\
    Kimi-Audio & none & 45.6 & 47.9 & 42.8 & 47.6 & 45.7 \\
    MiMo-Audio & none & 47.6 & 43.1 & 46.6 & 48.4 & 46.1 \\
    Fun-Audio-Chat & none & 45.9 & 45.9 & 45.2 & 49.2 & 46.1 \\
    Step-Audio 2 & none & 43.5 & 46.9 & 44.5 & 49.2 & 45.5 \\
    \midrule
    \multicolumn{7}{@{}l}{\emph{Commercial models (API), no training}} \\
    GPT-Audio-mini & none & 49.7 & 50.7 & 49.0 & 54.0 & 50.3 \\
    GPT-Audio-1.5 & none & 50.0 & 47.6 & 53.8 & 47.6 & 50.1 \\
    Gemini 2.5 Pro & none & 22.4 & 17.2 & 32.5 & 46.8 & 26.9 \\
    Gemini 3.8 Flash & none & 29.9 & 36.6 & 46.2 & 50.0 & 39.1 \\
    \bottomrule
  \end{tabular}
\end{table}

    \begin{table}[!htbp]
      \caption{False-flip rate (\%) by task and \STAY{} change; lower is better. Room: room echo; Tr.: transposition. Melody and harmony: 63 EQ, 62 room-echo, 63 hiss and 62 transposition pairs. Timbre and rhythm: 84 EQ, 83 room-echo and 83 hiss pairs, and no transposition. $^\dagger$Step 500. }
      \label{tab:app-ffr-task}
      \centering
      \footnotesize
      \setlength{\tabcolsep}{3pt}
      \fitwidth{%
      \begin{tabular}{@{}ll rrrr rrrr rrr rrr@{}}
        \toprule
              &      & \multicolumn{4}{c}{Melody} & \multicolumn{4}{c}{Harmony} & \multicolumn{3}{c}{Timbre} & \multicolumn{3}{c}{Rhythm} \\
        \cmidrule(lr){3-6}\cmidrule(lr){7-10}\cmidrule(lr){11-13}\cmidrule(l){14-16}
        Model & GRPO & EQ & Room & Hiss & Tr. & EQ & Room & Hiss & Tr. & EQ & Room & Hiss & EQ & Room & Hiss \\
        \midrule
        Qwen2-Audio & none & 49.2 & 58.1 & 60.3 & 48.4 & 47.6 & 51.6 & 47.6 & 51.6 & 52.4 & 53.0 & 53.0 & 46.4 & 44.6 & 50.6 \\
         & LoRA & 20.6 & 56.5 & 60.3 & 43.5 & 38.1 & 51.6 & 47.6 & 50.0 & 8.3 & 37.3 & 56.6 & 51.2 & 42.2 & 41.0 \\
         & full & 25.4 & 58.1 & 60.3 & 43.5 & 36.5 & 48.4 & 47.6 & 48.4 & 25.0 & 41.0 & 56.6 & 54.8 & 45.8 & 45.8 \\
        \midrule
        Qwen2.5-Omni & none & 34.9 & 32.3 & 54.0 & 46.8 & 33.3 & 45.2 & 54.0 & 50.0 & 22.6 & 34.9 & 43.4 & 44.0 & 37.3 & 47.0 \\
         & LoRA & 0.0 & 29.0 & 60.3 & 40.3 & 4.8 & 35.5 & 47.6 & 50.0 & 1.2 & 8.4 & 53.0 & 0.0 & 1.2 & 49.4 \\
         & full & 0.0 & 35.5 & 60.3 & 40.3 & 19.0 & 46.8 & 47.6 & 51.6 & 1.2 & 10.8 & 56.6 & 0.0 & 4.8 & 48.2 \\
        \midrule
        Audio Flamingo 3 & none & 49.2 & 51.6 & 34.9 & 50.0 & 57.1 & 51.6 & 54.0 & 46.8 & 41.7 & 42.2 & 39.8 & 52.4 & 53.0 & 45.8 \\
         & LoRA & 41.3 & 56.5 & 60.3 & 45.2 & 28.6 & 33.9 & 46.0 & 56.5 & 3.6 & 18.1 & 39.8 & 53.6 & 55.4 & 50.6 \\
         & full$^\dagger$ & 38.1 & 51.6 & 60.3 & 43.5 & 17.5 & 43.5 & 46.0 & 48.4 & 2.4 & 12.0 & 47.0 & 32.1 & 39.8 & 45.8 \\
        \midrule
        Phi-4-multimodal & none & 50.8 & 43.5 & 41.3 & 51.6 & 50.8 & 51.6 & 46.0 & 53.2 & 42.9 & 43.4 & 43.4 & 53.6 & 55.4 & 49.4 \\
         & LoRA  & 52.4 & 45.2 & 39.7 & 51.6 & 52.4 & 48.4 & 52.4 & 48.4 & 42.9 & 43.4 & 43.4 & 53.6 & 55.4 & 49.4 \\
         & full & 46.0 & 43.5 & 54.0 & 50.0 & 50.8 & 41.9 & 42.9 & 48.4 & 8.3 & 7.2 & 14.5 & 53.6 & 55.4 & 49.4 \\
        \midrule
        \multicolumn{16}{@{}l}{\emph{Other open models, no training}} \\
        Kimi-Audio & none & 41.3 & 43.5 & 39.7 & 40.3 & 54.0 & 50.0 & 50.8 & 54.8 & 42.9 & 47.0 & 38.6 & 45.2 & 50.6 & 43.4 \\
        MiMo-Audio & none & 50.8 & 41.9 & 39.7 & 51.6 & 52.4 & 53.2 & 49.2 & 45.2 & 39.3 & 42.2 & 43.4 & 50.0 & 37.3 & 53.0 \\
        Fun-Audio-Chat & none & 52.4 & 50.0 & 47.6 & 51.6 & 49.2 & 46.8 & 50.8 & 46.8 & 35.7 & 34.9 & 38.6 & 48.8 & 53.0 & 45.8 \\
        Step-Audio 2 & none & 55.6 & 45.2 & 47.6 & 48.4 & 42.9 & 43.5 & 49.2 & 50.0 & 35.7 & 41.0 & 37.3 & 42.9 & 56.6 & 45.8 \\
        \midrule
        \multicolumn{16}{@{}l}{\emph{Commercial models (API), no training}} \\
        GPT-Audio-mini & none & 55.6 & 51.6 & 49.2 & 58.1 & 49.2 & 50.0 & 55.6 & 50.0 & 41.7 & 45.8 & 43.4 & 53.6 & 55.4 & 49.4 \\
        GPT-Audio-1.5 & none & 49.2 & 56.5 & 60.3 & 46.8 & 44.4 & 45.2 & 47.6 & 48.4 & 50.0 & 45.8 & 55.4 & 54.8 & 44.6 & 51.8 \\
        Gemini 2.5 Pro & none & 22.2 & 17.7 & 34.9 & 46.8 & 42.9 & 25.8 & 33.3 & 46.8 & 13.1 & 14.5 & 20.5 & 16.7 & 13.3 & 42.2 \\
        Gemini 3.8 Flash & none & 34.9 & 51.6 & 52.4 & 50.0 & 50.8 & 50.0 & 49.2 & 50.0 & 10.7 & 9.6 & 34.9 & 29.8 & 42.2 & 50.6 \\
        \bottomrule
      \end{tabular}}
    \end{table}
\paragraph{Models near chance show no pattern.} For the untrained open models other than Qwen2.5-Omni, and for both GPT-Audio models, every pooled rate lies between 42.8\% and 54.0\%, whatever the change. These models barely tell the clips apart, so the kind of change makes little difference.

\paragraph{Models that listen ignore EQ and room echo best.} The three models that listen without training err most on hiss and transposition pairs (32.5\% to 50.0\%). Qwen2.5-Omni and Gemini 3.8 Flash err least on EQ pairs (33.7\% and 29.9\%), and Gemini 2.5 Pro on room-echo pairs (17.2\%, against 22.4\% on EQ). GRPO keeps this order and widens the gaps. In every run that learned, the EQ rate falls by 10.9 to 32.3 points, while the hiss rate changes by $-6.2$ to $+5.8$ points and the transposition rate by $-4.0$ to $+2.4$ points.

\paragraph{Hiss.} After GRPO, the four Qwen runs err on 51.0\% to 53.1\% of hiss pairs, about as often as before training. Both Qwen2.5-Omni runs err on exactly 38 of the 63 melody and 30 of the 63 harmony hiss pairs. These are the numbers of hiss pairs whose correct answer is `same' (the rest are `different'), so the counts are what a model gives if it answers `different' to every hiss pair of these two tasks, whatever the music. On timbre and rhythm, hiss causes 85 of the 94 false flips of Qwen2.5-Omni with LoRA and 87 of the 101 with full fine-tuning. On the EQ and room-echo pairs of these two tasks, the two runs err on 2.7\% and 4.2\%. Phi-4-multimodal with full fine-tuning is the exception. On timbre, the only task it learned, it errs on 14.5\% of hiss pairs, against 8.3\% for EQ and 7.2\% for room echo. It is also the one run in which misses and false flips fell by similar amounts (Section~\ref{sec:res-asym}).

\paragraph{Transposition.} No model learns to ignore transposition: with or without training, every run errs on 45.2\% to 54.0\% of transposition pairs, including Gemini 2.5 Pro (46.8\%), which errs on only 17.2\% of room-echo pairs. After GRPO, Qwen2.5-Omni detects a single moved note in 99.6\% (LoRA) and 100.0\% (full) of melody \FLIP{} pairs, but errs on 40.3\% of transposed melody pairs in both runs.

\paragraph{Room echo depends on the task.} After GRPO, Qwen2.5-Omni with LoRA errs on 29.0\% and 35.5\% of room-echo pairs in melody and harmony, but on 8.4\% and 1.2\% in timbre and rhythm.

\paragraph{\rev{Hiss level.}} \rev{In the test, EQ and hiss are each applied at one strength. The hiss sweep in Appendix~\ref{app:encoder} varies the hiss level: the trained Qwen2.5-Omni already calls 42.9\% of `same' pairs different at $-80$\,dBFS.}

\FloatBarrier
\section{Per-item analysis of Gemini 2.5 Pro and GPT-Audio-mini}
\label{app:api-items}

\begin{figure}[!htbp]
  \centering
  \includegraphics[width=\linewidth]{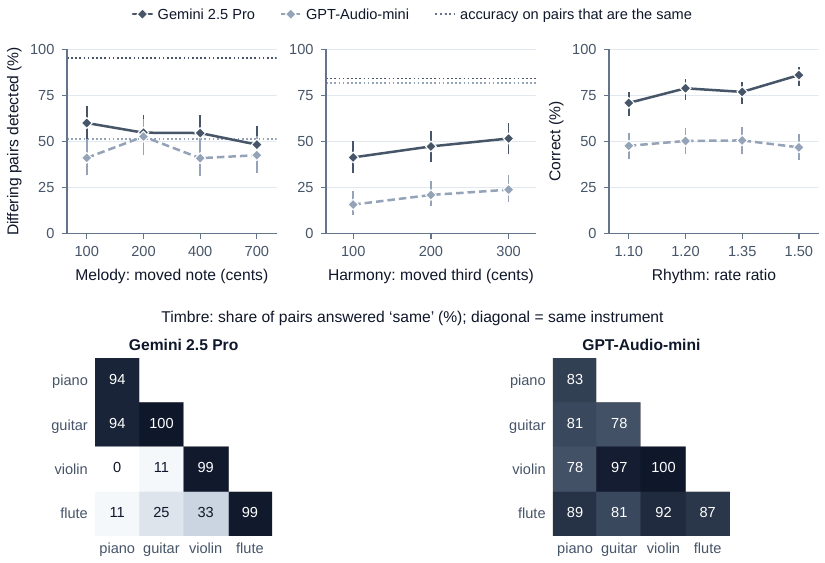}
  \caption{Gemini 2.5 Pro and GPT-Audio-mini, item by item, pooled over the clean, \FLIP{} and \STAY{} versions of each item. \textbf{Top:} share of differing melody and harmony pairs that the model calls different, by the size of the change (dotted: accuracy on pairs that are the same), and rhythm accuracy by the rate ratio of the two click tracks; bars are 95\% Wilson intervals over 87 to 189 items per point, and the dashed line is 50\%. \textbf{Bottom:} share of timbre pairs answered `same' for each instrument pair; diagonal cells are same-instrument pairs, where `same' is correct. Gemini 2.5 Pro detects about half of the moved notes whatever the size of the move, improves with the size of the tempo difference, and hears piano and guitar as the same instrument. GPT-Audio-mini answers `same' to most timbre pairs and stays near 50\% on melody and rhythm.}
  \label{fig:api-items}
\end{figure}

For these two commercial models we analyze every answer, so we can split their errors by the correct answer and by the size of the difference (Table~\ref{tab:app-api-split} and Figure~\ref{fig:api-items}). The items are the same 1{,}000 clean and 2{,}000 \FLIP/\STAY{} items as in Table~\ref{tab:base}.

\begin{table}[!htbp]
  \caption{Errors (\%) on melody, harmony and timbre, split by the correct answer. \emph{Same}: pairs whose answer is `same', where an error is a `different' answer. \emph{Diff.}: pairs whose answer is `different', where an error is a `same' answer. $n$: number of pairs.}
  \label{tab:app-api-split}
  \centering
  \small
  \begin{tabular}{@{}l rr rr rr@{}}
    \toprule
                          & \multicolumn{2}{c}{$n$} & \multicolumn{2}{c}{Gemini 2.5 Pro} & \multicolumn{2}{c}{GPT-Audio-mini} \\
    \cmidrule(lr){2-3}\cmidrule(lr){4-5}\cmidrule(l){6-7}
    Items                 & Same & Diff. & Same & Diff. & Same & Diff. \\
    \midrule
    Clean                 & 401 & 349 & 3.5  & 41.0 & 25.7 & 75.4 \\
    \FLIP                 & 349 & 401 & 4.3  & 44.6 & 24.1 & 73.1 \\
    \STAY, EQ             & 109 & 101 & 6.4  & 44.6 & 22.0 & 76.2 \\
    \STAY, room echo      & 115 & 92  & 2.6  & 39.1 & 28.7 & 73.9 \\
    \STAY, hiss           & 115 & 94  & 10.4 & 51.1 & 27.8 & 74.5 \\
    \STAY, transposition  & 62  & 62  & 51.6 & 41.9 & 41.9 & 66.1 \\
    \bottomrule
  \end{tabular}
\end{table}

\paragraph{\rev{GPT-Audio-mini mostly answers `A' or gives no letter.}} It answers `A' to 81.1\% of clean items and to all 500 rhythm \FLIP/\STAY{} items, so its rhythm \FLIP/\STAY{} accuracy is 47.8\%, the score of always answering `A' (Section~\ref{sec:flipstay}). \rev{Many of these `A' answers are replies without an answer letter (1{,}385 of its 3{,}000 replies), which Table~\ref{tab:base} records as `A' (Appendix~\ref{app:models}).} On timbre it answers `same' to 78\% to 100\% of every instrument pair, whether the instruments differ or not. On melody and rhythm its accuracy does not change with the size of the difference.

\paragraph{Gemini 2.5 Pro leans toward `same'.} On the three same/different tasks, 91\% of its clean errors and 83\% of its \FLIP/\STAY{} errors are `same' answers to pairs that differ. On pairs that are the same it is almost always right, except after a transposition (Table~\ref{tab:app-api-split}). Its detection of a moved note does not improve as the note moves further: it detects 60.0\% of melody pairs with a 100-cent move and 48.3\% with a 700-cent move. On harmony, detection rises from 41.3\% to 51.6\% as the third moves further. Only on rhythm, the one task that is not a same/different question, does accuracy clearly rise with the size of the difference, from 70.9\% at a rate ratio of 1.10 to 86.0\% at 1.50. On timbre it tells piano from guitar in only 2 of 36 pairs, although it tells piano from violin in all 36 and piano from flute in 32.

\paragraph{Sound changes rarely change its answer, except transposition and, on rhythm, hiss.} Most of its false flips on the same/different tasks are `same' answers to \STAY{} pairs that differ (39.1\% to 51.1\% of such pairs), the same error it makes on clean and \FLIP{} pairs (41.0\% and 44.6\%). On \STAY{} pairs that are the same, EQ, room echo and hiss make it answer `different' in only 6.4\%, 2.6\% and 10.4\% of cases, but transposition does so in 51.6\%. So on these tasks Gemini 2.5 Pro mostly ignores the three sound effects, yet it treats a transposed melody or chord as a different one. On rhythm, hiss lowers its accuracy to 57.8\%, against 83.3\% with EQ and 86.7\% with room echo, and its errors go both ways (17 and 18 of the 83 hiss pairs). Its false-flip rate and that of the trained Qwen2.5-Omni are therefore similar in size (26.9\% and 26.1\%) but differ in kind: Gemini 2.5 Pro mostly misses differences that are there, while the trained model reacts to hiss.

\begingroup

\FloatBarrier
\section{Slakh2100 test}
\label{app:slakh}

\paragraph{Construction.}
This test asks whether our findings hold on real songs with richer sounds, while the labels stay exact. Slakh2100~\citep{manilow2019slakh} renders songs from MIDI with commercial sample-based instruments, so the MIDI of each stem gives every label. We build the test like the rendered set (Appendix~\ref{app:construction}). The tasks, clip lengths, file format, question wordings and letter rules are the same. Each task has 250 clean test items from Slakh's test tracks, and each item has one \FLIP{} and one \STAY{} version. The \STAY{} changes follow Table~\ref{tab:app-flipstay} and use the code of our generator. The training split has 2{,}500 items per task from Slakh's training and validation tracks. Training and test items never share a track, and melody items also never share an instrument plugin.
\begin{itemize}
\item \textbf{Melody.} A phrase of 4 to 6 notes from a stem that plays one note at a time (checked against the MIDI), 2.5\,s long. In `different' pairs, the second note is pitch-shifted by 100, 200, 400 or 700 cents with Rubber Band, keeping the formants. In `same' pairs, the same note passes through the same shifter at 0 cents, so both kinds of pair are processed alike.
\item \textbf{Harmony.} A single recorded note from a piano, guitar, organ or strings stem. We build the triad from it with the same shifter, 2.0\,s long. In `different' pairs the third moves by 100, 200 or 300 cents, as in our generator.
\item \textbf{Timbre.} Passages from piano, acoustic guitar, violin or string ensemble, and flute stems, 2.5\,s long. The two clips are always different passages. A `same' pair takes two passages of one stem, and a `different' pair takes passages of stems from two instrument families.
\item \textbf{Rhythm.} A drum stem, time-stretched by 1.10, 1.20, 1.35 or 1.50, 4.0\,s long. The slower clip also passes through the stretcher, at 1.0. Both clips come from the same part of the song, and \FLIP{} swaps them.
\end{itemize}

\paragraph{Scoring.}
Scoring is the same as for the rendered test set, with one difference. Here a reply without an answer letter counts as wrong, while Table~\ref{tab:base} records it as `A' (Appendix~\ref{app:models}). On the rendered test this changes little: Gemini 2.5 Pro gives 16 such replies on the \FLIP/\STAY{} items, and its accuracy there becomes 73.7\% instead of 74.2\%. On the Slakh2100 \FLIP/\STAY{} items such replies are more common: 54.6\% for GPT-Audio-mini, 8.4\% for GPT-Audio-1.5 and 5.5\% for Gemini 2.5 Pro (45.2\%, 6.9\% and 0.8\% on the rendered test). On the items they do answer, these models score 51.5\%, 50.7\% and 79.7\%. If such replies count as `A', as in Table~\ref{tab:base}, Gemini 2.5 Pro scores 78.2\%. The open models always give a letter.

\paragraph{Results.}
Without training (Table~\ref{tab:slakh_untrained}), the picture matches the rendered test. The ranking of the 12 models is almost the same (Spearman $\rho = 0.93$, $p < 0.001$; 0.86 if replies without a letter count as `A'; 0.76 over the eight open models). Only Qwen2.5-Omni (56.5\%), Fun-Audio-Chat (56.6\%) and the two Gemini models are clearly above chance on \FLIP/\STAY{} (two-sided binomial test, $p < 0.001$). Audio Flamingo 3 scores 52.2\% ($p = 0.05$), and the other open models 49.8\% to 51.4\%.

Table~\ref{tab:slakh_trained} adds every trained configuration. It includes the checkpoints trained on our rendered split, and new GRPO runs on the Slakh2100 training split (settings of Table~\ref{tab:grpo_details}, 1{,}000 steps, or 500 for Audio Flamingo 3 with full fine-tuning). Training on our rendered pairs carries over for Qwen2.5-Omni (72.3\% with LoRA, 70.9\% with full fine-tuning) and for Audio Flamingo 3 (59.0\% and 61.7\%). Training on Slakh2100 itself lifts Qwen2.5-Omni to 86.0\% (LoRA) and 80.8\% (full). For both runs, misses again fall more than false flips (Table~\ref{tab:slakh_drop}). The weaker runs, of Qwen2-Audio and Audio Flamingo 3, reach only 56.2\% to 58.1\%, and for them false flips fall slightly more than misses. As on the rendered test, transposition is the \STAY{} change that trained models fail most. Qwen2.5-Omni trained with LoRA on Slakh2100, for example, makes 7.8\% false flips under EQ but 39.5\% under transposition. Tables~\ref{tab:slakh_pertask} and~\ref{tab:slakh_letter} give per-task error rates and letter bias. 

\begin{table}[!htbp]
\centering
\small
\caption{Slakh2100 test without training (\%; Tables~\ref{tab:base} and~\ref{tab:app-task-fs} on this set). Clean: 250 items per task (1,000 in All). \FLIP/\STAY{}: 500 items per task (2,000 in $\accfs$). A reply without an answer letter counts as an error.}
\label{tab:slakh_untrained}
\setlength{\tabcolsep}{3.5pt}
\resizebox{\linewidth}{!}{%
\begin{tabular}{lrrrrrrrrrrrr}
\toprule
 & \multicolumn{5}{c}{Clean accuracy} & \multicolumn{4}{c}{\FLIP/\STAY{} accuracy} & & & \\
\cmidrule(lr){2-6}\cmidrule(lr){7-10}
Model & Mel. & Har. & Tim. & Rhy. & All & Mel. & Har. & Tim. & Rhy. & $\accfs$ & MR & FFR \\
\midrule
Qwen2-Audio & 50.0 & 50.4 & 50.4 & 50.0 & 50.2 & 50.0 & 50.0 & 50.8 & 54.0 & 51.2 & 48.0 & 49.6 \\
Qwen2.5-Omni & 54.4 & 50.8 & 66.0 & 52.0 & 55.8 & 55.0 & 51.2 & 63.6 & 56.0 & 56.5 & 43.7 & 43.4 \\
Audio Flamingo 3 & 48.8 & 50.8 & 60.8 & 52.0 & 53.1 & 49.6 & 49.0 & 58.8 & 51.4 & 52.2 & 48.6 & 47.0 \\
Phi-4-multimodal & 50.0 & 48.4 & 50.8 & 50.0 & 49.8 & 51.2 & 51.8 & 50.0 & 46.0 & 49.8 & 51.2 & 49.3 \\
Kimi-Audio & 51.6 & 51.2 & 51.2 & 51.2 & 51.3 & 49.2 & 49.6 & 51.4 & 51.8 & 50.5 & 50.7 & 48.3 \\
MiMo-Audio & 50.0 & 51.6 & 52.8 & 57.6 & 53.0 & 50.0 & 53.2 & 52.4 & 50.0 & 51.4 & 48.5 & 48.7 \\
Fun-Audio-Chat & 54.4 & 55.2 & 60.0 & 54.0 & 55.9 & 52.2 & 53.6 & 63.8 & 56.8 & 56.6 & 43.3 & 43.5 \\
Step-Audio 2 & 48.8 & 54.0 & 56.4 & 49.2 & 52.1 & 48.4 & 51.8 & 56.0 & 46.6 & 50.7 & 49.8 & 48.8 \\
\midrule
GPT-Audio-mini & 38.0 & 29.6 & 19.6 & 5.2 & 23.1 & 37.8 & 31.2 & 18.4 & 6.0 & 23.4 & 76.7 & 76.6 \\
GPT-Audio-1.5 & 47.6 & 42.0 & 52.0 & 44.0 & 46.4 & 48.8 & 43.8 & 50.6 & 42.4 & 46.4 & 53.9 & 53.3 \\
Gemini 2.5 Pro & 68.8 & 68.8 & 89.6 & 79.2 & 76.6 & 63.2 & 64.6 & 90.8 & 82.6 & 75.3 & 24.2 & 25.2 \\
Gemini 3.8 Flash & 58.0 & 54.0 & 84.0 & 63.6 & 64.9 & 52.8 & 51.2 & 83.2 & 59.0 & 61.6 & 38.1 & 38.8 \\
Always `A' &  &  &  &  & 50.2 &  &  &  &  & 49.0 & 51.9 & 50.1 \\
\bottomrule
\end{tabular}}
\end{table}

\begin{table}[!htbp]
\centering
\footnotesize
\caption{All open-model configurations on the Slakh2100 test (\%; Tables~\ref{tab:grpo} and~\ref{tab:app-ffr-change} on this set). Mel.\ to All: clean accuracy. EQ to Tr.: false-flip rate per \STAY{} change (Tr.: transposition). Slakh2100 split: 10,000 training items, settings of Table~\ref{tab:grpo_details}. R1: GRPO on clean + sound-only pairs; R2: SFT on clean pairs. $^\dagger$Step 500; all other runs step 1,000. }
\label{tab:slakh_trained}
\setlength{\tabcolsep}{3.5pt}
\resizebox{\linewidth}{!}{%
\begin{tabular}{llrrrrrrrrrrrr}
\toprule
Model & Training & Mel. & Har. & Tim. & Rhy. & All & $\accfs$ & MR & FFR & EQ & Room & Hiss & Tr. \\
\midrule
\multicolumn{14}{l}{\emph{No training}} \\
Qwen2-Audio & none & 50.0 & 50.4 & 50.4 & 50.0 & 50.2 & 51.2 & 48.0 & 49.6 & 54.6 & 53.1 & 44.3 & 41.9 \\
Qwen2.5-Omni & none & 54.4 & 50.8 & 66.0 & 52.0 & 55.8 & 56.5 & 43.7 & 43.4 & 45.1 & 40.4 & 39.1 & 56.5 \\
Audio Flamingo 3 & none & 48.8 & 50.8 & 60.8 & 52.0 & 53.1 & 52.2 & 48.6 & 47.0 & 42.4 & 43.2 & 49.8 & 60.5 \\
Phi-4-multimodal & none & 50.0 & 48.4 & 50.8 & 50.0 & 49.8 & 49.8 & 51.2 & 49.3 & 44.4 & 46.2 & 54.0 & 57.3 \\
\multicolumn{14}{l}{\emph{GRPO on the rendered split (Table~\ref{tab:grpo})}} \\
Qwen2-Audio & LoRA & 50.4 & 55.2 & 52.0 & 48.0 & 51.4 & 52.6 & 46.3 & 48.5 & 52.9 & 52.7 & 42.6 & 41.9 \\
 & full & 50.0 & 53.6 & 54.0 & 50.4 & 52.0 & 52.8 & 46.6 & 47.8 & 51.5 & 50.3 & 43.6 & 42.7 \\
Qwen2.5-Omni & LoRA & 80.0 & 86.4 & 64.4 & 72.0 & 75.7 & 72.3 & 25.3 & 30.2 & 24.1 & 30.8 & 32.2 & 38.7 \\
 & full & 82.8 & 84.4 & 65.2 & 72.4 & 76.2 & 70.9 & 27.0 & 31.3 & 25.4 & 32.2 & 33.6 & 37.9 \\
Audio Flamingo 3 & LoRA & 57.6 & 69.2 & 72.4 & 50.0 & 62.3 & 59.0 & 40.0 & 42.0 & 34.9 & 45.5 & 44.3 & 45.2 \\
 & full$^\dagger$ & 60.0 & 64.4 & 80.0 & 53.2 & 64.4 & 61.7 & 36.4 & 40.3 & 34.6 & 43.5 & 41.2 & 44.4 \\
Phi-4-multimodal &  LoRA & 49.6 & 50.0 & 50.8 & 50.0 & 50.1 & 49.1 & 52.0 & 49.8 & 43.4 & 46.6 & 54.7 & 61.3 \\
 & full & 50.8 & 53.6 & 56.4 & 49.2 & 52.5 & 51.5 & 48.9 & 48.2 & 45.4 & 47.3 & 49.8 & 53.2 \\
\multicolumn{14}{l}{\emph{GRPO on the Slakh2100 split}} \\
Qwen2-Audio & LoRA & 50.8 & 58.4 & 72.8 & 49.6 & 57.9 & 56.8 & 43.7 & 42.8 & 43.4 & 47.3 & 38.8 & 40.3 \\
 & full & 51.2 & 54.4 & 73.2 & 49.2 & 57.0 & 56.2 & 43.7 & 44.0 & 46.8 & 47.9 & 39.1 & 39.5 \\
Qwen2.5-Omni & LoRA & 89.2 & 88.8 & 96.4 & 95.6 & 92.5 & 86.0 & 10.4 & 17.6 & 7.8 & 18.2 & 17.6 & 39.5 \\
 & full & 83.6 & 88.4 & 95.2 & 75.2 & 85.6 & 80.8 & 16.7 & 21.8 & 13.9 & 20.2 & 24.2 & 38.7 \\
Audio Flamingo 3 & LoRA & 55.2 & 60.8 & 78.8 & 56.0 & 62.7 & 58.1 & 43.5 & 40.4 & 35.9 & 42.8 & 42.9 & 39.5 \\
 & full$^\dagger$ & 54.0 & 54.8 & 78.4 & 50.0 & 59.3 & 57.2 & 45.8 & 39.9 & 32.5 & 38.0 & 43.6 & 53.2 \\
Phi-4-multimodal &  LoRA & 49.6 & 50.0 & 50.8 & 50.0 & 50.1 & 49.1 & 52.0 & 49.8 & 43.4 & 46.6 & 54.7 & 61.3 \\
 & full & 48.8 & 55.2 & 53.2 & 50.0 & 51.8 & 51.7 & 47.9 & 48.7 & 44.4 & 50.7 & 51.9 & 46.8 \\
\multicolumn{14}{l}{\emph{Baselines of Appendix~\ref{app:trainbase} (rendered pairs)}} \\
Qwen2.5-Omni & R1 & 74.0 & 79.6 & 96.4 & 76.4 & 81.6 & 77.1 & 22.1 & 23.7 & 17.3 & 18.8 & 26.0 & 45.2 \\
 & R2 & 84.8 & 86.8 & 64.0 & 72.8 & 77.1 & 70.9 & 26.9 & 31.3 & 25.1 & 32.5 & 33.6 & 37.9 \\
\bottomrule
\end{tabular}}
\end{table}

\begin{table}[!htbp]
\centering
\footnotesize
\caption{Per-task miss rate and false-flip rate on the Slakh2100 test (\%; Tables~\ref{tab:app-base-task} and~\ref{tab:app-grpo-task} on this set). Columns in each block: melody, harmony, timbre, rhythm. Untrained rows: the models of Table~\ref{tab:app-base-task}. $^\dagger$Step 500.}
\label{tab:slakh_pertask}
\setlength{\tabcolsep}{3.5pt}
\begin{tabular}{llrrrrrrrr}
\toprule
& & \multicolumn{4}{c}{MR} & \multicolumn{4}{c}{FFR} \\
\cmidrule(lr){3-6}\cmidrule(lr){7-10}
Model & GRPO & Mel. & Har. & Tim. & Rhy. & Mel. & Har. & Tim. & Rhy. \\
\midrule
\multicolumn{10}{l}{\emph{No training}} \\
Qwen2-Audio & none & 50.0 & 50.0 & 48.8 & 43.2 & 50.0 & 50.0 & 49.6 & 48.8 \\
Qwen2.5-Omni & none & 44.0 & 50.0 & 40.4 & 40.4 & 46.0 & 47.6 & 32.4 & 47.6 \\
Audio Flamingo 3 & none & 48.4 & 51.6 & 44.0 & 50.4 & 52.4 & 50.4 & 38.4 & 46.8 \\
Phi-4-multimodal & none & 47.6 & 49.6 & 50.8 & 56.8 & 50.0 & 46.8 & 49.2 & 51.2 \\
Kimi-Audio & none & 52.8 & 49.6 & 49.6 & 50.8 & 48.8 & 51.2 & 47.6 & 45.6 \\
MiMo-Audio & none & 50.0 & 46.4 & 48.0 & 49.6 & 50.0 & 47.2 & 47.2 & 50.4 \\
Fun-Audio-Chat & none & 44.0 & 47.6 & 37.2 & 44.4 & 51.6 & 45.2 & 35.2 & 42.0 \\
Step-Audio 2 & none & 49.2 & 49.2 & 45.2 & 55.6 & 54.0 & 47.2 & 42.8 & 51.2 \\
GPT-Audio-mini & none & 60.8 & 69.6 & 81.2 & 95.2 & 63.6 & 68.0 & 82.0 & 92.8 \\
Gemini 2.5 Pro & none & 34.4 & 36.8 & 8.0 & 17.6 & 39.2 & 34.0 & 10.4 & 17.2 \\
\multicolumn{10}{l}{\emph{GRPO on the rendered split}} \\
Qwen2-Audio & LoRA & 44.0 & 48.4 & 48.4 & 44.4 & 50.4 & 47.6 & 48.8 & 47.2 \\
 & full & 48.4 & 46.4 & 46.4 & 45.2 & 50.0 & 47.6 & 47.6 & 46.0 \\
Qwen2.5-Omni & LoRA & 21.6 & 17.6 & 34.4 & 27.6 & 31.2 & 26.8 & 34.8 & 28.0 \\
 & full & 22.4 & 20.4 & 35.6 & 29.6 & 28.0 & 32.0 & 35.6 & 29.6 \\
Audio Flamingo 3 & LoRA & 42.0 & 40.8 & 20.4 & 56.8 & 46.0 & 41.2 & 29.6 & 51.2 \\
 & full$^\dagger$ & 46.8 & 33.2 & 15.2 & 50.4 & 46.0 & 43.6 & 24.8 & 46.8 \\
Phi-4-multimodal &  LoRA & 50.8 & 49.2 & 51.2 & 56.8 & 48.4 & 49.6 & 50.0 & 51.2 \\
 & full & 46.4 & 49.2 & 44.8 & 55.2 & 51.6 & 47.2 & 43.6 & 50.4 \\
\multicolumn{10}{l}{\emph{GRPO on the Slakh2100 split}} \\
Qwen2-Audio & LoRA & 46.8 & 44.0 & 40.8 & 43.2 & 47.6 & 46.4 & 28.4 & 48.8 \\
 & full & 48.8 & 44.4 & 38.8 & 42.8 & 50.0 & 46.8 & 30.4 & 48.8 \\
Qwen2.5-Omni & LoRA & 13.2 & 15.2 & 9.6 & 3.6 & 26.4 & 29.2 & 6.0 & 8.8 \\
 & full & 18.0 & 16.0 & 11.6 & 21.2 & 28.8 & 28.8 & 6.4 & 23.2 \\
Audio Flamingo 3 & LoRA & 47.2 & 38.4 & 37.6 & 50.8 & 49.6 & 42.4 & 21.2 & 48.4 \\
 & full$^\dagger$ & 45.2 & 42.8 & 38.4 & 56.8 & 42.0 & 46.8 & 21.2 & 49.6 \\
Phi-4-multimodal &  LoRA & 50.8 & 49.2 & 51.2 & 56.8 & 48.4 & 49.6 & 50.0 & 51.2 \\
 & full & 43.2 & 41.6 & 50.0 & 56.8 & 49.2 & 46.4 & 48.0 & 51.2 \\
\bottomrule
\end{tabular}
\end{table}

\begin{table}[!htbp]
\centering
\footnotesize
\caption{Letter bias on the Slakh2100 clean test (\%; Table~\ref{tab:app-bias} on this set). A rate: share of items answered `A' (50.2\% of correct answers are `A'). Balanced: mean of Acc$\mid$A and Acc$\mid$B. Gap: Acc$\mid$A $-$ Acc$\mid$B in absolute value. $^\dagger$Step 500.}
\label{tab:slakh_letter}
\setlength{\tabcolsep}{3.5pt}
\begin{tabular}{llrrrrr}
\toprule
Model & GRPO & A rate & Acc$\mid$A & Acc$\mid$B & Balanced & Gap \\
\midrule
\multicolumn{7}{l}{\emph{None, and GRPO on the rendered split}} \\
Qwen2-Audio & none & 0.6 & 1.0 & 99.8 & 50.4 & 98.8 \\
 & LoRA & 6.0 & 7.6 & 95.6 & 51.6 & 88.0 \\
 & full & 5.6 & 7.8 & 96.6 & 52.2 & 88.8 \\
Qwen2.5-Omni & none & 32.6 & 38.4 & 73.3 & 55.9 & 34.8 \\
 & LoRA & 44.5 & 70.1 & 81.3 & 75.7 & 11.2 \\
 & full & 40.2 & 66.3 & 86.1 & 76.2 & 19.8 \\
Audio Flamingo 3 & none & 83.1 & 86.1 & 19.9 & 53.0 & 66.2 \\
 & LoRA & 59.3 & 71.5 & 53.0 & 62.3 & 18.5 \\
 & full$^\dagger$ & 55.0 & 69.3 & 59.4 & 64.4 & 9.9 \\
Phi-4-multimodal & none & 98.0 & 97.6 & 1.6 & 49.6 & 96.0 \\
 &  LoRA & 96.9 & 96.8 & 3.0 & 49.9 & 93.8 \\
 & full & 87.3 & 89.6 & 15.1 & 52.4 & 74.6 \\
GPT-Audio-mini & none & 35.0 & 34.7 & 11.4 & 23.1 & 23.2 \\
Gemini 2.5 Pro & none & 57.2 & 86.7 & 66.5 & 76.6 & 20.2 \\
\multicolumn{7}{l}{\emph{GRPO on the Slakh2100 split}} \\
Qwen2-Audio & LoRA & 12.1 & 20.1 & 96.0 & 58.1 & 75.9 \\
 & full & 12.2 & 19.3 & 95.0 & 57.2 & 75.7 \\
Qwen2.5-Omni & LoRA & 48.5 & 90.8 & 94.2 & 92.5 & 3.3 \\
 & full & 49.0 & 84.5 & 86.7 & 85.6 & 2.3 \\
Audio Flamingo 3 & LoRA & 42.1 & 54.8 & 70.7 & 62.7 & 15.9 \\
 & full$^\dagger$ & 69.9 & 79.1 & 39.4 & 59.2 & 39.7 \\
Phi-4-multimodal &  LoRA & 96.9 & 96.8 & 3.0 & 49.9 & 93.8 \\
 & full & 84.6 & 86.3 & 17.1 & 51.7 & 69.2 \\
\bottomrule
\end{tabular}
\end{table}

\begin{table}[!htbp]
\centering
\footnotesize
\caption{Drop in miss rate and false-flip rate from no training to GRPO on the Slakh2100 test, in points (Table~\ref{tab:app-drop} on this set). $z$ and $p$: two-sided test of `the miss rate fell more than the false-flip rate'. $^\dagger$Step 500.}
\label{tab:slakh_drop}
\setlength{\tabcolsep}{3.5pt}
\begin{tabular}{llrrrrr}
\toprule
Model & GRPO & $\Delta$MR & $\Delta$FFR & Ratio & $z$ & $p$ \\
\midrule
\multicolumn{7}{l}{\emph{GRPO on the rendered split}} \\
Qwen2-Audio & LoRA & 1.7 & 1.1 & 1.5 & 0.2 & 0.85 \\
 & full & 1.4 & 1.8 & 0.8 & -0.1 & 0.90 \\
Qwen2.5-Omni & LoRA & 18.4 & 13.2 & 1.4 & 1.7 & 0.08 \\
 & full & 16.7 & 12.1 & 1.4 & 1.5 & 0.13 \\
Audio Flamingo 3 & LoRA & 8.6 & 5.0 & 1.7 & 1.1 & 0.25 \\
 & full$^\dagger$ & 12.2 & 6.7 & 1.8 & 1.8 & 0.08 \\
Phi-4-multimodal &  LoRA & -0.8 & -0.5 & n/a & -0.1 & 0.92 \\
 & full & 2.3 & 1.1 & 2.1 & 0.4 & 0.70 \\
\multicolumn{7}{l}{\emph{GRPO on the Slakh2100 split}} \\
Qwen2-Audio & LoRA & 4.3 & 6.8 & 0.6 & -0.8 & 0.43 \\
 & full & 4.3 & 5.6 & 0.8 & -0.4 & 0.68 \\
Qwen2.5-Omni & LoRA & 33.3 & 25.8 & 1.3 & 2.8 & 0.0055 \\
 & full & 27.0 & 21.6 & 1.2 & 1.9 & 0.06 \\
Audio Flamingo 3 & LoRA & 5.1 & 6.6 & 0.8 & -0.5 & 0.63 \\
 & full$^\dagger$ & 2.8 & 7.1 & 0.4 & -1.4 & 0.17 \\
Phi-4-multimodal &  LoRA & -0.8 & -0.5 & n/a & -0.1 & 0.92 \\
 & full & 3.3 & 0.6 & 5.5 & 0.9 & 0.39 \\
\bottomrule
\end{tabular}
\end{table}

\FloatBarrier
\section{Training with sound-only pairs, and SFT}
\label{app:trainbase}

Our main runs train on clean pairs only (Section~\ref{sec:res-asym}). This appendix asks two questions. Does the pattern of Section~\ref{sec:res-asym}, with misses falling much more than false flips, depend on GRPO, or does it appear with any training on clean pairs? And does it help to add pairs where only the sound changes?

All runs use Qwen2.5-Omni with the LoRA settings of Table~\ref{tab:grpo_details}, the reward of Appendix~\ref{app:grpo} and 1{,}000 steps. R0 is the run of Table~\ref{tab:grpo}. \textbf{R1} draws half of every batch from the clean training pairs and half from their sound-only versions: codec, soundfont swap, transposition and, for timbre, a new melody and key (Table~\ref{tab:app-nuis}). Each version keeps the question and label of its item. \textbf{R2} is SFT. It replaces the GRPO loss with cross-entropy on the correct letter, and keeps the prompts, batch, steps, LoRA and optimizer, so it sees as many training examples. Room echo, EQ and hiss are never in training.

Table~\ref{tab:trainbase} shows the results. SFT shows the same pattern as GRPO. Misses fall from 36.7\% to 3.0\% (33.7 points), but false flips only from 41.0\% to 27.8\% (13.2 points). For GRPO the drops are 34.8 and 14.9 points (Table~\ref{tab:app-drop}). The false flips per sound change are also nearly the same as for R0. So the pattern does not depend on the training algorithm: it appears whenever the model trains on clean pairs only.

Adding sound-only pairs (R1) lowers false flips under room echo (16.6\% to 4.5\%) and hiss (52.4\% to 37.7\%), although R1 never sees either. It does not lower them under transposition, although R1 trains on it (45.2\% for both runs). R1 still calls 25 of the 30 transposed `same' melody pairs different (R0: also 25), and 30 of the 32 transposed `same' chord pairs (R0: 31). Its \FLIP/\STAY{} accuracy rises from 86.0\% to 88.2\%. The cost is 3.4 points of clean accuracy, mostly on harmony (96.8\% to 84.4\%). On the Slakh2100 test (Table~\ref{tab:slakh_trained}), R1 scores 77.1\% on \FLIP/\STAY{}, against 72.3\% for R0 and 70.9\% for R2, and on the real test set (Table~\ref{tab:real_trained}) 70.1\%, against 62.8\% and 62.9\%. On both, most of the gain is on timbre (clean timbre accuracy 64.4\% to 96.4\% on Slakh2100 and 63.3\% to 84.7\% on real recordings), and on real recordings false flips under hiss do not fall (20.9\% to 22.8\%). The gain on rendered clips under unseen echo and hiss therefore does not show that the model ignores sound changes on real audio.

\begin{table}[!htbp]
\centering
\footnotesize
\caption{Training baselines for Qwen2.5-Omni (LoRA, Table~\ref{tab:grpo_details} settings, 1,000 steps, rendered test, \%). Mel.\ to All: clean accuracy; EQ to Tr.: false-flip rate per \STAY{} change (Table~\ref{tab:app-ffr-change} format). Room echo, EQ and hiss are never in training. R1 adds the training split's sound-only versions (codec, soundfont swap, transposition and, for timbre, a new melody and key; Table~\ref{tab:app-nuis}), half of every batch, each with its item's label. R2 replaces the GRPO loss with cross-entropy on the gold letter; same prompts, batch, steps, LoRA and optimizer.}
\label{tab:trainbase}
\setlength{\tabcolsep}{3.5pt}
\resizebox{\linewidth}{!}{%
\begin{tabular}{lrrrrrrrrrrrr}
\toprule
Run & Mel. & Har. & Tim. & Rhy. & All & $\accfs$ & MR & FFR & EQ & Room & Hiss & Tr. \\
\midrule
No training & 68.8 & 59.2 & 66.8 & 50.0 & 61.2 & 61.2 & 36.7 & 41.0 & 33.7 & 37.2 & 49.0 & 48.4 \\
R0: GRPO, clean (Table~\ref{tab:grpo}) & 99.6 & 96.8 & 99.6 & 100.0 & 99.0 & 86.0 & 1.9 & 26.1 & 1.4 & 16.6 & 52.4 & 45.2 \\
R1: GRPO, clean + sound-only & 100.0 & 84.4 & 98.0 & 100.0 & 95.6 & 88.2 & 4.9 & 18.7 & 2.7 & 4.5 & 37.7 & 45.2 \\
R2: SFT, clean & 100.0 & 89.6 & 100.0 & 100.0 & 97.4 & 84.6 & 3.0 & 27.8 & 2.4 & 20.7 & 52.7 & 46.0 \\
\bottomrule
\end{tabular}}
\end{table}

\FloatBarrier
\section{The frozen audio encoder and hiss}
\label{app:encoder}
After GRPO, Qwen2.5-Omni scores 99.0\% on clean pairs but still reacts to a quiet hiss. This appendix relates that behaviour to the frozen audio encoder (Section~\ref{sec:res-asym}).

\paragraph{Clean pairs.}
As in the standard same/different design, clip A and clip B of every clean `same' pair of melody, harmony and timbre are identical (401 pairs: 135, 124 and 142), and they differ in all 349 `different' pairs. Clean accuracy therefore cannot tell a model that compares the music from one that reacts to any difference in sound. The \FLIP/\STAY{} items make this distinction, and the analysis below asks which of the two the trained model does.

\paragraph{A rule built on the frozen encoder.}
The audio encoder is frozen in every run (Table~\ref{tab:grpo_details}). Training can therefore only change how the language model reads the encoder output. To see what it learns to read, we built a rule on the encoder alone. We encode clip A and clip B separately with the Qwen2.5-Omni encoder and take the mean cosine distance between the two outputs, frame by frame. The rule answers `different' when this distance is above a threshold, with one threshold per task fitted on clean training pairs. It scores 100\% on those pairs and 99.7\% on the clean test pairs (Table~\ref{tab:encoder_rule}). On \STAY{} pairs, the trained model gives the same answer as this rule on 85.1\% of items, against 56.0\% before training. The agreement is 98.6\% on hiss pairs and 98.4\% on transposition pairs, but only 81.4\% on EQ and 67.1\% on room-echo pairs. On these two changes the trained model does far better than the rule (false flips 1.9\% against 18.6\% under EQ, and 22.7\% against 55.6\% under room echo; Table~\ref{tab:encoder_rule}), so training teaches more than a threshold on encoder distance. Under hiss and transposition, which move the encoder output far (below), it answers like the rule.

\paragraph{How far each change moves the encoder.}
Table~\ref{tab:encoder_ratio} compares the distance caused by each \STAY{} change with the distance caused by a real musical change. Hiss moves the output 2.75 times as far as a real change, and 76 times as far as EQ does. On melody alone, hiss moves it 12 times as far as a moved note. To the encoder, a quiet hiss is a bigger change than the music itself. This distance also predicts which `same' \STAY{} pairs the trained model gets wrong: AUROC 0.92 (218 false flips on 401 pairs), against 0.74 before training.

The encoder does hold some of what is needed to ignore hiss. A linear probe on the frozen features (Table~\ref{tab:encoder_rule}) makes 55.0\% false flips under hiss when it is trained on clean pairs. When room-echo, EQ and hiss versions of training items are added, this falls to 32.5\% (EQ: 0.0\% in both cases). A simple readout can learn to ignore part of the hiss, but only once hiss is in its training data.

\paragraph{Hiss sweep.}
How quiet must hiss be before the model ignores it? We added hiss at $-80$ to $-30$\,dBFS to all 1{,}000 clean test items, in the same way as the hiss \STAY{} change of the rendered test (Table~\ref{tab:hiss_sweep}). We added it either to clip B only, or the same hiss to both clips. With hiss on clip B only, the trained model calls 42.9\% of `same' pairs different already at $-80$\,dBFS, the quietest level we tested (2.0\% without hiss). From $-50$\,dBFS up, it calls 98\% to 100\% of them different. With the same hiss on both clips, the two clips of a `same' pair are again identical, yet the model still calls 7.5\% to 51.6\% of them different. So most false flips come from the difference in sound between the two clips, but hiss on its own also pushes the model toward `different'. On rhythm, accuracy with hiss on clip B falls from 98.8\% at $-80$\,dBFS to 53.2\% at $-50$\,dBFS. R1, which never sees hiss, holds up longer: it calls 0.7\% of `same' pairs different at $-80$\,dBFS and 43.9\% at $-60$\,dBFS.

We also ran Gemini 2.5 Pro on a subset of 115 `same' pairs (Table~\ref{tab:hiss_gemini}). With hiss on clip B at $-70$ or $-50$\,dBFS, it errs about as often as without hiss. At $-30$\,dBFS it errs on 40.0\%, where the trained Qwen2.5-Omni errs on 100\%. So Gemini 2.5 Pro tolerates far louder hiss than the trained model, but it too starts to flip.

\begin{table}[!htbp]
\centering
\footnotesize
\caption{Rules built on the frozen Qwen2.5-Omni audio encoder, against the model itself, on melody, harmony and timbre of the rendered test (750 clean, 1,500 \FLIP/\STAY{} items; \%). Distance: mean frame-by-frame cosine distance between the encoder outputs of clip A and clip B, each encoded on its own. Threshold: one per task, fitted on clean training pairs (`different' above it). Probe: logistic regression on the mean absolute difference of the frame features; the second probe adds room-echo, EQ and hiss versions of training items (7,500 clean and 9,000 such pairs). EQ to Tr.: false-flip rate per \STAY{} change.}
\label{tab:encoder_rule}
\setlength{\tabcolsep}{3.5pt}
\begin{tabular}{lrrrrrrrr}
\toprule
 & Clean & $\accfs$ & MR & FFR & EQ & Room & Hiss & Tr. \\
\midrule
Threshold on encoder distance (no training) & 99.7 & 77.9 & 0.5 & 43.6 & 18.6 & 55.6 & 55.0 & 46.8 \\
Linear probe, clean training pairs & 99.7 & 80.7 & 0.5 & 38.0 & 0.0 & 54.1 & 55.0 & 46.8 \\
Linear probe, + echo, EQ, hiss versions & 100.0 & 88.5 & 0.0 & 23.1 & 0.0 & 23.7 & 32.5 & 45.2 \\
Qwen2.5-Omni, no training & 64.9 & 63.3 & 32.9 & 40.4 & 29.5 & 37.2 & 49.8 & 48.4 \\
Qwen2.5-Omni, GRPO LoRA & 98.7 & 84.1 & 2.5 & 29.2 & 1.9 & 22.7 & 53.6 & 45.2 \\
\bottomrule
\end{tabular}
\end{table}

\begin{table}[!htbp]
\centering
\small
\caption{How far each \STAY{} change moves the frozen encoder output, relative to a real change: median encoder distance on `same' \STAY{} pairs with this change, divided by the median distance on \FLIP{} pairs that differ (same task set). Values above 1 mean the sound change moves the output more than the musical change. Timbre has no transposition \STAY{} pairs.}
\label{tab:encoder_ratio}
\setlength{\tabcolsep}{3.5pt}
\begin{tabular}{lrrrr}
\toprule
\STAY{} change & All three tasks & Melody & Harmony & Timbre \\
\midrule
EQ & 0.04 & 0.14 & 0.03 & 0.02 \\
Room echo & 0.93 & 3.39 & 0.97 & 0.50 \\
Hiss & 2.75 & 11.98 & 2.76 & 1.28 \\
Transposition & 0.65 & 2.11 & 1.18 & -- \\
\bottomrule
\end{tabular}
\end{table}

\begin{table}[!htbp]
\centering
\small
\caption{Hiss sweep on all 1,000 clean test items (\%), processed as the hiss \STAY{} change of the rendered test. Left and middle: share of the 401 `same' pairs of melody, harmony and timbre answered `different' (0\% is correct), with hiss on clip B only and with the same hiss on both clips. Right: rhythm accuracy (250 items) with hiss on clip B. Base: Qwen2.5-Omni without training; R0: GRPO LoRA (Table~\ref{tab:grpo}); R1: GRPO with sound-only pairs (Appendix~\ref{app:trainbase}), which never sees hiss.}
\label{tab:hiss_sweep}
\setlength{\tabcolsep}{3.5pt}
\begin{tabular}{lrrrrrrrrr}
\toprule
& \multicolumn{3}{c}{Hiss on B: `same' pairs} & \multicolumn{3}{c}{Hiss on both: `same' pairs} & \multicolumn{3}{c}{Hiss on B: rhythm acc.} \\
\cmidrule(lr){2-4}\cmidrule(lr){5-7}\cmidrule(lr){8-10}
Hiss (dBFS) & Base & R0 & R1 & Base & R0 & R1 & Base & R0 & R1 \\
\midrule
none & 17.7 & 2.0 & 0.0 & 17.7 & 2.0 & 0.0 & 50.0 & 100.0 & 100.0 \\
$-80$ & 48.6 & 42.9 & 0.7 & 35.7 & 7.5 & 0.0 & 49.6 & 98.8 & 99.2 \\
$-70$ & 60.1 & 51.1 & 10.2 & 42.4 & 23.7 & 4.7 & 51.6 & 92.8 & 98.8 \\
$-60$ & 66.3 & 86.5 & 43.9 & 55.4 & 50.6 & 12.2 & 56.0 & 66.4 & 89.2 \\
$-50$ & 70.6 & 98.3 & 73.6 & 53.9 & 29.4 & 7.5 & 55.6 & 53.2 & 68.0 \\
$-40$ & 72.1 & 100.0 & 77.3 & 60.1 & 46.1 & 10.5 & 54.0 & 51.6 & 63.2 \\
$-30$ & 72.8 & 100.0 & 83.3 & 64.8 & 51.6 & 30.7 & 52.8 & 53.6 & 64.0 \\
\bottomrule
\end{tabular}
\end{table}

\begin{table}[!htbp]
\centering
\small
\caption{Gemini 2.5 Pro on a subset of the hiss sweep: the same 115 `same' pairs of melody, harmony and timbre at every level (\% answered `different' or without an answer letter; 0\% is correct). No letter: replies without an answer letter among them, counted as errors. $p$: exact McNemar test against the same pairs without hiss (the run of Table~\ref{tab:base}). The two Qwen2.5-Omni columns are on exactly the same pairs.}
\label{tab:hiss_gemini}
\setlength{\tabcolsep}{3.5pt}
\begin{tabular}{lrrrrr}
\toprule
Hiss (dBFS) & Gemini 2.5 Pro & no letter & $p$ vs none & Omni, GRPO & Omni, base \\
\midrule
none & 2.6 &  & -- & 2.6 & 18.3 \\
$-70$, on B & 4.3 & 1 & 0.69 & 52.2 & 66.1 \\
$-50$, on B & 7.8 & 4 & 0.15 & 100.0 & 76.5 \\
$-30$, on B & 40.0 & 2 & $<10^{-10}$ & 100.0 & 76.5 \\
$-70$, on both & 4.3 & 4 & -- & 20.0 & 43.5 \\
$-50$, on both & 0.9 & 0 & -- & 28.7 & 56.5 \\
$-30$, on both & 9.6 & 3 & -- & 51.3 & 69.6 \\
\bottomrule
\end{tabular}
\end{table}

\FloatBarrier
\section{Paired metrics and per-task significance}
\label{app:paired}

Section~\ref{sec:flipstay} notes that FFR also counts errors a model already makes on the clean pair. Table~\ref{tab:paired} recomputes the paired metrics for every model and run of Tables~\ref{tab:base} and~\ref{tab:grpo}.

The difference matters for Gemini 2.5 Pro. It makes 209 errors on the 750 \STAY{} items of melody, harmony and timbre, and 155 of them are `same' answers to pairs that do differ (Table~\ref{tab:app-api-split}). These are missed musical changes, the same kind of error it makes on 41.0\% of clean pairs that differ, not reactions to the sound. If we count only items whose clean version is right, its false-flip rate falls from 26.9\% to 20.8\%. The same step barely moves the rate of the trained Qwen2.5-Omni (26.1\% to 25.7\%). Its false flips are new errors on items it answers correctly without the sound change. It changes its answer on 97.1\% of \FLIP{} versions, as it should, but also on 25.7\% of \STAY{} versions. Split by the correct answer, it is wrong on 54.4\% of \STAY{} pairs whose answer is `same' and on 0.3\% of those whose answer is `different’. Gemini 2.5 Pro is wrong on 13.5\% and 44.4\%. Counting Gemini 2.5 Pro's replies without a letter as errors moves its row by at most 1.1 points.

Table~\ref{tab:mcnemar} tests each task before and after GRPO. Of the 32 pairs of run and task, 22 are significant, and all of them are gains. Rhythm improves only for Qwen2.5-Omni (both runs) and for Audio Flamingo 3 with full fine-tuning, as Section~\ref{sec:res-grpo} states. We make no claim about the other changes on rhythm.

\begin{table}[!htbp]
\centering
\footnotesize
\caption{Paired metrics on the rendered test, recomputed from the runs of Tables~\ref{tab:base} and~\ref{tab:grpo} (\%). FFR$_{\text{same}}$ / FFR$_{\text{diff}}$: false-flip rate on \STAY{} pairs whose answer is `same' (401) / `different' (349), melody, harmony and timbre. FFR$_c$ = P(\STAY{} wrong $\mid$ clean right), MR$_c$ = P(\FLIP{} wrong $\mid$ clean right). $\Delta$\STAY{} / $\Delta$\FLIP{}: share of items whose answer changes from the clean to the \STAY{} / \FLIP{} version (ideal 0\% / 100\%). Item: clean, \FLIP{} and \STAY{} versions all right (guessing: 12.5\%). Commercial models: `as A' scores a reply without an answer letter as `A', as in Table~\ref{tab:base}; `no letter = error' counts it as wrong. Such replies: GPT-Audio-mini 1385, GPT-Audio-1.5 241, Gemini 2.5 Pro 21, Gemini 3.8 Flash 0 of 3,000. Open models always give a letter. $^\dagger$Step 500.}
\label{tab:paired}
\setlength{\tabcolsep}{3.5pt}
\resizebox{\linewidth}{!}{%
\begin{tabular}{llrrrrrrrrr}
\toprule
Model & GRPO / scoring & FFR & FFR$_{\text{same}}$ & FFR$_{\text{diff}}$ & FFR$_c$ & MR & MR$_c$ & $\Delta$\STAY{} & $\Delta$\FLIP{} & Item \\
\midrule
Qwen2-Audio & none & 50.9 & 97.3 & 0.3 & 15.0 & 47.7 & 85.4 & 16.1 & 14.3 & 3.3 \\
 & LoRA & 42.7 & 77.6 & 1.1 & 30.5 & 29.7 & 34.3 & 30.4 & 51.4 & 28.1 \\
 & full & 45.4 & 82.0 & 0.9 & 29.3 & 31.7 & 39.1 & 27.2 & 46.1 & 24.4 \\
Qwen2.5-Omni & none & 41.0 & 40.4 & 40.4 & 40.4 & 36.7 & 42.6 & 47.2 & 45.7 & 21.0 \\
 & LoRA & 26.1 & 54.4 & 0.3 & 25.7 & 1.9 & 1.9 & 25.7 & 97.1 & 72.2 \\
 & full & 28.9 & 60.8 & 0.3 & 26.1 & 4.2 & 4.4 & 25.1 & 92.0 & 67.1 \\
Audio Flamingo 3 & none & 47.6 & 5.7 & 93.7 & 17.8 & 49.6 & 79.9 & 18.5 & 18.1 & 5.8 \\
 & LoRA & 41.4 & 60.3 & 11.2 & 31.7 & 29.7 & 33.0 & 32.9 & 53.6 & 31.8 \\
 & full$^\dagger$ & 36.7 & 64.3 & 3.2 & 28.3 & 22.5 & 23.6 & 28.9 & 64.5 & 41.9 \\
Phi-4-multimodal & none & 48.3 & 5.5 & 94.3 & 19.7 & 53.5 & 83.6 & 20.0 & 18.4 & 3.7 \\
 &  LoRA & 48.4 & 1.0 & 99.7 & 14.3 & 53.4 & 88.1 & 14.3 & 13.5 & 3.0 \\
 & full & 39.3 & 24.7 & 46.4 & 30.8 & 42.9 & 48.3 & 36.8 & 45.0 & 22.6 \\
Kimi-Audio & none & 45.7 & 8.2 & 88.3 & 22.6 & 52.3 & 76.5 & 25.4 & 24.4 & 7.5 \\
MiMo-Audio & none & 46.1 & 5.2 & 92.6 & 18.1 & 49.4 & 79.6 & 17.8 & 15.9 & 5.8 \\
Fun-Audio-Chat & none & 46.1 & 34.9 & 56.7 & 34.3 & 46.6 & 60.4 & 37.3 & 35.6 & 13.0 \\
Step-Audio 2 & none & 45.5 & 34.9 & 55.6 & 33.3 & 49.1 & 61.0 & 36.3 & 37.1 & 11.9 \\
GPT-Audio-mini & as `A' & 50.3 & 28.7 & 73.4 & 41.1 & 50.6 & 64.0 & 40.6 & 36.3 & 9.1 \\
 & no letter = error & 73.5 & 60.3 & 73.4 & 63.4 & 72.2 & 67.7 & 33.4 & 63.1 & 2.8 \\
GPT-Audio-1.5 & as `A' & 50.1 & 89.3 & 4.9 & 20.5 & 45.8 & 74.0 & 22.0 & 23.1 & 6.5 \\
 & no letter = error & 52.3 & 92.8 & 4.9 & 16.9 & 51.3 & 85.7 & 19.4 & 20.8 & 3.2 \\
Gemini 2.5 Pro & as `A' & 26.9 & 13.5 & 44.4 & 20.8 & 24.7 & 28.5 & 26.9 & 58.7 & 43.6 \\
 & no letter = error & 27.4 & 14.2 & 44.4 & 21.4 & 25.2 & 29.1 & 27.6 & 58.2 & 42.5 \\
Gemini 3.8 Flash & as `A' & 39.1 & 52.1 & 22.9 & 32.8 & 33.0 & 31.7 & 38.0 & 57.3 & 30.1 \\
 & no letter = error & 39.1 & 52.1 & 22.9 & 32.8 & 33.0 & 31.7 & 38.0 & 57.3 & 30.1 \\
\bottomrule
\end{tabular}}
\end{table}

\begin{table}[!htbp]
\centering
\footnotesize
\caption{Per-task clean accuracy before $\to$ after GRPO (\%, 250 items per cell). $^{*}$: significant in an exact McNemar test, Holm-corrected over the 32 run and task pairs ($p_{\text{Holm}} < 0.05$). Of the 32, 22 are significant, all of them gains. $^\dagger$Step 500.}
\label{tab:mcnemar}
\setlength{\tabcolsep}{3.5pt}
\begin{tabular}{llrrrr}
\toprule
Model & GRPO & Mel. & Har. & Tim. & Rhy. \\
\midrule
Qwen2-Audio & LoRA & 46.0$\to$70.8$^{*}$ & 50.4$\to$60.8$^{*}$ & 43.2$\to$92.4$^{*}$ & 49.2$\to$51.6 \\
 & full & 46.0$\to$70.4$^{*}$ & 50.4$\to$61.6$^{*}$ & 43.2$\to$80.8$^{*}$ & 49.2$\to$52.0 \\
Qwen2.5-Omni & LoRA & 68.8$\to$99.6$^{*}$ & 59.2$\to$96.8$^{*}$ & 66.8$\to$99.6$^{*}$ & 50.0$\to$100.0$^{*}$ \\
 & full & 68.8$\to$100.0$^{*}$ & 59.2$\to$85.2$^{*}$ & 66.8$\to$100.0$^{*}$ & 50.0$\to$99.6$^{*}$ \\
Audio Flamingo 3 & LoRA & 51.2$\to$57.2 & 48.8$\to$74.8$^{*}$ & 58.0$\to$97.6$^{*}$ & 52.8$\to$50.8 \\
 & full$^\dagger$ & 51.2$\to$68.0$^{*}$ & 48.8$\to$80.8$^{*}$ & 58.0$\to$99.6$^{*}$ & 52.8$\to$69.2$^{*}$ \\
Phi-4-multimodal &  LoRA & 54.4$\to$52.0 & 47.2$\to$49.6 & 56.8$\to$56.8 & 50.8$\to$50.8 \\
 & full & 54.4$\to$55.6 & 47.2$\to$54.0$^{*}$ & 56.8$\to$88.8$^{*}$ & 50.8$\to$50.8 \\
\bottomrule
\end{tabular}
\end{table}

\FloatBarrier
\section{Edited clips played twice, and detection by size of change}
\label{app:twice}

A large harmony edit can produce an unusual chord. Moving the third by 300 cents gives root, minor second and fifth, or root, tritone and fifth. A model could then answer `different' because the edited clip sounds odd on its own, not because it compared the two clips. To test this, we built two kinds of `same' pair for every melody and harmony test item: the edited clip played twice, and the original clip played twice (Table~\ref{tab:twice}). The two clips of each pair are identical, so the right answer is always `same'. A model that reacts to oddness would call the edited pairs different.

The trained Qwen2.5-Omni shows no clear effect. It answers `same' about as often for the edited clip as for the original (harmony at 300 cents: 100.0\% and 96.4\%, and 90.4\% and 94.8\% over all harmony levels; melody at 700 cents: 100.0\% and 100.0\%). So the edited clip alone does not make it answer `different'. The control cannot show which cues the model uses when the two clips differ. Both clips here are identical, so a model can pass it by recognizing identical clips and still use cues in the edited clip on other pairs. Other trained models do show an effect. After GRPO, Qwen2-Audio and Audio Flamingo 3 answer `same' less often when the edited clip is played twice, on melody and on harmony. Audio Flamingo 3 (full) answers `same' to 27.7\% of the edited 300-cent chords against 72.3\% of the originals, and to 17.7\% against 46.8\% of the 700-cent melodies. Qwen2-Audio (LoRA) gives 0.0\% against 30.1\%, and 25.8\% against 50.0\%. For these two models, part of the gain may come from the edited clip itself. Gemini 2.5 Pro shows a smaller effect of the same kind on harmony, and none on melody. At 300 cents it answers `same' to 83.1\% of the edited chords against 96.4\% of the originals (69 against 80 of 83 items, exact McNemar $p = 0.007$). Over all levels the shares are 88.4\% and 98.4\% ($p < 10^{-5}$). On melodies at 700 cents it answers `same' to 96.8\% in both cases.

At 100 cents nothing sounds odd: a major triad becomes minor, or the reverse, and both are ordinary chords. Table~\ref{tab:detect} gives the detection rate by size of change. The trained Qwen2.5-Omni detects every 100-cent harmony change (84 of 84), and 98.4\% to 100.0\% of melody changes at every size. Before training it detected 25.8\% to 46.0\%. Because nothing sounds odd at 100 cents, the 84 harmony detections at that size cannot come from an odd-sounding chord. For larger edits, cues in the edited clip may also contribute.

\begin{table}[!htbp]
\centering
\footnotesize
\caption{Edited clip played twice (Ed.) and original clip played twice (Orig.): share answered `same' (\%; both clips are always identical, so 100\% is correct). One pair of each kind per melody and harmony test item (250 per task: 62 or 63 per melody level, 83 or 84 per harmony level). For Gemini 2.5 Pro a reply without an answer letter counts as not `same'. $^\dagger$Step 500.}
\label{tab:twice}
\setlength{\tabcolsep}{3.5pt}
\begin{tabular}{llrrrrrrrr}
\toprule
& & \multicolumn{2}{c}{Melody, all} & \multicolumn{2}{c}{Melody, 700 c} & \multicolumn{2}{c}{Harmony, all} & \multicolumn{2}{c}{Harmony, 300 c} \\
\cmidrule(lr){3-4}\cmidrule(lr){5-6}\cmidrule(lr){7-8}\cmidrule(lr){9-10}
Model & GRPO & Ed. & Orig. & Ed. & Orig. & Ed. & Orig. & Ed. & Orig. \\
\midrule
Qwen2-Audio & none & 0.0 & 0.0 & 0.0 & 0.0 & 0.0 & 0.0 & 0.0 & 0.0 \\
 & LoRA & 29.6 & 54.4 & 25.8 & 50.0 & 9.2 & 20.8 & 0.0 & 30.1 \\
 & full & 26.4 & 51.2 & 19.4 & 48.4 & 10.0 & 21.2 & 0.0 & 31.3 \\
Qwen2.5-Omni & none & 95.6 & 97.2 & 91.9 & 96.8 & 77.2 & 81.6 & 67.5 & 80.7 \\
 & LoRA & 100.0 & 100.0 & 100.0 & 100.0 & 90.4 & 94.8 & 100.0 & 96.4 \\
 & full & 99.6 & 100.0 & 98.4 & 100.0 & 64.8 & 68.0 & 86.7 & 69.9 \\
Audio Flamingo 3 & none & 87.2 & 88.0 & 85.5 & 85.5 & 99.2 & 96.8 & 100.0 & 97.6 \\
 & LoRA & 12.0 & 24.4 & 6.5 & 19.4 & 86.8 & 97.2 & 67.5 & 96.4 \\
 & full$^\dagger$ & 30.4 & 48.4 & 17.7 & 46.8 & 64.8 & 68.4 & 27.7 & 72.3 \\
Phi-4-multimodal & none & 98.8 & 98.4 & 100.0 & 96.8 & 85.2 & 87.6 & 86.7 & 86.7 \\
 &  LoRA & 97.2 & 97.6 & 96.8 & 95.2 & 100.0 & 100.0 & 100.0 & 100.0 \\
 & full & 65.2 & 71.2 & 66.1 & 75.8 & 82.0 & 85.2 & 77.1 & 81.9 \\
Kimi-Audio & none & 89.2 & 89.2 & 91.9 & 88.7 & 95.2 & 93.2 & 96.4 & 94.0 \\
MiMo-Audio & none & 100.0 & 100.0 & 100.0 & 100.0 & 92.0 & 96.0 & 90.4 & 92.8 \\
Fun-Audio-Chat & none & 55.6 & 58.0 & 48.4 & 58.1 & 59.2 & 61.2 & 47.0 & 59.0 \\
Step-Audio 2 & none & 84.0 & 84.0 & 79.0 & 85.5 & 40.8 & 45.6 & 34.9 & 49.4 \\
Gemini 2.5 Pro & none & 97.2 & 93.2 & 96.8 & 96.8 & 88.4 & 98.4 & 83.1 & 96.4 \\
\bottomrule
\end{tabular}
\end{table}

\begin{table}[!htbp]
\centering
\footnotesize
\caption{Detection of real changes by size (\% of clean and \FLIP{} pairs that differ answered `different'; 63, 63, 62, 62 melody and 84, 83, 83 harmony pairs per level). A reply without an answer letter is not a detection, under either scoring rule. 100-cent harmony changes turn a major triad into a minor one or the reverse, so both chords are ordinary. Figure~\ref{fig:api-items} pools the clean, \FLIP{} and \STAY{} versions, so its values for Gemini 2.5 Pro differ from these, which have no sound change. $^\dagger$Step 500.}
\label{tab:detect}
\setlength{\tabcolsep}{3.5pt}
\begin{tabular}{llrrrrrrr}
\toprule
& & \multicolumn{4}{c}{Melody, cents} & \multicolumn{3}{c}{Harmony, cents} \\
\cmidrule(lr){3-6}\cmidrule(lr){7-9}
Model & GRPO & 100 & 200 & 400 & 700 & 100 & 200 & 300 \\
\midrule
Qwen2-Audio & none & 100.0 & 100.0 & 100.0 & 100.0 & 100.0 & 100.0 & 100.0 \\
 & LoRA & 77.8 & 84.1 & 96.8 & 98.4 & 100.0 & 100.0 & 100.0 \\
 & full & 84.1 & 85.7 & 96.8 & 95.2 & 100.0 & 100.0 & 100.0 \\
Qwen2.5-Omni & none & 46.0 & 39.7 & 25.8 & 29.0 & 38.1 & 47.0 & 42.2 \\
 & LoRA & 98.4 & 100.0 & 98.4 & 100.0 & 100.0 & 100.0 & 100.0 \\
 & full & 100.0 & 100.0 & 100.0 & 100.0 & 100.0 & 100.0 & 100.0 \\
Audio Flamingo 3 & none & 15.9 & 15.9 & 9.7 & 9.7 & 3.6 & 0.0 & 2.4 \\
 & LoRA & 90.5 & 90.5 & 96.8 & 100.0 & 34.5 & 42.2 & 86.7 \\
 & full$^\dagger$ & 73.0 & 81.0 & 95.2 & 98.4 & 94.0 & 78.3 & 100.0 \\
Phi-4-multimodal & none & 0.0 & 0.0 & 1.6 & 0.0 & 9.5 & 12.0 & 10.8 \\
 &  LoRA & 0.0 & 1.6 & 0.0 & 3.2 & 0.0 & 0.0 & 0.0 \\
 & full & 34.9 & 36.5 & 32.3 & 30.6 & 19.0 & 10.8 & 28.9 \\
Kimi-Audio & none & 11.1 & 17.5 & 14.5 & 16.1 & 6.0 & 8.4 & 8.4 \\
MiMo-Audio & none & 0.0 & 0.0 & 0.0 & 0.0 & 11.9 & 13.3 & 10.8 \\
Fun-Audio-Chat & none & 49.2 & 47.6 & 38.7 & 41.9 & 48.8 & 48.2 & 44.6 \\
Step-Audio 2 & none & 15.9 & 12.7 & 14.5 & 11.3 & 53.6 & 61.4 & 56.6 \\
GPT-Audio-mini & none & 39.7 & 55.6 & 46.8 & 38.7 & 16.7 & 21.7 & 20.5 \\
GPT-Audio-1.5 & none & 100.0 & 95.2 & 100.0 & 100.0 & 78.6 & 62.7 & 89.2 \\
Gemini 2.5 Pro & none & 66.7 & 58.7 & 50.0 & 51.6 & 39.3 & 41.0 & 55.4 \\
Gemini 3.8 Flash & none & 71.4 & 84.1 & 80.6 & 79.0 & 64.3 & 61.4 & 61.4 \\
\bottomrule
\end{tabular}
\end{table}

\FloatBarrier
\section{Longer melodies}
\label{app:long}

Our melodies have only four notes. Do longer melodies change the picture? We built melodies of 12 and 24 notes (0.4\,s per note) with one or two changed notes, 200 base items per setting. They use the test pitch range and soundfont, and the same \FLIP/\STAY{} construction as the melody task. Table~\ref{tab:long} tests the existing checkpoints, without retraining.

The picture stays the same. The order of models stays close to the 4-note order (Spearman $\rho$ 0.61 to 0.79 over the 16 open configurations, $p < 0.02$ in every setting). The untrained open models other than Qwen2.5-Omni stay near chance (48.5\% to 56.8\% on \FLIP/\STAY{}). One changed note in a longer melody is harder for the trained Qwen2.5-Omni (LoRA: 69.0\% with 12 notes and 61.8\% with 24, against 83.6\% with 4). The drop comes mostly from misses (22.0\% and 38.0\%, against 0.4\% with 4 notes). This fits Appendix~\ref{app:encoder}: one changed note among 24 changes the clip less, and the trained model mostly reacts to how much the clip changes. False flips stay high (40.0\% and 38.5\%). Two changed notes are easier for most models that listen. We did not run Gemini 2.5 Pro: in a first batch of 75 answers on 12-note items, 25 replies had no answer letter, so its scores would measure missing answers rather than hearing.

\begin{table}[!htbp]
\centering
\footnotesize
\caption{Longer melodies (\%). 4 notes: melody \FLIP/\STAY{} accuracy from Tables~\ref{tab:app-grpo-task} and~\ref{tab:app-task-fs}. Other columns: new items with 12 or 24 notes (0.4\,s each) and one or two changed notes, 200 base items per setting (200 clean, 400 \FLIP/\STAY{}), built like the melody task (test pitch range and soundfont, same \FLIP/\STAY{} construction, \STAY{} changes EQ, room echo, hiss, transposition). FS: \FLIP/\STAY{} accuracy. Existing checkpoints, no retraining. n/r: not run (in a first batch of 75 answers on 12-note items, 25 Gemini 2.5 Pro replies had no answer letter). $^\dagger$Step 500.}
\label{tab:long}
\setlength{\tabcolsep}{3.5pt}
\resizebox{\linewidth}{!}{%
\begin{tabular}{llrrrrrrrrr}
\toprule
& & & \multicolumn{2}{c}{12 notes, 1 change} & \multicolumn{2}{c}{12 notes, 2 changes} & \multicolumn{2}{c}{24 notes, 1 change} & \multicolumn{2}{c}{24 notes, 2 changes} \\
\cmidrule(lr){4-5}\cmidrule(lr){6-7}\cmidrule(lr){8-9}\cmidrule(lr){10-11}
Model & GRPO & 4 notes & Clean & FS & Clean & FS & Clean & FS & Clean & FS \\
\midrule
Qwen2-Audio & none & 50.0 & 46.5 & 50.0 & 53.0 & 50.0 & 47.5 & 50.0 & 54.0 & 50.0 \\
 & LoRA & 64.4 & 65.0 & 55.5 & 71.5 & 65.8 & 61.5 & 57.3 & 67.5 & 64.5 \\
 & full & 61.8 & 56.0 & 51.3 & 64.0 & 60.5 & 55.5 & 56.3 & 60.5 & 60.5 \\
Qwen2.5-Omni & none & 60.0 & 58.5 & 49.8 & 60.5 & 63.8 & 54.0 & 53.5 & 64.5 & 66.0 \\
 & LoRA & 83.6 & 83.0 & 69.0 & 95.0 & 83.5 & 67.5 & 61.8 & 84.0 & 79.0 \\
 & full & 83.0 & 89.0 & 73.8 & 96.5 & 85.0 & 77.0 & 70.5 & 92.0 & 84.3 \\
Audio Flamingo 3 & none & 51.6 & 52.5 & 49.0 & 49.0 & 50.3 & 56.5 & 49.5 & 48.5 & 49.3 \\
 & LoRA & 54.4 & 56.0 & 53.8 & 59.0 & 57.3 & 56.0 & 52.3 & 62.5 & 54.0 \\
 & full$^\dagger$ & 58.6 & 65.5 & 57.3 & 68.5 & 60.5 & 60.0 & 60.0 & 64.0 & 60.0 \\
Phi-4-multimodal & none & 49.6 & 53.5 & 50.0 & 47.0 & 50.0 & 52.5 & 50.0 & 46.0 & 50.0 \\
 &  LoRA & 48.6 & 53.5 & 50.3 & 46.5 & 50.0 & 52.0 & 49.8 & 48.5 & 49.8 \\
 & full & 51.4 & 51.0 & 52.3 & 48.5 & 50.8 & 49.5 & 49.3 & 44.0 & 48.5 \\
Kimi-Audio & none & 52.8 & 55.0 & 48.5 & 48.5 & 49.5 & 51.0 & 52.5 & 50.0 & 51.8 \\
MiMo-Audio & none & 50.0 & 53.5 & 50.0 & 47.0 & 50.0 & 52.5 & 50.0 & 46.0 & 50.0 \\
Fun-Audio-Chat & none & 49.8 & 53.5 & 48.8 & 51.0 & 56.8 & 50.5 & 51.3 & 56.5 & 51.8 \\
Step-Audio 2 & none & 47.8 & 51.0 & 49.5 & 47.0 & 53.0 & 48.5 & 51.3 & 47.0 & 50.5 \\
Gemini 2.5 Pro & none & 73.2 & n/r & n/r & n/r & n/r & n/r & n/r & n/r & n/r \\
\multicolumn{3}{l}{Spearman $\rho$ with 4 notes (16 open configurations)} & \multicolumn{2}{c}{0.61 ($p=0.011$)} & \multicolumn{2}{c}{0.77 ($p<0.001$)} & \multicolumn{2}{c}{0.78 ($p<0.001$)} & \multicolumn{2}{c}{0.79 ($p<0.001$)} \\
\bottomrule
\end{tabular}}
\end{table}

\FloatBarrier
\section{Graded timbre and close instrument pairs}
\label{app:timbre}

Our timbre task uses four instruments from four different families, and the trained Qwen2.5-Omni scores 99.6\% on it. Does it also hear finer differences in timbre? Three new sub-tasks test this. Each has 250 base items (half `same' with byte-identical clips, half `different') and the \FLIP/\STAY{} construction of the rendered set. Both clips play the same 4-note motif from the test pitch range in a 2.5\,s window.

\paragraph{Brightness.} A band-limited sawtooth (additive synthesis) through a second-order Butterworth low-pass filter. The cutoff is 1\,kHz in one clip and 1/3, 2/3, 1 or 2 octaves higher in the other, with a 5\,ms attack. The \STAY{} changes are room echo and hiss only, since EQ changes brightness.

\paragraph{Attack.} The same synthesizer with a 2\,kHz cutoff. The attack is 5\,ms in one clip and 20, 50, 100 or 200\,ms in the other. Attack time and spectral centroid are two of the main dimensions found in timbre studies~\citep{mcadams1995perceptual}. Brightness and attack cover only a small part of what a synthesizer can change, so these two sub-tasks are a first step toward graded timbre tests, not a full test of timbre.

\paragraph{Close pairs.} Violin and viola, viola and cello, trumpet and trombone, acoustic and electric piano, and nylon and steel guitar, all from the test soundfont.

\paragraph{Results.} Without training, Qwen2.5-Omni answers `different' to 19\% to 47\% of changed pairs at every step, but also to 23.2\% of identical pairs (Table~\ref{tab:graded}). Its answers do not follow the size of the change. After GRPO, it never calls identical pairs different. But it also calls almost no graded change a different instrument: no brightness step up to one octave, 14.5\% at two octaves, and almost no attack step. Every timbre question asks whether the instrument is the same. These answers may therefore mean that the model judges a brighter or softer version of one synthesizer to be the same instrument. The close pairs avoid this ambiguity. Among them, the trained model hears acoustic against electric piano (62\%) and nylon against steel guitar (56\%), but rarely violin against viola (10\%), trumpet against trombone (4\%) or viola against cello (0\%) (Table~\ref{tab:close}). Its 99.6\% on the timbre task thus reflects telling instrument families apart, not finer timbre. Gemini 2.5 Pro (100 base items per sub-task) errs on one of 150 identical pairs, a reply without a letter. It hears 2 of 25 pairs at two octaves brighter, and none at any other step.

\begin{table}[!htbp]
\centering
\footnotesize
\caption{Graded synth timbre (\%). Brightness: cutoff of a 2nd-order low-pass on a band-limited sawtooth, 1\,kHz against 1/3 to 2 octaves higher. Attack: 5\,ms against 20 to 200\,ms. Step columns: share of changed pairs (clean and \FLIP{}) answered `different'. Ident.: share of identical clean pairs answered wrongly, pooled over brightness, attack and close pairs (0\% is correct). 250 base items per sub-task (62 to 64 changed pairs per step, 375 identical pairs); Gemini 2.5 Pro on 100 (25 or 26 changed pairs per step, 150 identical pairs), a reply without an answer letter counting as an error. $^\dagger$Step 500.}
\label{tab:graded}
\setlength{\tabcolsep}{3.5pt}
\resizebox{\linewidth}{!}{%
\begin{tabular}{llrrrrrrrrr}
\toprule
& & \multicolumn{4}{c}{Brightness, octaves} & \multicolumn{4}{c}{Attack, ms} & \\
\cmidrule(lr){3-6}\cmidrule(lr){7-10}
Model & GRPO & 1/3 & 2/3 & 1 & 2 & 20 & 50 & 100 & 200 & Ident. \\
\midrule
Qwen2-Audio & none & 100.0 & 100.0 & 100.0 & 100.0 & 100.0 & 100.0 & 100.0 & 100.0 & 97.9 \\
 & LoRA & 34.4 & 45.2 & 56.5 & 54.8 & 9.4 & 24.2 & 40.3 & 80.6 & 29.9 \\
 & full & 35.9 & 41.9 & 51.6 & 61.3 & 28.1 & 46.8 & 71.0 & 98.4 & 38.9 \\
Qwen2.5-Omni & none & 29.7 & 22.6 & 25.8 & 41.9 & 34.4 & 46.8 & 19.4 & 40.3 & 23.2 \\
 & LoRA & 0.0 & 0.0 & 0.0 & 14.5 & 0.0 & 0.0 & 0.0 & 0.0 & 0.0 \\
 & full & 0.0 & 0.0 & 1.6 & 19.4 & 0.0 & 0.0 & 0.0 & 3.2 & 0.0 \\
Audio Flamingo 3 & none & 1.6 & 0.0 & 0.0 & 1.6 & 1.6 & 0.0 & 4.8 & 11.3 & 1.9 \\
 & LoRA & 0.0 & 0.0 & 0.0 & 0.0 & 0.0 & 0.0 & 0.0 & 0.0 & 0.0 \\
 & full$^\dagger$ & 0.0 & 0.0 & 0.0 & 24.2 & 0.0 & 0.0 & 0.0 & 0.0 & 0.0 \\
Phi-4-multimodal & none & 0.0 & 0.0 & 0.0 & 0.0 & 0.0 & 0.0 & 0.0 & 0.0 & 0.0 \\
 &  LoRA & 0.0 & 0.0 & 0.0 & 0.0 & 0.0 & 0.0 & 0.0 & 0.0 & 0.0 \\
 & full & 0.0 & 11.3 & 27.4 & 48.4 & 0.0 & 0.0 & 6.5 & 4.8 & 1.1 \\
Kimi-Audio & none & 0.0 & 0.0 & 0.0 & 0.0 & 0.0 & 0.0 & 0.0 & 0.0 & 1.1 \\
MiMo-Audio & none & 0.0 & 0.0 & 0.0 & 0.0 & 0.0 & 0.0 & 0.0 & 0.0 & 0.0 \\
Fun-Audio-Chat & none & 6.3 & 4.8 & 6.5 & 16.1 & 4.7 & 1.6 & 4.8 & 1.6 & 6.7 \\
Step-Audio 2 & none & 28.1 & 33.9 & 22.6 & 25.8 & 26.6 & 24.2 & 22.6 & 37.1 & 24.5 \\
Gemini 2.5 Pro & none & 0.0 & 0.0 & 0.0 & 8.0 & 0.0 & 0.0 & 0.0 & 0.0 & 0.7 \\
\bottomrule
\end{tabular}}
\end{table}

\begin{table}[!htbp]
\centering
\footnotesize
\caption{Close instrument pairs from the test soundfont (\% of changed clean and \FLIP{} pairs answered `different', 50 per pair) and \FLIP/\STAY{} accuracy of each new sub-task (\%). \STAY{} changes: room echo and hiss for brightness (EQ changes brightness), room echo, EQ and hiss otherwise. Gemini 2.5 Pro: 100 base items per sub-task; `--': pair not in its subset. $^\dagger$Step 500.}
\label{tab:close}
\setlength{\tabcolsep}{3.5pt}
\begin{tabular}{llrrrrrrrr}
\toprule
& & \multicolumn{5}{c}{Close pairs: detected} & \multicolumn{3}{c}{\FLIP/\STAY{} accuracy} \\
\cmidrule(lr){3-7}\cmidrule(lr){8-10}
Model & GRPO & Vn/Va & Va/Vc & Tpt/Tbn & Pno/E.pno & Nyl/Steel & Bright. & Attack & Close \\
\midrule
Qwen2-Audio & none & 86.0 & 86.0 & 88.0 & 100.0 & 100.0 & 50.0 & 50.2 & 49.4 \\
 & LoRA & 100.0 & 96.0 & 96.0 & 100.0 & 92.0 & 54.6 & 52.8 & 79.2 \\
 & full & 100.0 & 94.0 & 100.0 & 100.0 & 82.0 & 53.6 & 54.2 & 71.6 \\
Qwen2.5-Omni & none & 42.0 & 24.0 & 20.0 & 50.0 & 30.0 & 53.2 & 58.4 & 55.0 \\
 & LoRA & 10.0 & 0.0 & 4.0 & 62.0 & 56.0 & 54.4 & 51.4 & 60.0 \\
 & full & 14.0 & 0.0 & 4.0 & 82.0 & 64.0 & 55.8 & 52.2 & 61.6 \\
Audio Flamingo 3 & none & 0.0 & 0.0 & 0.0 & 0.0 & 0.0 & 49.4 & 52.0 & 50.2 \\
 & LoRA & 56.0 & 4.0 & 88.0 & 84.0 & 76.0 & 50.2 & 50.4 & 78.4 \\
 & full$^\dagger$ & 58.0 & 0.0 & 64.0 & 90.0 & 96.0 & 55.8 & 50.6 & 77.8 \\
Phi-4-multimodal & none & 0.0 & 0.0 & 0.0 & 0.0 & 0.0 & 50.0 & 50.0 & 50.0 \\
 &  LoRA & 0.0 & 0.0 & 0.0 & 0.0 & 0.0 & 50.4 & 50.0 & 49.8 \\
 & full & 26.0 & 68.0 & 18.0 & 48.0 & 22.0 & 59.2 & 52.0 & 64.8 \\
Kimi-Audio & none & 2.0 & 0.0 & 0.0 & 2.0 & 2.0 & 50.4 & 50.0 & 49.8 \\
MiMo-Audio & none & 0.0 & 0.0 & 0.0 & 2.0 & 0.0 & 50.0 & 50.0 & 50.8 \\
Fun-Audio-Chat & none & 2.0 & 2.0 & 0.0 & 10.0 & 22.0 & 52.6 & 48.0 & 50.8 \\
Step-Audio 2 & none & 14.0 & 22.0 & 62.0 & 22.0 & 30.0 & 47.8 & 50.8 & 51.6 \\
Gemini 2.5 Pro & none & 22.0 & -- & 36.0 & 4.0 & -- & 50.5 & 49.0 & 52.5 \\
\bottomrule
\end{tabular}
\end{table}

\clearpage
\FloatBarrier
\section{Real recordings}
\label{app:real}

\paragraph{Construction.}
The real test set is the second test set of MusicListenBench (Section~\ref{sec:splits}). It measures listening on real recordings, with real rooms, microphones and players, and is used for evaluation only. We build it like the rendered set: the same four tasks, clip lengths, 1\,s gap, loudness ($-23$\,LUFS per clip), questions and letter rule. The \STAY{} changes are also the same (EQ, room echo, hiss and, for melody and harmony, transposition by 0 to 200 cents), made with the code of our generator. Each task has 150 base items, and each base item has a \FLIP{} and a \STAY{} version (600 clean and 1{,}200 \FLIP/\STAY{} items in all).
\begin{itemize}
\item \textbf{Melody.} Phrases of 4 to 6 notes (2.5\,s) from the microphone recordings of the solo excerpts in GuitarSet~\citep{xi2018guitarset}: acoustic guitar, six players. The note annotations of each string confirm that one note plays at a time. In `different' pairs, the second note is pitch-shifted by 100, 200, 400 or 700 cents with Rubber Band. In `same' pairs, the same note passes through the shifter at 0 cents.
\item \textbf{Harmony.} Windows of 2\,s from the accompaniment (comping) excerpts of GuitarSet. These are recorded with a hexaphonic pickup, which gives one cleaned track per string. Each window starts at a strum, every annotated note in it belongs to one major or minor triad, and no earlier note sounds into it. We shift only the strings that play the third, by 100, 200 or 300 cents (a major third down, a minor third up, as in our generator). The other strings stay untouched. A `different' pair is thus a real strummed chord against the same strum with its third moved. In `same' pairs, the strings of the third pass through the shifter at 0 cents.
\item \textbf{Timbre.} Excerpts of 2.5\,s from Medley-solos-DB~\citep{lostanlen2018medley}: clarinet, distorted electric guitar, flute, piano, tenor saxophone, trumpet and violin (we leave out the singer class). The two clips of every pair come from different recordings, in `same' pairs as well as `different' ones. So the room or the microphone cannot give the answer away.
\item \textbf{Rhythm.} Windows of 6\,s from MUSDB18-HQ drum stems~\citep{rafii2019musdb18hq}, with an onset in every second. We time-stretch them by 1.10 to 1.50 with Rubber Band and cut them to 4.0\,s. The slower clip also passes through Rubber Band, at 1.0, so both clips are processed alike. Both clips come from the same window.
\end{itemize}

\paragraph{Label checks.}
The labels come from the annotations of each dataset, and we checked them on the audio. In every `different' melody pair, the edited note lies within 20 cents of its target (median deviation 0 cents, measured with pYIN). In `same' pairs it is unchanged. For harmony, the share of chroma energy at the pitch class of the third falls to a median of 43\%, 52\% and 22\% of its original value at 100, 200 and 300 cents. In `same' pairs it stays at 100\%. For rhythm, the ratio of onset counts between the faster and the slower clip has a median of 1.08, 1.19, 1.31 and 1.40 at stretch factors 1.10, 1.20, 1.35 and 1.50.

\paragraph{Scoring.}
As in Appendix~\ref{app:slakh}, a reply without an answer letter counts as wrong. On the \FLIP/\STAY{} items, Gemini 2.5 Pro gives such replies on 3.7\% of items (five of them are calls that timed out three times), and scores 76.6\% on the rest. GPT-Audio-1.5 gives them on 6.1\% and GPT-Audio-mini on 45.5\% of items, mostly refusals to process the audio. On the items they answer, these two score 49.7\% and 51.5\%. We ran Gemini 3.8 Flash through Google's batch interface, with the same request as for the rendered test set (Appendix~\ref{app:models}).

\paragraph{Results.}
Without training (Table~\ref{tab:real_untrained}), the open models are close to chance on \FLIP/\STAY{} (48.3\% to 55.5\%). Only Audio Flamingo 3 (55.5\%) and Fun-Audio-Chat (54.6\%) are significantly above it. Gemini 2.5 Pro is again the best model (73.8\%, against 73.7\% on the rendered test and 75.3\% on Slakh2100). Qwen2.5-Omni falls to chance (52.1\%, against 61.2\% on the rendered test). Over all 12 untrained models, the ranking agrees with both other tests. The eight open models alone all lie within 7.2 points of each other, near chance, so their order among themselves carries little information (Spearman $\rho = 0.32$ with the rendered test). Melody is the hardest task: no open model or trained run exceeds 55.3\% clean accuracy on it, against 72.7\% for Gemini 2.5 Pro.

Of the runs trained on our rendered pairs (Table~\ref{tab:real_trained}), only those of Qwen2.5-Omni carry over. With LoRA, it rises from 52.1\% to 62.8\% on \FLIP/\STAY{} (exact McNemar test on the same 1{,}200 items, $p < 0.001$), and with full fine-tuning to 61.7\%. Qwen2-Audio, Audio Flamingo 3 and Phi-4-multimodal stay within 6 points of chance. Adding sound-only pairs to training (R1) raises Qwen2.5-Omni to 70.1\% ($p < 0.001$ against R0). SFT on clean pairs (R2) gives the same as GRPO (62.9\%). Trained on Slakh2100 pairs, Qwen2.5-Omni reaches 75.0\% with LoRA, level with Gemini 2.5 Pro ($p = 0.50$). Here the two error rates fall by similar amounts (GRPO with LoRA on rendered pairs: misses from 46.5\% to 37.0\%, false flips from 49.3\% to 37.3\%), and, as on the other two test sets, false flips stay at or above misses in every trained Qwen2.5-Omni run. The check below looks at false flips on natural differences between two recordings.

\paragraph{Second takes.}
Real recordings allow a check that rendered audio cannot give. In the GuitarSet accompaniment excerpts, players strum the same chord several times. Two strums of the same chord always differ a little, but a listener should call them the same (Table~\ref{tab:real_take}). Every run of Qwen2.5-Omni trained on clean pairs detects almost every moved third (37 to 40 of 41). But each run also calls 27 to 35 of the 42 same-chord pairs different. After training on clean pairs, any difference between two recordings tends to count as a change. With sound-only pairs in training (R1), the model calls only 7 of 42 same-chord pairs different (exact McNemar test against R0, $p < 0.001$), and detects 26 of 41 moved thirds. Most untrained models sit at one extreme: Qwen2-Audio answers `different' to all 83 pairs, and Phi-4-multimodal answers `same' to 82. Gemini 2.5 Pro errs the other way from the trained model: it calls 37 of 42 same-chord pairs the same, but detects only 12 of 41 moved thirds. Gemini 3.8 Flash calls 27 of 42 same-chord pairs different.

\begin{table}[!htbp]
\centering
\small
\caption{Real-recordings test without training (\%; Tables~\ref{tab:base} and~\ref{tab:app-task-fs} on this set). Clean: 150 items per task (600 in All). \FLIP/\STAY{}: 300 items per task (1,200 in $\accfs$). A reply without an answer letter counts as an error; such replies on \FLIP/\STAY{} items: GPT-Audio-mini 45.5\%, GPT-Audio-1.5 6.1\%, Gemini 2.5 Pro 3.7\%, Gemini 3.8 Flash 0.0\%. Spearman $\rho$ of \FLIP/\STAY{} accuracy over the 12 models: 0.80 ($p=0.002$) with the rendered test, 0.91 ($p<0.001$) with the Slakh2100 test.}
\label{tab:real_untrained}
\setlength{\tabcolsep}{3.5pt}
\resizebox{\linewidth}{!}{%
\begin{tabular}{lrrrrrrrrrrrr}
\toprule
 & \multicolumn{5}{c}{Clean accuracy} & \multicolumn{4}{c}{\FLIP/\STAY{} accuracy} & & & \\
\cmidrule(lr){2-6}\cmidrule(lr){7-10}
Model & Mel. & Har. & Tim. & Rhy. & All & Mel. & Har. & Tim. & Rhy. & $\accfs$ & MR & FFR \\
\midrule
Qwen2-Audio & 49.3 & 50.0 & 50.0 & 44.0 & 48.3 & 50.0 & 50.0 & 50.0 & 48.3 & 49.6 & 48.5 & 52.3 \\
Qwen2.5-Omni & 44.7 & 52.7 & 62.7 & 48.7 & 52.2 & 47.7 & 49.7 & 58.7 & 52.3 & 52.1 & 46.5 & 49.3 \\
Audio Flamingo 3 & 55.3 & 56.0 & 64.0 & 52.7 & 57.0 & 54.0 & 53.0 & 65.3 & 49.7 & 55.5 & 44.5 & 44.5 \\
Phi-4-multimodal & 50.7 & 49.3 & 50.0 & 56.0 & 51.5 & 50.0 & 52.3 & 50.0 & 51.7 & 51.0 & 50.2 & 47.8 \\
Kimi-Audio & 50.0 & 48.0 & 50.0 & 52.0 & 50.0 & 48.3 & 52.3 & 50.7 & 50.7 & 50.5 & 50.3 & 48.7 \\
MiMo-Audio & 50.7 & 46.0 & 54.0 & 60.0 & 52.7 & 50.0 & 51.3 & 52.7 & 54.3 & 52.1 & 49.3 & 46.5 \\
Fun-Audio-Chat & 48.0 & 58.0 & 56.0 & 56.0 & 54.5 & 49.3 & 53.0 & 57.7 & 58.3 & 54.6 & 45.2 & 45.7 \\
Step-Audio 2 & 49.3 & 53.3 & 63.3 & 58.0 & 56.0 & 47.7 & 47.0 & 54.7 & 43.7 & 48.3 & 53.8 & 49.7 \\
\midrule
GPT-Audio-mini & 48.0 & 26.7 & 36.0 & 4.7 & 28.8 & 45.3 & 30.7 & 32.3 & 4.0 & 28.1 & 71.5 & 72.3 \\
GPT-Audio-1.5 & 49.3 & 42.0 & 54.7 & 38.0 & 46.0 & 47.3 & 43.7 & 53.0 & 42.7 & 46.7 & 53.2 & 53.5 \\
Gemini 2.5 Pro & 72.7 & 64.0 & 76.0 & 88.7 & 75.3 & 67.3 & 60.3 & 75.7 & 91.7 & 73.8 & 25.3 & 27.2 \\
Gemini 3.8 Flash & 65.3 & 52.0 & 65.3 & 60.7 & 60.8 & 54.3 & 52.0 & 68.0 & 54.7 & 57.3 & 42.0 & 43.5 \\
Always `A' &  &  &  &  & 51.7 &  &  &  &  & 50.4 & 51.5 & 47.7 \\
\bottomrule
\end{tabular}}
\end{table}

\begin{table}[!htbp]
\centering
\footnotesize
\caption{All open-model configurations on the real-recordings test (\%; no training on real recordings). Mel.\ to All: clean accuracy. EQ to Tr.: false-flip rate per \STAY{} change (Tr.: transposition). R1: GRPO on clean + sound-only rendered pairs; R2: SFT on clean rendered pairs. $^\dagger$Step 500.}
\label{tab:real_trained}
\setlength{\tabcolsep}{3.5pt}
\resizebox{\linewidth}{!}{%
\begin{tabular}{llrrrrrrrrrrrr}
\toprule
Model & Training & Mel. & Har. & Tim. & Rhy. & All & $\accfs$ & MR & FFR & EQ & Room & Hiss & Tr. \\
\midrule
\multicolumn{14}{l}{\emph{No training}} \\
Qwen2-Audio & none & 49.3 & 50.0 & 50.0 & 44.0 & 48.3 & 49.6 & 48.5 & 52.3 & 74.3 & 48.7 & 33.5 & 49.3 \\
Qwen2.5-Omni & none & 44.7 & 52.7 & 62.7 & 48.7 & 52.2 & 52.1 & 46.5 & 49.3 & 62.1 & 43.4 & 40.3 & 48.0 \\
Audio Flamingo 3 & none & 55.3 & 56.0 & 64.0 & 52.7 & 57.0 & 55.5 & 44.5 & 44.5 & 34.5 & 50.4 & 52.4 & 41.3 \\
Phi-4-multimodal & none & 50.7 & 49.3 & 50.0 & 56.0 & 51.5 & 51.0 & 50.2 & 47.8 & 25.2 & 52.2 & 66.0 & 53.3 \\
\multicolumn{14}{l}{\emph{GRPO on the rendered split (Table~\ref{tab:grpo})}} \\
Qwen2-Audio & LoRA & 49.3 & 52.7 & 50.0 & 45.3 & 49.3 & 49.6 & 48.5 & 52.3 & 75.2 & 48.7 & 32.5 & 49.3 \\
 & full & 49.3 & 52.0 & 50.0 & 44.0 & 48.8 & 50.1 & 48.0 & 51.8 & 73.8 & 49.6 & 32.0 & 49.3 \\
Qwen2.5-Omni & LoRA & 50.7 & 76.7 & 63.3 & 67.3 & 64.5 & 62.8 & 37.0 & 37.3 & 46.6 & 43.4 & 20.9 & 48.0 \\
 & full & 48.0 & 76.0 & 64.0 & 66.7 & 63.7 & 61.7 & 38.0 & 38.7 & 46.6 & 46.9 & 22.3 & 49.3 \\
Audio Flamingo 3 & LoRA & 50.0 & 60.7 & 60.0 & 56.0 & 56.7 & 53.9 & 47.8 & 44.3 & 53.9 & 49.6 & 29.6 & 50.7 \\
 & full$^\dagger$ & 48.7 & 57.3 & 67.3 & 46.0 & 54.8 & 55.7 & 43.3 & 45.3 & 51.0 & 46.0 & 35.0 & 57.3 \\
Phi-4-multimodal &  LoRA & 52.7 & 50.7 & 50.7 & 56.0 & 52.5 & 50.9 & 50.7 & 47.5 & 26.2 & 50.4 & 65.5 & 52.0 \\
 & full & 50.7 & 47.3 & 52.7 & 57.3 & 52.0 & 51.7 & 49.8 & 46.8 & 31.1 & 52.2 & 55.8 & 57.3 \\
\multicolumn{14}{l}{\emph{GRPO on the Slakh2100 split (Appendix~\ref{app:slakh})}} \\
Qwen2-Audio & LoRA & 46.0 & 52.7 & 50.0 & 44.0 & 48.2 & 50.0 & 48.7 & 51.3 & 71.8 & 49.6 & 32.0 & 50.7 \\
 & full & 44.0 & 49.3 & 51.3 & 44.0 & 47.2 & 49.6 & 49.0 & 51.8 & 71.8 & 48.7 & 34.0 & 50.7 \\
Qwen2.5-Omni & LoRA & 54.0 & 76.7 & 84.7 & 84.7 & 75.0 & 75.0 & 23.0 & 27.0 & 22.8 & 38.1 & 18.4 & 45.3 \\
 & full & 51.3 & 72.0 & 82.7 & 70.0 & 69.0 & 68.8 & 31.0 & 31.5 & 27.2 & 42.5 & 23.8 & 48.0 \\
Audio Flamingo 3 & LoRA & 45.3 & 58.7 & 68.0 & 46.7 & 54.7 & 56.8 & 41.7 & 44.7 & 46.6 & 50.4 & 35.9 & 54.7 \\
 & full$^\dagger$ & 50.0 & 50.7 & 67.3 & 54.0 & 55.5 & 56.9 & 43.0 & 43.2 & 42.7 & 51.3 & 36.4 & 50.7 \\
Phi-4-multimodal &  LoRA & 52.7 & 50.7 & 50.7 & 56.0 & 52.5 & 50.9 & 50.7 & 47.5 & 26.2 & 50.4 & 65.5 & 52.0 \\
 & full & 42.7 & 46.7 & 48.7 & 56.0 & 48.5 & 51.9 & 48.3 & 47.8 & 27.7 & 46.0 & 65.5 & 57.3 \\
\multicolumn{14}{l}{\emph{Baselines of Appendix~\ref{app:trainbase} (rendered pairs)}} \\
Qwen2.5-Omni & R1 & 54.0 & 70.7 & 84.7 & 68.7 & 69.5 & 70.1 & 28.2 & 31.7 & 30.6 & 38.9 & 22.8 & 48.0 \\
 & R2 & 49.3 & 74.0 & 61.3 & 65.3 & 62.5 & 62.9 & 35.8 & 38.3 & 48.5 & 43.4 & 21.4 & 49.3 \\
\bottomrule
\end{tabular}}
\end{table}

\begin{table}[!htbp]
\centering
\footnotesize
\caption{Second takes: 83 pairs of two real strums of the same chord voicing in one GuitarSet excerpt, at least 2\,s apart. `Same' (42): take~1 against take~2. `Different' (41): take~1 against take~2 with the strings that play the third shifted by 100, 200 or 300 cents. Acc.: accuracy (\%). Same: `different': same-chord pairs answered `different' (0 is correct). Changed: detected: moved thirds answered `different'. 100 to 300: share detected per size (\%; 14, 14 and 13 pairs). Training: GRPO (LoRA or full) on the rendered or the Slakh2100 split; R1: GRPO with sound-only pairs; R2: SFT on clean pairs. A reply without an answer letter counts as an error.}
\label{tab:real_take}
\setlength{\tabcolsep}{3.5pt}
\begin{tabular}{llrrrrrr}
\toprule
Model & Training & Acc. & Same: `different' & Changed: detected & 100 & 200 & 300 \\
\midrule
Gemini 2.5 Pro & none & 59.0 & 5 of 42 & 12 of 41 & 35.7 & 35.7 & 15.4 \\
Gemini 3.8 Flash & none & 44.6 & 27 of 42 & 22 of 41 & 35.7 & 64.3 & 61.5 \\
Qwen2.5-Omni & none & 62.7 & 18 of 42 & 28 of 41 & 64.3 & 71.4 & 69.2 \\
 & LoRA, rendered & 63.9 & 27 of 42 & 38 of 41 & 100.0 & 78.6 & 100.0 \\
 & full, rendered & 59.0 & 32 of 42 & 39 of 41 & 100.0 & 85.7 & 100.0 \\
 & LoRA, Slakh2100 & 63.9 & 29 of 42 & 40 of 41 & 100.0 & 92.9 & 100.0 \\
 & full, Slakh2100 & 62.7 & 27 of 42 & 37 of 41 & 100.0 & 71.4 & 100.0 \\
 & R1, rendered & 73.5 & 7 of 42 & 26 of 41 & 64.3 & 57.1 & 69.2 \\
 & R2 (SFT), rendered & 55.4 & 35 of 42 & 39 of 41 & 100.0 & 85.7 & 100.0 \\
Qwen2-Audio & none & 49.4 & 42 of 42 & 41 of 41 & 100.0 & 100.0 & 100.0 \\
Audio Flamingo 3 & none & 54.2 & 6 of 42 & 9 of 41 & 14.3 & 14.3 & 38.5 \\
Phi-4-multimodal & none & 49.4 & 1 of 42 & 0 of 41 & 0.0 & 0.0 & 0.0 \\
Kimi-Audio & none & 54.2 & 6 of 42 & 9 of 41 & 21.4 & 21.4 & 23.1 \\
MiMo-Audio & none & 49.4 & 6 of 42 & 5 of 41 & 21.4 & 7.1 & 7.7 \\
Fun-Audio-Chat & none & 63.9 & 18 of 42 & 29 of 41 & 71.4 & 64.3 & 76.9 \\
Step-Audio 2 & none & 48.2 & 30 of 42 & 28 of 41 & 57.1 & 78.6 & 69.2 \\
\bottomrule
\end{tabular}
\end{table}

\endgroup

\end{document}